\documentclass[10pt,journal,compsoc]{IEEEtran}
\usepackage{amsmath,amsfonts}
\usepackage{algorithmic}
\usepackage{balance}
\usepackage{algorithm,enumitem}
\usepackage{array}
\usepackage{textcomp}
\usepackage{stfloats}
\usepackage{url}
\usepackage{verbatim}
\usepackage{graphicx}
\usepackage{subcaption}
\usepackage{caption}
\usepackage{color}
\usepackage{multirow}
\usepackage{adjustbox}
\usepackage{arydshln}
\usepackage{forest}
\usepackage{booktabs}
\usepackage{makecell}
\definecolor{urlcolor}{RGB}{55,118,185}
\usepackage[bookmarksnumbered,unicode,colorlinks,linkcolor=urlcolor,citecolor=urlcolor,urlcolor=urlcolor]{hyperref}

\newcommand{\remove}[1]{}

\newcommand{\junwei}[1]{\textcolor{black}{#1}}
\newcommand{\junweim}[1]{\textcolor{black}{#1}}

\newcommand{\kaixin}[1]{\textcolor[RGB]{255,0,0}{Kaixin: #1}}

\newcommand{\et}{\textit{et al.}}
\newcommand{\etc}{etc.}
\newcommand{\ie}{\textit{i.e.,}}
\newcommand{\eg}{\textit{e.g.,}}
\newcommand{\hhline}{\hline\noalign{\hrule height 0.5pt}}

\newcommand{\tabincell}[2]{\begin{tabular}{@{}#1@{}}#2\end{tabular}}

\newcommand{\headt}[1]{\textbf{#1}}

\newcommand{\headfi}[1]{\textit{#1}}

\newcommand{\highlight}[1]{\textit{#1}}

\newcommand{\papern}{124}

\definecolor{verylightgray}{RGB}{240,240,240} 
\definecolor{darkblue}{RGB}{56,84,146}

\begin{document}

\title{Large Language Model-Based Agents for Software Engineering: A Survey}








\author{Junwei~Liu, Kaixin~Wang, Yixuan~Chen, Xin~Peng, Zhenpeng~Chen, Lingming~Zhang, Yiling~Lou
\IEEEcompsocitemizethanks{\IEEEcompsocthanksitem J. Liu, K. Wang, Y. Chen, and X. Peng are with the Department of Computer Science, Fudan University, China.
E-mails:  \{jwliu24, kxwang23, yixuanchen23\}@m.fudan.edu.cn, pengxin@fudan.edu.cn
\IEEEcompsocthanksitem Z. Chen is with the School of Computer Science and Engineering, Nanyang Technological University, Singapore.
E-mail: zhenpeng.chen@ntu.edu.sg
\IEEEcompsocthanksitem L. Zhang and Y. Lou is with the Department of Computer Science, University of Illinois Urbana-Champaign, USA.
E-mail: {lingming,yilingl}@illinois.edu
\IEEEcompsocthanksitem Y. Lou is the corresponding author.
}
}

\markboth{September~2024}{Liu \MakeLowercase{\textit{et al.}}: Large Language Model-Based Agents for Software Engineering: A Survey}

\IEEEtitleabstractindextext{%
\begin{abstract}
The recent advance in Large Language Models (LLMs) has shaped a new paradigm of AI agents, \ie{} LLM-based agents. Compared to standalone LLMs, LLM-based agents substantially extend the versatility and expertise of LLMs by enhancing LLMs with the capabilities of perceiving and utilizing external resources and tools. To date, LLM-based agents have been applied and shown remarkable effectiveness in Software Engineering (SE). The synergy between multiple agents and human interaction brings further promise in tackling complex real-world SE problems. In this work, we present a comprehensive and systematic survey on LLM-based agents for SE. We collect \papern{} papers and categorize them from two perspectives, \ie{} the SE and agent perspectives. In addition, we discuss open challenges and future directions in this critical domain. The repository of this survey is at \url{https://github.com/FudanSELab/Agent4SE-Paper-List}.
\end{abstract}

}

\maketitle

\section{Introduction}
Large Language Models (LLMs) have achieved remarkable progress and demonstrated potential of human-like intelligence~\cite{DBLP:journals/corr/abs-2303-18223}. In recent years, LLMs have been widely applied in Software Engineering (SE). As shown by recent surveys~\cite{LLM4SE1, LLM4SE2}, LLMs have been adopted and shown promising performance in various software development and maintenance tasks, such as program generation~\cite{self-collaboration, DBLP:journals/corr/abs-2304-10778, DBLP:journals/corr/abs-2303-17780, stall, liu2024your}, software testing~\cite{TitanFuzz, DBLP:journals/tse/WangHCLWW24, lemieux2023codamosa}, debugging~\cite{xia2022less, AUTOFL, JoshiSG0VR23, AdbGPT, xia2023automated, jiang2023impact, linuxflplus, vulrag}, and program improvement~\cite{ShypulaMZ0GYHNR24, DBLP:journals/corr/abs-2304-13187, DBLP:journals/corr/abs-2309-07062, transagent}. 

AI Agents are artificial entities that can autonomously perceive and act on surrounding environments so as to achieve specific goals~\cite{agents_fudannlp}. The concept of AI agents has been evolving for a long time (\eg{} early agents are constructed on symbolic logic or reinforcement learning~\cite{ribeiro2002reinforcement, kaelbling1996reinforcement,minsky1961steps,isbell2001social}). Recently, the remarkable progress in LLMs has further shaped a new paradigm of AI agents, \ie{} LLM-based agents, which leverage LLMs as the central agent controller. Different from standalone LLMs, LLM-based agents extend the versatility and expertise of LLMs by equipping LLMs with the capabilities of perceiving and utilizing external resources and tools, which can tackle more complex real-world goals via collaboration between multiple agents or involvement of human interaction.   

In this work, we present a comprehensive and systematic survey on LLM-based agents for SE. We collect \papern{} papers and categorize them from two perspectives, \ie{} both the SE and agent perspectives. Additionally, we discuss the open challenges and future directions in this domain. 

From the \emph{SE} perspective, we analyze how LLM-based agents are applied across different software development and maintenance activities, including individual tasks (\eg{} requirements engineering, code generation, static code checking, testing, and debugging) as well as the end-to-end procedure of software development and maintenance. From this perspective, we provide a comprehensive overview of how SE tasks are tackled by LLM-based agents. 

From the \emph{agent} perspective, we focus on the design of components in LLM-based agents for SE.  Specifically, we analyze foundation LLMs and key components, including planning, memory, perception, and action, in these agents. Beyond basic agent construction, we also analyze multi-agent systems, including their agent roles, collaboration mechanisms, information flows, real-world applications, and human-agent collaboration. From this perspective, we summarize the characteristics of different components of LLM-based agents when applied to the SE domain.

In summary, this survey makes the following contributions:

\begin{itemize}[leftmargin=*,label=-]

\item It provides a comprehensive survey of \papern{} papers that apply LLM-based agents to SE.

\item It analyzes how existing LLM-based agents are designed and applied for software development and maintenance from both the SE and agent perspectives.

\item It discusses research opportunities and future directions in this critical domain. 
\end{itemize}

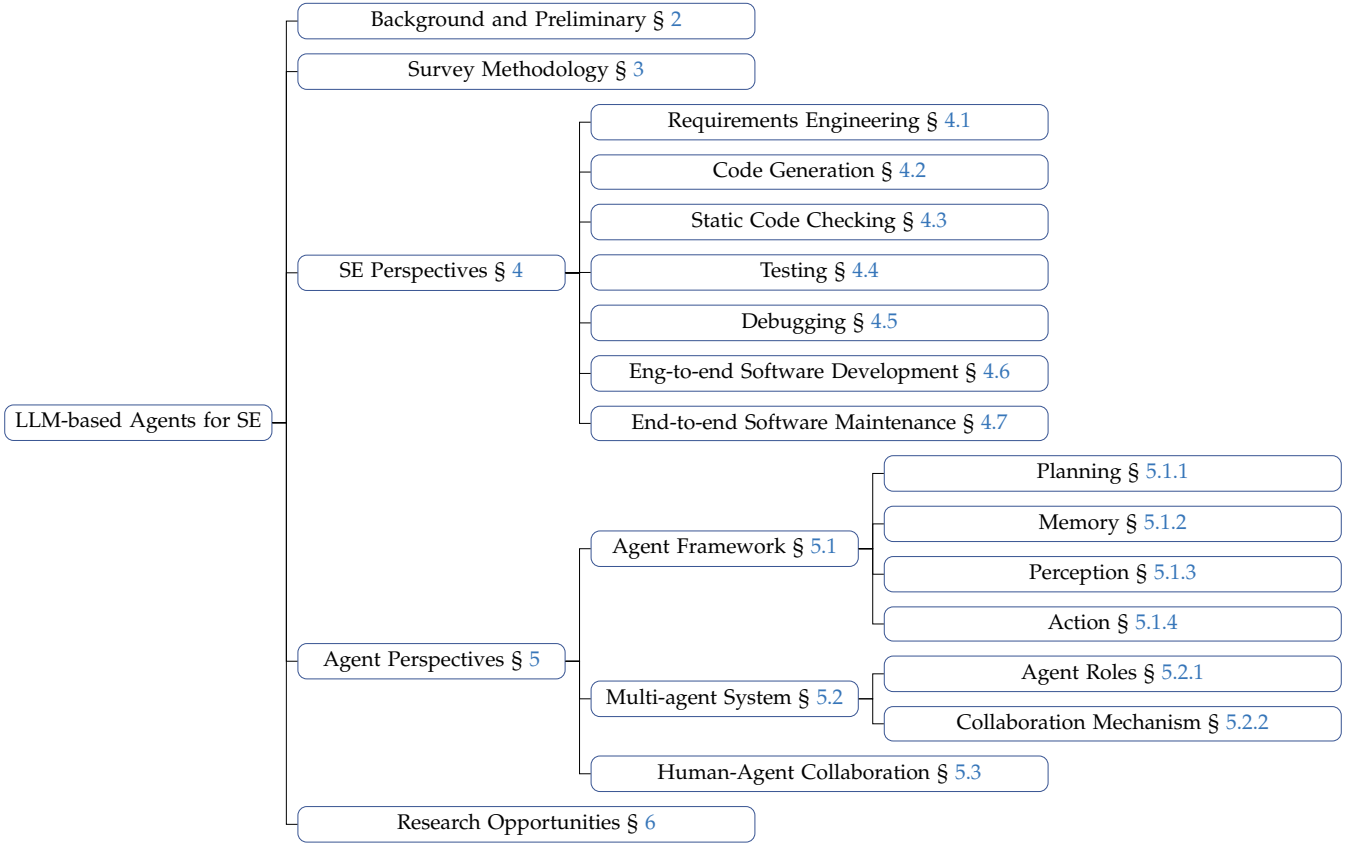
\begin{figure*}[!htbp]
  \centering 
  \begin{adjustbox}{width=1.0
  \textwidth}
    \begin{forest}
for tree={
    rounded corners,
    child anchor=west,
    parent anchor=east,
    grow'=east,  
    text width=4cm,%
    draw=darkblue,
    anchor=west,
    node options={align=center},
    edge path={
      \noexpand\path[\forestoption{edge}]
        (.child anchor) -| +(-5pt,0) -- +(-5pt,0) |-
        (!u.parent anchor)\forestoption{edge label};
    },
    where n children=0{text width=7cm}{}
  },
  [LLM-based Agents for SE
    [Background and Preliminary \S ~\ref{sec:background}],
    [Survey Methodology \S ~\ref{sec:method}],
    [SE Perspectives \S~\ref{sec:se}
        [Requirements Engineering \S~\ref{sec:se:requirement}],
        [Code Generation \S~\ref{sec:se:codegen}],
        [Static Code Checking \S~\ref{sec:se:static}],
        [Testing \S~\ref{sec:se:test}],
        [Debugging \S~\ref{sec:se:debugging}],
        [IT Operations \S~\ref{sec:se:operations}]
        [Eng-to-end Software Development \S~\ref{sec:se:development}],
        [End-to-end Software Maintenance \S~\ref{sec:se:improvement}]
    ],
    [Agent Perspectives \S~\ref{sec:agent}
        [Agent Framework \S~\ref{sec:agent:framework}
            [Planning \S~\ref{sec:agent:planning}],
            [Memory \S~\ref{sec:agent:memory}],
            [Perception \S~\ref{sec:agent:perception}],
            [Action \S~\ref{sec:agent:action}]
            [Foundation LLM \S~\ref{sec:agent:basellm}]
        ],
        [Multi-agent System \S~\ref{sec:agent:multiagent}
            [Agent Roles \S~\ref{sec:agent:role}],
            [Collaboration Mechanism \S~\ref{sec:agent:cooperate}]
            [Information Flow \S~\ref{sec:agent:information}]
            [Real-world Applications \S~\ref{sec:agent:application}]
        ],
        [Human-Agent Collaboration \S~\ref{sec:agent:human}]
    ],
    [Research Opportunities \S~\ref{sec:oppo}]
  ]
\end{forest}
\end{adjustbox}
\caption{Structure of This Survey} 
\label{fig:structure}
\end{figure*}

\textbf{Survey Structure.} 
Figure~\ref{fig:structure} summarizes the structure of this survey. Section~\ref{sec:background} introduces background knowledge, while Section~\ref{sec:method} presents the methodology. Section~\ref{sec:se} and Section~\ref{sec:agent} present the relevant work from the SE perspective and the agent perspective, respectively. Finally, Section~\ref{sec:oppo} discusses the potential research opportunities.

\section{Background and Preliminary}\label{sec:background}
In this section, we first introduce the background about the basic and advanced LLM-based agents, and then we discuss the related surveys.

\subsection{Basic Framework of LLM-based Agents}\label{sec:related:framework}
LLM-based agents are typically composed of four key components: \textit{planning}, \textit{memory}, \textit{perception}, and \textit{action}~\cite{agents_fudannlp}. The planning and memory serve as the key components of the \textit{LLM-controlled brain}, which interacts with the environment through the perception and action components to achieve specific goals. Figure \ref{fig:components} illustrates the basic framework of LLM-based agents.

\textbf{Planning.} The planning component decomposes complex tasks into multiple sub-tasks and schedules the sub-tasks to achieve final goals. In particular, agents can (i) generate an initial plan by different reasoning strategies, or (ii) adjust a generated plan with the external feedback (\eg{} environmental feedback or human feedback). 

\textbf{Memory.} The memory component records the historical thoughts, actions, and environmental observations generated during the agent execution~\cite{agents_fudannlp,agents_Renmin, memory_survey}. Based on accumulated memory, agents can revisit and leverage previous records and experience to tackle complex tasks more effectively. The memory management (\ie{} how to represent the memory) and utilization (\ie{} how to read/write or retrieve the memory) are essential, which directly impact the efficiency and effectiveness of the agent system.  

\textbf{Perception.} The perception component receives the information from the environment, which can facilitate better planning. In particular, agents can perceive multi-modal inputs, \eg{} textual inputs, visual inputs, and auditory inputs. 

\textbf{Action.}  Based on the planning and decisions made by the brain, the action component conducts concrete actions to interact with and impact the environment. One essential mechanism in action is to control and utilize external tools, which can extend the inherent capabilities of LLMs by accessing more external resources and extending the action space beyond textual-alone interaction.   

\textbf{\begin{figure}[!htb]
    \centering
    \includegraphics[width=1.0\columnwidth]{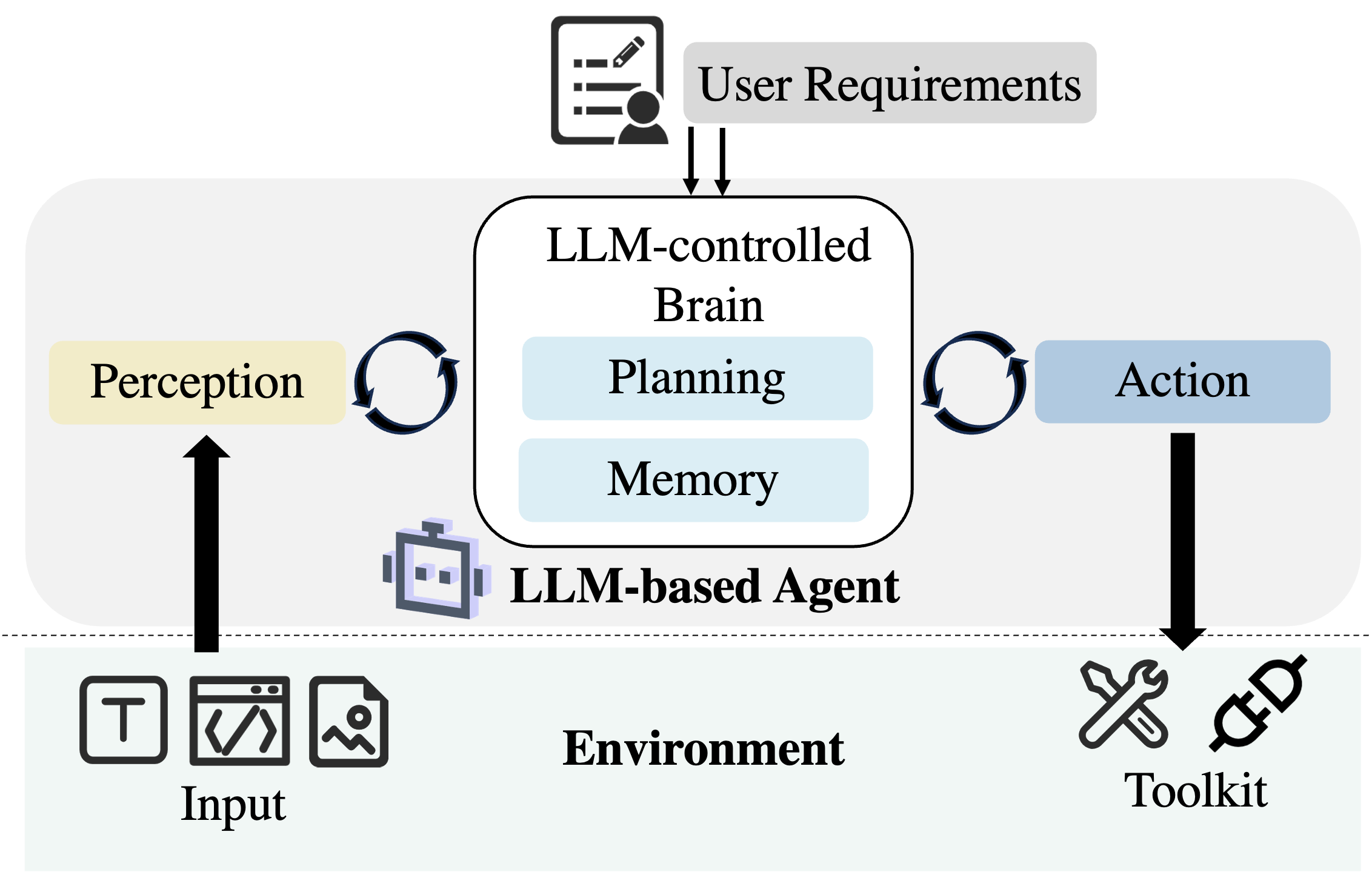}
    \caption{Basic Framework of LLM-based Agents} 
    \label{fig:components}
\end{figure}
}

\subsection{Advanced LLM-based Agent Systems}~\label{sec:related:advance}
 
\textbf{Multi-agent Systems.} 
While a single-agent system can be specialized to solve one certain task, enabling the collaboration between multiple agents (\ie{} \textit{multi-agent systems}) can further solve more complex tasks associated with diverse knowledge domains. In particular, in a multi-agent system, each agent is assigned a distinct role and relevant expertise, making it specifically responsible for different tasks; in addition, the agents can communicate with each other and share progress/information as the task proceeds. Typically, agents can work collaboratively (\ie{} by working on different sub-tasks to achieve a final goal) or competitively (\ie{} by working on the same task while debating adversarially).

\textbf{Human-agent Coordination.} 
Agent systems can further incorporate the instructions from humans and then proceed with tasks under human guidance. This human-agent coordination paradigm facilitates better alignment with human preferences and uses human expertise. In particular, during human-agent interaction, humans can not only provide agents with task requirements and feedback on the current task status, but also cooperate with agents to achieve goals together. 

\junwei{
\subsection{Software Engineering}
Since the 1960s, the discipline of ``Software Engineering'' has evolved, focusing on developing high-quality software in an efficient and cost-effective manner~\cite{naur1969software}. Generally, software engineering involves applying engineering principles throughout the entire software development and maintenance life cycle~\cite{laplante2017dictionary}. This process encompasses a variety of activities, from understanding user requirements to ensuring the software remains reliable and efficient over time. The key phases in software engineering are as follows:
\begin{itemize}[label=-, leftmargin=*]
    \item \textit{Requirements Engineering}: Gathering, analyzing, and documenting functional and non-functional requirements to define the software’s goals and scope.
    \item \textit{Software Design}: Planning the system architecture, components, and interactions to ensure scalability, maintainability, and performance.
    \item \textit{Coding}: Writing the actual code based on the design specifications, following best practices like modularization, version control, and coding standards.
    \item \textit{Static Checking}: Using static analysis techniques to analyze code for errors and vulnerabilities before execution. Code reviews also play a crucial role in identifying issues early by having peers manually inspect the code.
    \item \textit{Software Testing}: Conducting testing activities to ensure the software meets its requirements and functions correctly. There are various types of testing, such as unit, integration, and system testing.
    \item \textit{Maintenance}: Handling fault localization and repair, feature updates, and operational monitoring to ensure continued software stability, security, and alignment with evolving user needs.
\end{itemize}}

\junwei{
These activities share some key characteristics and challenges, including but not limited to: (i) \textit{Complexity}: Software systems typically comprise multiple modules and functionalities, requiring thorough analysis and careful design. (ii) \textit{Adapting to changes}: Requirements evolve due to market trends, customer needs, or regulatory changes, while codebases continuously adapt and expand. (iii) \textit{Iterating through development}: Software development follows an iterative approach, progressively refining through multiple versions. (iv) \textit{Collaboration}: Development and maintenance require teamwork involving product managers, developers, testers, operations engineers, and other stakeholders.}

\junwei{To address these challenges, early-stage software engineering is predominantly human-centric, relying on methodologies and tools to enhance the productivity of individual engineers, reduce the complexity of team collaboration, and standardize software development and maintenance practices~\cite{boehm2006view}. In recent years, advancements in artificial intelligence, particularly the emergence of LLM-based agents, have significantly accelerated the automation of software engineering tasks. Leveraging the aforementioned capabilities, LLM-based agents can address the challenges of software activities, including handling evolving requirements and codebases through perception and iterative mechanisms, decomposing and implementing complex modules via planning and action modules, and simulating human-team collaboration through multi-agent systems. These capabilities not only enhance their performance in single-phase tasks but also enable them to tackle complex multi-phase tasks, such as end-to-end software development and maintenance.
In Section~\ref{sec:se}, we will elaborate on the specific applications of LLM-based agents across different software engineering tasks.
}

\subsection{Related Surveys}~\label{sec:relatedsurvey}
LLM-based agents in general domains have been widely discussed and surveyed~\cite{agents_fudannlp, agents_NotreDame, agents_Renmin, agent_hongkongchinese, DBLP:journals/tmlr/MialonDLNPRRSDC23, agent_design_pattern, DBLP:journals/corr/abs-2404-04442}. Different from these surveys, this survey focuses on the design and application of LLM-based agents specifically for the software engineering domain. 
In the software engineering domain, there have been several surveys or literature reviews on the general application of LLMs in software engineering~\cite{LLM4SE1, LLM4SE2, DBLP:journals/tse/WangHCLWW24, LLM4CodeGeneration_survey, DBLP:journals/corr/abs-2404-04442}.
\junwei{As agents extend the capabilities of standalone LLMs with action, perception, planning, and memory, they can handle more complex and multi-turn tasks than standalone LLMs. Therefore, this survey differs from existing surveys on LLMs for SE by (i) covering wider range of SE tasks, \eg{}  the end-to-end software development or end-to-end software maintenance; (ii) summarizing from the perspective of agent architectures, \eg{} building taxonomy of memory, planning, action components, multi-agent collaboration modes, and human–agent interaction modes.}
In addition, He~\et{}~\cite{he2024llm} present a vision paper on the potential applications and emerging challenges of multi-agent systems for software engineering. Different from the vision paper, this work focuses on conducting a comprehensive survey of existing agent systems (including both single-agent and multi-agent systems). 
In summary,  to the best of our knowledge, this is a comprehensive survey specifically focusing on the literature on LLM-based agents for software engineering.

\section{Survey Methodology}~\label{sec:method}
This section defines the scope of this survey and describes our approach to collecting and analyzing papers within the scope.

\subsection{Survey Scope}
We focus on the papers that apply \textit{LLM-based agents} to tackle \textit{SE tasks}. In the following, we specify the terms. 
\begin{itemize}[leftmargin=*,label=-]
\item \textit{SE tasks}. Following previous surveys on the application of LLMs in SE~\cite{LLM4SE1, LLM4SE2}, we focus on all SE tasks along the software life cycle, including requirements engineering, software design, code generation, software quality assurance (\ie{} static checking and testing), and software improvement. 

\item \textit{LLM-based agents}. A standalone LLM can work as a naive ``agent'' since it can take textual inputs and produce textual outputs, leaving it no clear boundary between LLMs and LLM-based agents. 
However, this could result in an overly broad scope and significant overlap with existing surveys on LLM applications in SE~\cite{LLM4SE1, LLM4SE2}. Based on the widely-adopted consensus about AI agents,  the key characteristic of agents is their ability to autonomously and iteratively perceive feedback from, and act upon, a dynamic environment~\cite{agents_fudannlp}. To ensure a more focused discussion from the perspective of agents, this survey focuses on LLM-based agents that not only incorporate LLMs as the core of their ``brains'', but also have the capacity to iteratively interact with the environment, taking feedback and acting in real time. 
\end{itemize}

\junweim{In addition, we position this paper as a comprehensive survey rather than a systematic literature review, with the goal of providing researchers with a quick overview of this rapidly evolving field. Therefore, we focus on the organization and synthesis of existing research on LLM-based agents in the SE domain. While we include experimental results from the collected papers to offer comparative insights and enhance the understanding of various technical approaches, conducting extra experimental analysis is beyond the scope of this survey.}

\begin{table*}[]
\centering
\caption{Inclusion and Exclusion Criteria of Paper Collection}~\label{tab:paper_criteria}
\begin{adjustbox}{width=1.0\textwidth}
\renewcommand{\arraystretch}{1.3}
\begin{tabular}{l}
\toprule
\textbf{Inclusion Criteria}                                                                                                                                    \\ \hline
(1) The paper proposes a technique, framework, or tool that utilizes or enhances LLM-based agents for SE tasks.                                                 \\
(2) The paper presents an empirical study on the application of LLM-based agents for SE tasks.                                                                    \\ \hline
\textbf{Exclusion Criteria}                                                                                                                                    \\ \hline
(1) The agent framework is not based on LLM.\\
(2) The paper does not include any evaluation, or the evaluation does not involve any SE tasks.                                                                                                                  \\
(3) The paper only discusses LLM-based agents in the context of discussion or future work, without integrating them into the main approach.                    \\
(4) The paper merely employs a single LLM linear workflow, without any multi-agent setup or iterative interaction with the environment.\\
\junweim{(5) The paper is less than 2 pages.}\\
\junweim{(6) The paper is a grey literature, \eg{} a technical report or blog post.}\\
\junweim{(7) Duplicate papers or different versions of similar studies by the same authors.} \\
 \bottomrule
\end{tabular}
\end{adjustbox}
\end{table*}

\subsection{Paper Collection}
We apply the inclusion and exclusion criteria as shown in Table~\ref{tab:paper_criteria} for paper collection. 
\junweim{Based on the criteria, we employ a collaborative process to inspect each paper. In particular, the first two authors independently review and label each paper to determine its relevance to the scope of this survey. 
When disagreements arise, a third author serves as an arbiter until consensus is reached.}
Our paper collection process includes three steps: keyword searching, snowballing, and author feedback collection.

\subsubsection{Keyword Searching} 
We follow established practices in SE surveys ~\cite{tosemChenZHHS24,csurChenPPXZHZ20,tseMathewAM23,jcstZhangTJLPY18,tseZhangHML22} by using the DBLP database \cite{dblplink} for paper collection. Recent research~\cite{tseZhangHML22} has demonstrated that papers gathered from other prominent publication databases are typically a subset of those available on DBLP, which encompasses over 7 million publications from more than 6,500 academic conferences and 1,850 journals in computer science~\cite{dblpstatislink}. DBLP also covers arXiv~\cite{arxivlink}, a widely adopted open-access repository.

\junwei{
We employ an iterative trial-and-error approach, which is widely adopted in SE surveys \cite{tosemChenZHHS24, tosemLinCSBNL22}, to determine search keywords. Initially, all authors, with relevant research experience/publications in LLM and SE, convene to suggest papers relevant to our scope, yielding an initial set of relevant papers.
Subsequently, the first two authors review the titles, abstracts, and introductions of these papers to identify an initial list of keywords, which includes ``\textit{agent}'' \textit{AND} (``\textit{code}'' \textit{OR} ``\textit{software}'' \textit{OR} ``\textit{requirement}'' \textit{OR} ``\textit{verification}'' \textit{OR} ``\textit{test}'' \textit{OR} ``\textit{debug}'' \textit{OR} ``\textit{repair}'' \textit{OR} ``\textit{maintenance}'').
We then conduct brainstorming sessions to expand and refine our search strings, incorporating related terms (such as ``api'', ``deploy'', and ``evolution''), synonyms, and variations (such as ``coding'' in relation to ``code'').
This process enables the iterative enhancement of our search keyword list. For example, we observe that some studies, despite incorporating agent-based mechanisms, continue to refer to themselves as large language models. Therefore, we include ``\textit{llm}'' and ``\textit{language model}'' in our keyword list. With each newly added keyword, we perform an updated search and review the newly identified works to extract additional relevant keywords. If no new papers are found, we backtrack and explore alternative keywords, until we can no longer find any new papers. Through this iterative trial-and-error process, we identify the following additional keywords: ``\textit{api}'', ``\textit{bug}'', ``\textit{coding}'', ``\textit{defect}'', ``\textit{deploy}'', ``\textit{evolution}'',  ``\textit{fault}'', ``\textit{fix}'', ``\textit{program}'', ``\textit{refactor}'', and ``\textit{vulnerab}''. 
The final keywords include (``\textit{agent}'' \textit{OR} ``\textit{llm}'' \textit{OR} ``\textit{language model}'') \textit{AND} (``\textit{api}'' \textit{OR} ``\textit{bug}'' \textit{OR} ``\textit{code}'' \textit{OR} ``\textit{coding}'' \textit{OR} ``\textit{debug}'' \textit{OR} ``\textit{defect}'' \textit{OR} ``\textit{deploy}'' \textit{OR} ``\textit{evolution}'' \textit{OR} ``\textit{fault}'' \textit{OR} ``\textit{fix}'' \textit{OR} ``\textit{maintenance}'' \textit{OR} ``\textit{program}'' \textit{OR} ``\textit{refactor}'' \textit{OR} ``\textit{repair}'' \textit{OR} ``\textit{requirement}'' \textit{OR} ``\textit{software}'' \textit{OR} ``\textit{test}'' \textit{OR} ``\textit{verification}'' \textit{OR} ``\textit{vulnerab}'').
}

\begin{table}[]
\centering
\caption{Statistics of Paper Collection}~\label{keywords}
\begin{tabular}{lr}
\toprule
Keyword                                    & Hits \\ \midrule
agent \textbar\ llm \textbar\ language model + api         & 83   \\
agent \textbar\ llm \textbar\ language model + bug         & 98   \\
agent \textbar\ llm \textbar\ language model + code        & 915  \\
agent \textbar\ llm \textbar\ language model + coding      & 70   \\
agent \textbar\ llm \textbar\ language model + debug       & 95   \\
agent \textbar\ llm \textbar\ language model + defect      & 22   \\
agent \textbar\ llm \textbar\ language model + deploy      & 295  \\
agent \textbar\ llm \textbar\ language model + evolution   & 1,349 \\
agent \textbar\ llm \textbar\ language model + fault       & 685  \\
agent \textbar\ llm \textbar\ language model + fix         & 318  \\
agent \textbar\ llm \textbar\ language model + maintenance    & 64   \\
agent \textbar\ llm \textbar\ language model + program     & 1,969 \\
agent \textbar\ llm \textbar\ language model + refactor    & 15    \\
agent \textbar\ llm \textbar\ language model + repair      & 137  \\
agent \textbar\ llm \textbar\ language model + requirement & 451  \\
agent \textbar\ llm \textbar\ language model + software    & 2,151 \\
agent \textbar\ llm \textbar\ language model + test        & 976  \\
agent \textbar\ llm \textbar\ language model + verification& 525  \\
agent \textbar\ llm \textbar\ language model + vulnerab    & 144  \\ \hline
After manual inspection                    & 67   \\
After snowballing                          & 108  \\   
After author feedback collection & \papern{} \\
\bottomrule 
\end{tabular}
\end{table}

Based on the keywords, we conduct 57 searches on DBLP on July 1st, 2024, and obtain 10,362 hits. 
Table~\ref{keywords} presents the statistics of papers collected through keyword searching. 
The first three authors manually review each paper to filter out those not within the scope of this survey.
As a result, we identify 67 relevant papers through this process.

\subsubsection{Snowballing}
To enhance the comprehensiveness of our survey, we adopt snowballing approaches to identify papers that are transitively relevant and expand our paper collection~\cite{tosemChenZHHS24}. Specifically, between July 1 and July 10, 2024, we conduct both backward and forward snowballing. Backward snowballing involves examining references in each collected paper to identify relevant ones within our scope, while forward snowballing uses Google Scholar to find relevant papers citing the collected ones. This iterative process continues until no new relevant papers are found. In this process, we retrieve an additional 41 papers.

\subsubsection{Author Feedback Collection}
To further enhance the accuracy and comprehensiveness of our survey, we reach out to the authors of the papers gathered through keyword searches and snowballing after drafting the initial version. We invite these authors to review our descriptions of their work, ensuring correctness, and to recommend additional relevant papers. \junweim{In total, 321 authors were contacted via email, and we received 36 valid responses. Among them, eleven authors confirmed that our descriptions were accurate and required no changes; sixteen authors recommended 29 related papers, of which 16 papers were included in the survey after relevance filtering based on the inclusion and exclusion criteria presented in Table~\ref{tab:paper_criteria}; and thirteen authors suggested revisions to the survey, including five updates on paper publication status, six suggestions for improving method descriptions, and two formatting refinements. This feedback helps ensure that the survey accurately reflects the findings and perspectives of the original research.}

\subsection{Statistics of Collected Papers}
\begin{figure}[ht]
    \centering
    \begin{subfigure}[b]{0.45\textwidth}
        \includegraphics[width=\linewidth]{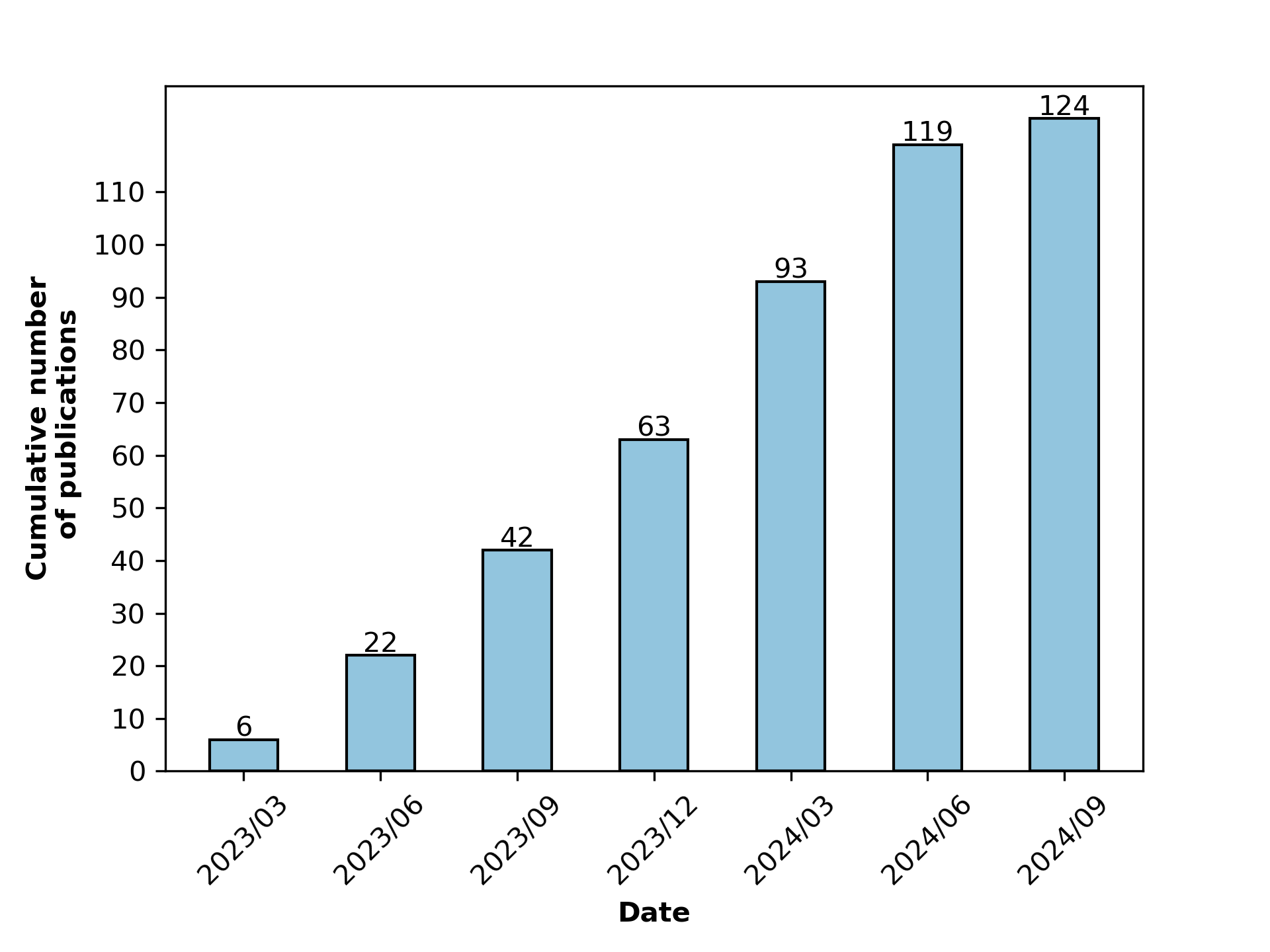}
        \caption{Cumulative Number of Papers Over Time}
        \label{fig:trend}
    \end{subfigure}\hfill
    \begin{subfigure}[b]{0.45\textwidth}
        \includegraphics[width=\linewidth]{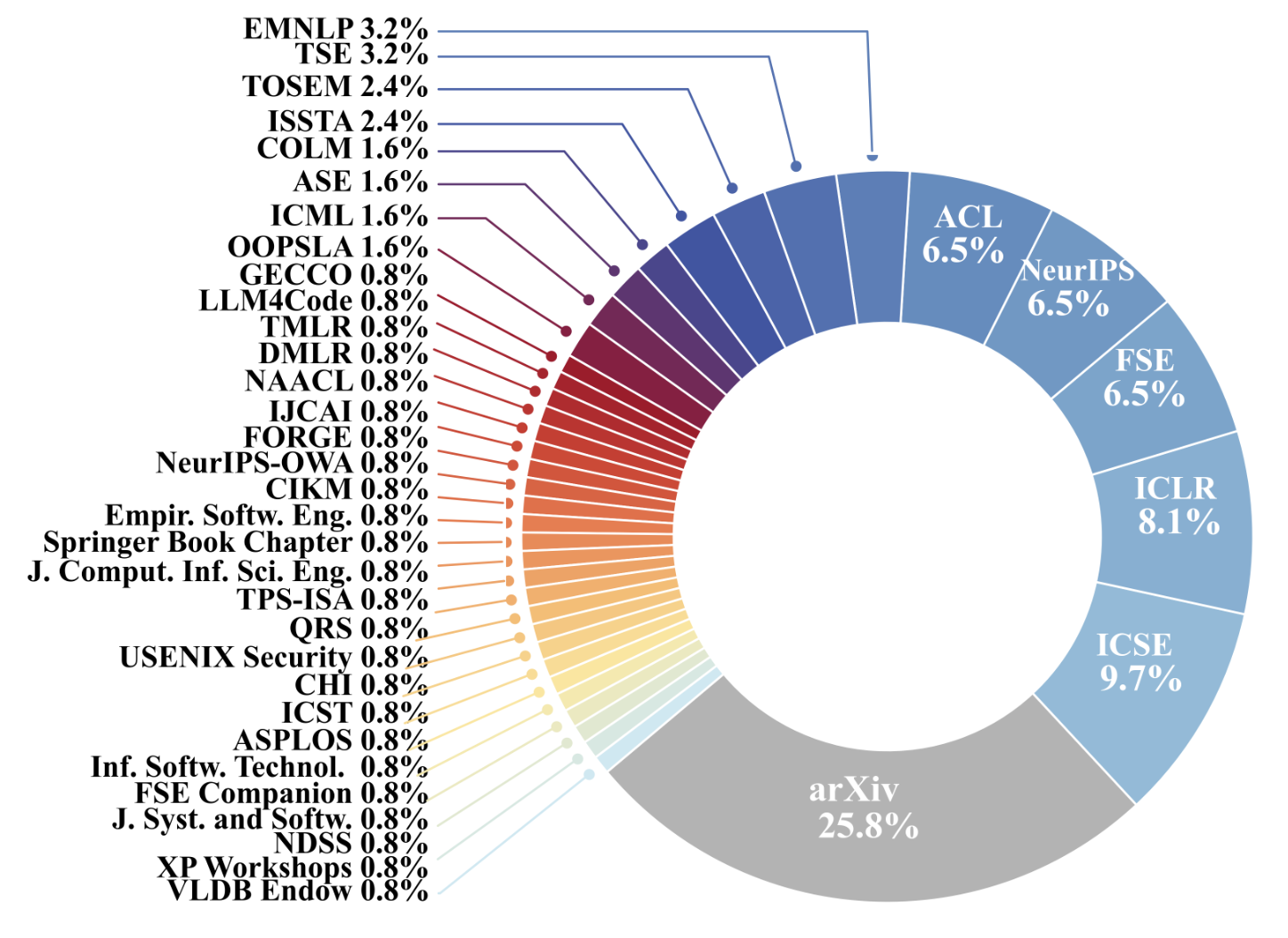}
        \caption{Distribution of Publication Venues of All Papers}
        \label{fig:venue}
    \end{subfigure}
    \caption{Statistics of Collected Papers}
    \label{fig:images}
\end{figure}

As shown in Table \ref{keywords}, we have collected a total of \papern{} papers for this survey. Figure~\ref{fig:trend} presents the cumulative number of papers published over time, up to September 11, 2024.\footnote{The most recent date of the papers collected during the author feedback process.} We observe that there is a continuous increase of research interest in this field, highlighting the necessity and relevance of this survey. Additionally, Figure~\ref{fig:venue} shows the distribution of publication venues for the papers, covering diverse research communities such as software engineering, artificial intelligence, and human-computer interaction. 
\junweim{In particular, approximately 75\% of the references are peer-reviewed publications from reputable journals and conferences, reflecting the academic rigor of our sources. The remaining citations are preprints from arXiv, which reference cutting-edge or emerging work not yet formally published. This mix provides both foundational support and timely insights, balancing scholarly reliability with the most recent developments in this field.}



\textbf{\begin{figure*}[htb]
    \centering
    \includegraphics[width=1\textwidth]{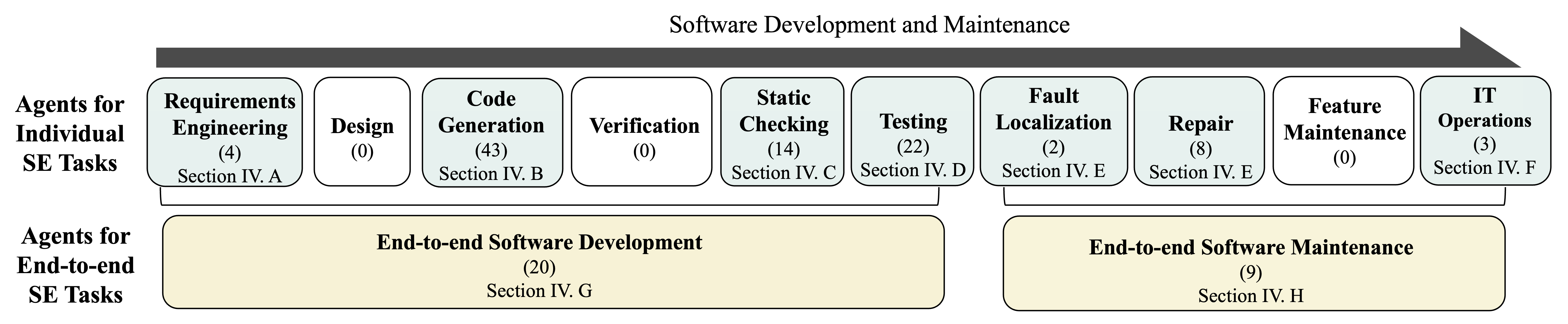}
    \caption{Agent  Distribution along Software Development and Maintenance Tasks} 
    \label{fig:SEprocedure}
\end{figure*}
}

\section{Analysis from SE Perspectives}\label{sec:se}
In this section, we organize the collected papers from the perspective of different SE tasks. Figure~\ref{fig:SEprocedure} presents the SE tasks along the common life cycle of software development and maintenance. 

\textbf{It is worth noting that, LLM-based agents can be designed not only to tackle individual SE tasks but also to support end-to-end software development or maintenance processes involving multiple SE activities.} From the collected papers, we observe LLM-based agents designed for (i) \textit{end-to-end software development} and (ii) \textit{end-to-end software maintenance}. Specifically, agents for end-to-end software development can generate a complete program based on requirements by performing multiple SE tasks, such as requirements engineering, design, code generation, and code quality assurance (\eg{} verification, static checking, and testing); agents for end-to-end software maintenance can generate patches for user-reported issues by supporting multiple SE maintenance activities, such as debugging (\eg{} fault localization and repair) and feature maintenance. As shown in previous papers~\cite{LLM4SE1, LLM4SE2}, standalone LLMs are primarily specialized in tackling single SE tasks and are generally inadequate for complex end-to-end software development and maintenance processes. In contrast, LLM-based agents, through their components (\ie{} planning, memory, perception, and action), coordination among multiple agents, and human interaction, provide the autonomy and flexibility necessary to tackle these complex tasks. 

\textbf{Distribution of LLM-based agents in different SE activities.} In Figure~\ref{fig:SEprocedure}, the numbers in brackets indicate the count of collected papers in each category. Notably, if LLM-based agents are designed for end-to-end software development or maintenance, they are only reported at the end-to-end level rather than at the level of individual tasks. Overall, we observe that the majority of LLM-based agents focus on individual-level SE tasks, especially for code generation and code quality assurance (\eg{} static checking and testing); in addition, a portion of agents are designed for end-to-end software development or maintenance tasks, indicating the promise of LLM-based agents in tackling more complex real-world SE tasks.

\begin{figure}[h]
    \centering
    \includegraphics[width=1.0\columnwidth]{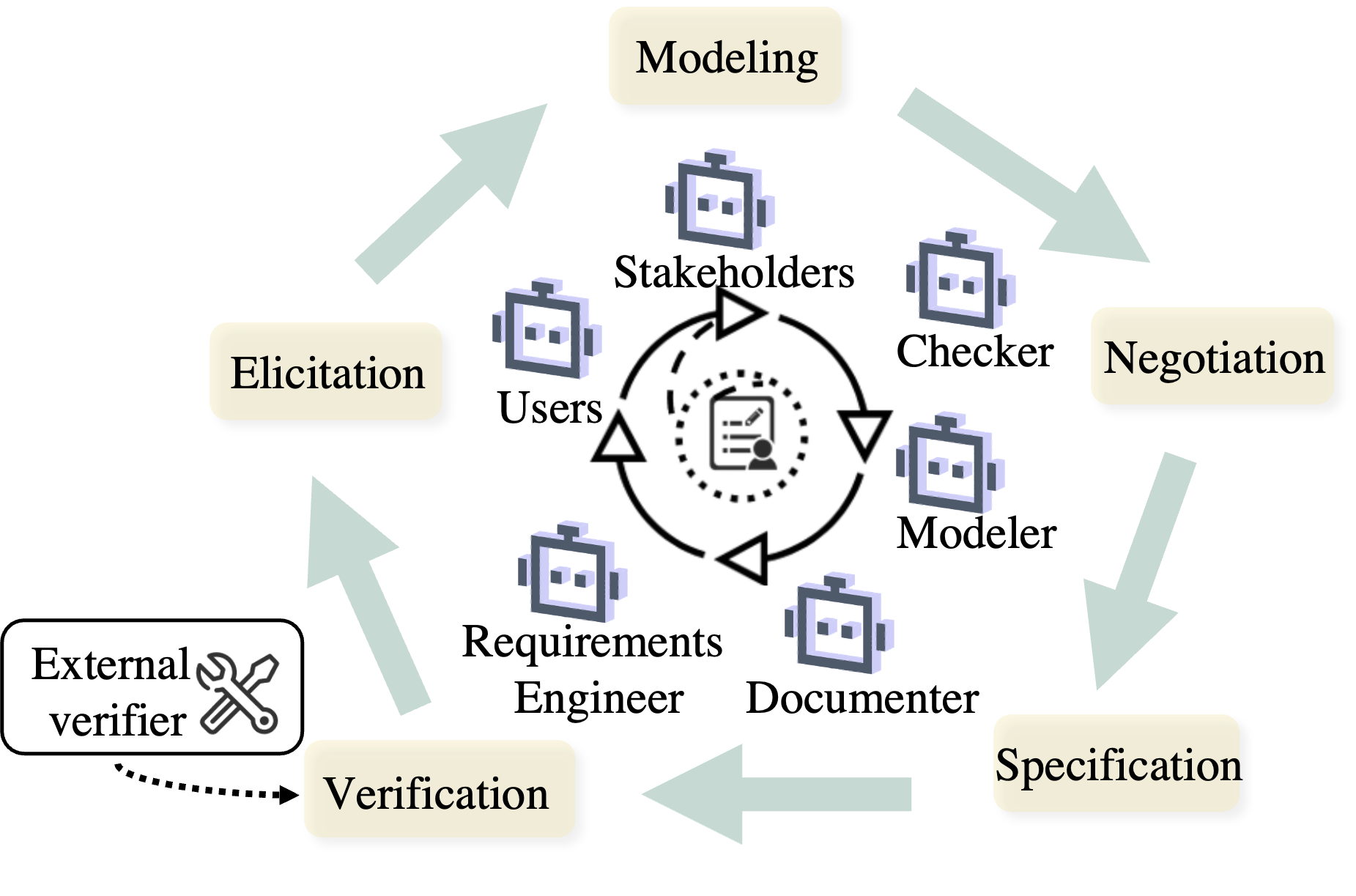}
    \caption{Pipeline of LLM-based Agents for Requirements Engineering} 
    \label{fig:RE}
\end{figure}
\subsection{Requirements Engineering}~\label{sec:se:requirement}
Requirements Engineering (RE) is a crucial phase for initiating the software development procedure. Generally, it covers the following phases~\cite{pohl1996requirements,nuseibeh2000requirements, shukla2015requirements}.

\begin{itemize}[leftmargin=*,label=-]
\item \textit{Elicitation}: New requirements are elicited and collected.

\item \textit{Modeling}: Abstract yet interpretable models, \eg{} Unified Modeling Language (UML)~\cite{booch1996unified} and Entity-Relationship-Attribute (ERA) model~\cite{johnson2002entity}, are constructed to describe the original requirements.

\item \textit{Negotiation}: Negotiation plays a crucial role in facilitating communication among different stakeholders and ensuring consistency, especially in conflicting requirements.

\item \textit{Specification}: Requirements are determined and documented in a formal format.

\item \textit{Verification}: Requirements and models are validated to ensure that they fully and unambiguously reflect the intent of stakeholders.

\item \textit{Evolution}: Requirements evolution refers to the ongoing process of refining and adapting requirements in response to changing needs and conditions. 
\end{itemize}

In real-world software development, RE takes a lot of manual effort due to the demand for massive interactions with different stakeholders. 
\junwei{To reduce the manual effort in RE, early works introduce various automation tools based on text mining~\cite{rago2013uncovering}, simple NLP techniques (\eg{} POS tagging and parsers)~\cite{nazir2017applications}, and machine learning~\cite{li2018automatically}. However, due to the limited natural language understanding capabilities of these methods, a prior study demonstrates that approximately 60\% of these tools remain semi-automated, still requiring human intervention~\cite{umar2024advances}. In recent years, the strong natural language understanding demonstrated by deep learning has opened up new possibilities for further automating requirements engineering.} Researchers have adopted deep learning models (including standalone LLMs) to enhance requirements engineering activities, but most of them still work on individual RE tasks, such as classification~\cite{DBLP:conf/kbse/LuoXXS22}, specification~\cite{krishna2024using}, information retrieval~\cite{zhang2023empirical}, evaluation~\cite{ronanki2022chatgpt}, and enhancement~\cite{luitel2024improving} of existing requirements. In comparison, the latest agent systems are designed to automate not only individual but also multiple RE phases. Table~\ref{tab:re} summarizes existing LLM-based agents specifically designed for RE.

\junwei{\textbf{Framework:} Figure~\ref{fig:RE} illustrates the common framework of LLM-based agents on requirements engineering. Current studies primarily leverage the role-playing and collaboration capabilities of LLM-based agents to simulate real-world requirements engineering roles, such as users, stakeholders, requirements engineers, modelers, checkers, and documenters, with the aim of producing a complete requirements document. This cycle includes requirements elicitation, modeling, negotiation, specification, and verification. In certain specific stages, such as requirements verification, external validation tools can be integrated as action modules to provide feedback, which can be achieved through the agent's tool usage and iterative refinement capabilities.}

\begin{table*}[ht]
\centering
\caption{Existing LLM-based Agents for Requirements Engineering}\label{tab:re}
\begin{adjustbox}{width=0.9\textwidth}
\renewcommand{\arraystretch}{1.2}
\begin{tabular}{lccccccc}
\hhline
\multirow{2}{*}{\textbf{Agents}} & \multirow{2}{*}{\textbf{Multi-Agent}} & \multicolumn{6}{c}{\textbf{Covered RE Phases}} \\ \cline{3-8}
 &  & \textbf{Elicitation} & \textbf{Modeling} & \textbf{Negotiation} & \textbf{Specification} & \textbf{Verification} & \textbf{Evolution} \\ 
 \hhline
 Elicitron~\cite{Elicitron} & \checkmark & \checkmark &  &  &  &  &  \\ 
SpecGen~\cite{ma2024specgen} & ×  &  &  &  & \checkmark &  &  \\ 
Arora \et{}~\cite{arora2024advancing} & \checkmark & \checkmark &  & \checkmark & \checkmark & \checkmark &  \\
MARE~\cite{MARE} & \checkmark & \checkmark & \checkmark &  & \checkmark & \checkmark &  \\
\hhline
\end{tabular}
\end{adjustbox}
\end{table*}

\junweim{
\subsubsection{Multi-agent Collaboration Strategy.}
Multi-agent collaboration approaches simulate real-world software engineering teams by assigning different roles and personas to multiple agents. One example is Elicitation~\cite{Elicitron}, which is a multi-agent framework designed for requirement elicitation. Within the designed context, Elicitation initializes multiple agents with different personas. These agents will simulate interactions with the target product from different user viewpoints and document the records (\ie{} actions, observations, and challenges). The latent requirements are then identified through the agent interviews and filtered by the provided criteria. Experimental results indicate that Elicitation can uncover and categorize hidden needs while reducing costs compared with conventional methodologies such as user studies. 
}

\junweim{
Similarly, Arora~\et{}~\cite{arora2024advancing} propose a multi-agent pipeline spanning four RE phases: elicitation, specification, analysis (negotiation), and validation. Their approach employs role-playing strategies in which agents assume roles such as actual users and software architects to collaboratively negotiate requirement priorities, thereby facilitating cross-phase coordination.
}

\junweim{
Along the same lines, MARE~\cite{MARE} also utilizes the role-playing strategy to construct a multi-agent framework that performs a different RE pipeline, including elicitation, modeling, verification, and specification. In the elicitation phase, a set of stakeholder agents expresses their needs, which would then be organized into a draft by the collector agent. Subsequently, the modeler agent identifies entities and relationships in the draft and constructs a requirement model. In the verification phase, the checker agent assesses the quality of the current requirements draft based on its criteria and hands it over to the document agent, who will write the requirement specifications or report errors. All of these agents are equipped with predefined actions and can communicate within a shared workspace, enabling the seamless exchange of intermediate information.
}

\junweim{
\subsubsection{Tool-enhanced Single-agent Strategy.}
In contrast, tool-enhanced single-agent approaches integrate external verification or analysis tools to iteratively improve the output quality generated by a single agent. SpecGen~\cite{ma2024specgen} exemplifies this approach by combining an LLM-based agent with the OpenJML verifier~\cite{cok2011openjml}. The agent generates Java Modeling Language (JML) specifications and refines them based on the error messages returned by OpenJML in an iterative process. Failed specifications undergo mutation and re-verification to produce a more diverse and accurate set of specifications. Experimental evidence shows that SpecGen significantly outperforms existing purely LLM-based methods and traditional specification generation tools such as Houdini~\cite{flanagan2001houdini} and Daikon~\cite{ernst2007daikon}, with improvements of 15.84\%, 47.01\%, and 53.76\%, respectively.
}

\junweim{
\subsubsection{Comparison of Multi-agent and Tool-enhanced Single-agent Strategy.}
In summary, multi-agent collaboration approaches enhance RE phases by simulating real-world team dynamics through role-playing and communication mechanisms. This allows coverage of individual or multiple RE phases but introduces challenges related to quality assurance in collaborative settings. On the other hand, tool-enhanced single-agent methods rely on feedback from external tools to iteratively improve the generated artifacts, yielding higher quality results at the cost of limited task scalability due to the single-agent setup.
}

\junweim{
\subsubsection{Challenges of LLM-based Agents in  RE}
LLM-based agents offer promising support for RE, but several challenges remain. First, the generated requirements may remain vague, irrelevant, or incorrect~\cite{arora2024advancing, Elicitron}. One reason is the lack of sufficient domain knowledge, which existing methods often struggle to incorporate effectively and consistently. Second, existing approaches often underemphasize human–agent interaction. RE depends on continuous communication with stakeholders, but current LLM-based agent systems simply replace human roles with LLM-based agents instead of supporting effective human-agent collaboration, reducing stakeholder involvement and trust. Lastly, as shown in Table~\ref{tab:re}, current agents lack mechanisms to support requirements evolution, limiting their usefulness in iterative and long-term development processes where change is constant.}
\subsection{Code Generation}\label{sec:se:codegen}
Code generation has been extensively explored with the development of AI technology~\cite{LLM4CodeGeneration_survey,guo-etal-2025-personality}. Due to being pre-trained on massive textual data (especially large code corpus), LLMs demonstrate promising effectiveness in generating code for given code contexts or natural language descriptions. Nevertheless, the code generated by LLMs can sometimes be unsatisfactory due to issues such as the notorious hallucination~\cite{hallucination_survey}. Therefore, beyond simply leveraging standalone LLMs for code generation, researchers also build LLM-based agents that can enhance the capabilities of LLMs via planning and iterative refinement.

\junwei{\textbf{Framework:} Figure~\ref{fig:codegen} illustrates how existing studies extend standalone LLMs to LLM-based agents in code generation. Overall, current studies primarily leverage the capabilities of agents to plan and take actions, thus transforming one-time code generation into a ``plan-generate-refine'' model to improve generation correctness. During the planning phase, in addition to the commonly used natural language form, some research also generates plans that use code as an intermediate representation, such as pseudocode, intermediate code, or code skeleton. In the iterative refinement process, various forms of feedback are utilized to further enhance code generation accuracy, which include model feedback (feedback from the LLM itself), tool feedback (feedback from external tools), human feedback (clarifications from humans), and hybrid feedback (combinations of different types of feedback).}

\begin{table*}[]
\centering
\caption{Existing LLM-based Agents for Code Generation}
\label{tab:codegen}

\begin{adjustbox}{width=1.0\textwidth}
\renewcommand{\arraystretch}{1.4} 

\begin{tabular}{
    >{\centering\arraybackslash}m{8cm}  
    c          
    cccc        
}
\hhline
\multirow{2}{*}{\textbf{Agents}} 
 & \multirow{2}{*}{\textbf{Multi-Agent}} 
 & \multicolumn{4}{c}{\textbf{Iterative Refinement}} 
\\ \cline{3-6}
 &  & \textbf{Model Feedback} & \textbf{Tool Feedback} & \textbf{Human Feedback} & \textbf{Hybrid Feedback} 
\\ \hhline

Reflexion~\cite{shinn2023reflexion}, SEIDR~\cite{liventsev2023fully}, Self-Repair~\cite{olausson2024selfrepair}, AutoGen~\cite{AutoGen}, 
INTERVENOR~\cite{INTERVENOR}, TGen~\cite{TGen}, AutoCoder~\cite{AutoCoder2405}, CodeChain~\cite{CodeChain}, RRR~\cite{deshpande2024classlevelcodegenerationnatural}
 & \checkmark & \checkmark & \checkmark &  & \checkmark 
\\ \hline

CAMEL~\cite{li2023camel}, AgentForest~\cite{li2024agents}, DyLAN~\cite{liu2023dynamic}
 & \checkmark & \checkmark &  &  &  
\\ \hline

Self-Debugging~\cite{chen2023teachinglargelanguagemodels}, $\mu$FiX~\cite{tian2024testcasedriven}, 
AlphaCodium~\cite{AlphaCodium}, LDB~\cite{LDB}, LATS~\cite{LATs}
 & $\times$ & \checkmark & \checkmark &  & \checkmark 
\\ \hline

ToolCoder~\cite{zhang2023toolcoder}, SelfEvolve~\cite{jiang2023selfevolve}, KPC~\cite{FromMisusetoMastery}, 
Lemur~\cite{xu2023lemur}, CodeAgent~\cite{zhang2024codeagent}, LLM4TDD~\cite{piya2023llm4tdd}, 
CodeCoT~\cite{huang2024codecot}, CodeAct~\cite{CodeAct}, 
InterCode~\cite{yang2023intercode}, CodePlan~\cite{bairi2023codeplan}, ToolGen~\cite{TOOLGEN}
 & $\times$  &  & \checkmark &  &  
\\ \hline

Self-Refine~\cite{self-refine} 
 & $\times$ & \checkmark &  &  &  
\\ \hline

Flows~\cite{Flows} 
 & \checkmark & \checkmark & \checkmark & \checkmark & \checkmark 
\\ \hline

MINT~\cite{mint} 
 & $\times$ & \checkmark & \checkmark & \checkmark &  
\\ \hline

ClarifyGPT~\cite{ClarifyGPT} 
 & $\times$ &  & \checkmark & \checkmark &  
\\ \hline

Self-Edit~\cite{self-edit}, AgentCoder~\cite{AgentCoder}, Gentopia~\cite{xu2023gentopia}, 
AutoDev~\cite{AutoDev}, SoA~\cite{SoA}, MapCoder~\cite{MapCoder}, 3DGen~\cite{3DGen}, CoCoST~\cite{he2024cocost}
 & \checkmark &  & \checkmark &  &  
\\ \hline

Parsel~\cite{parsel}, RAT~\cite{RAT}
 & \checkmark &  &  &  &  
\\ 
\hhline

\end{tabular}
\end{adjustbox}
\end{table*}

\textbf{\begin{figure}[h]
    \centering
    \includegraphics[width=1.0\columnwidth]{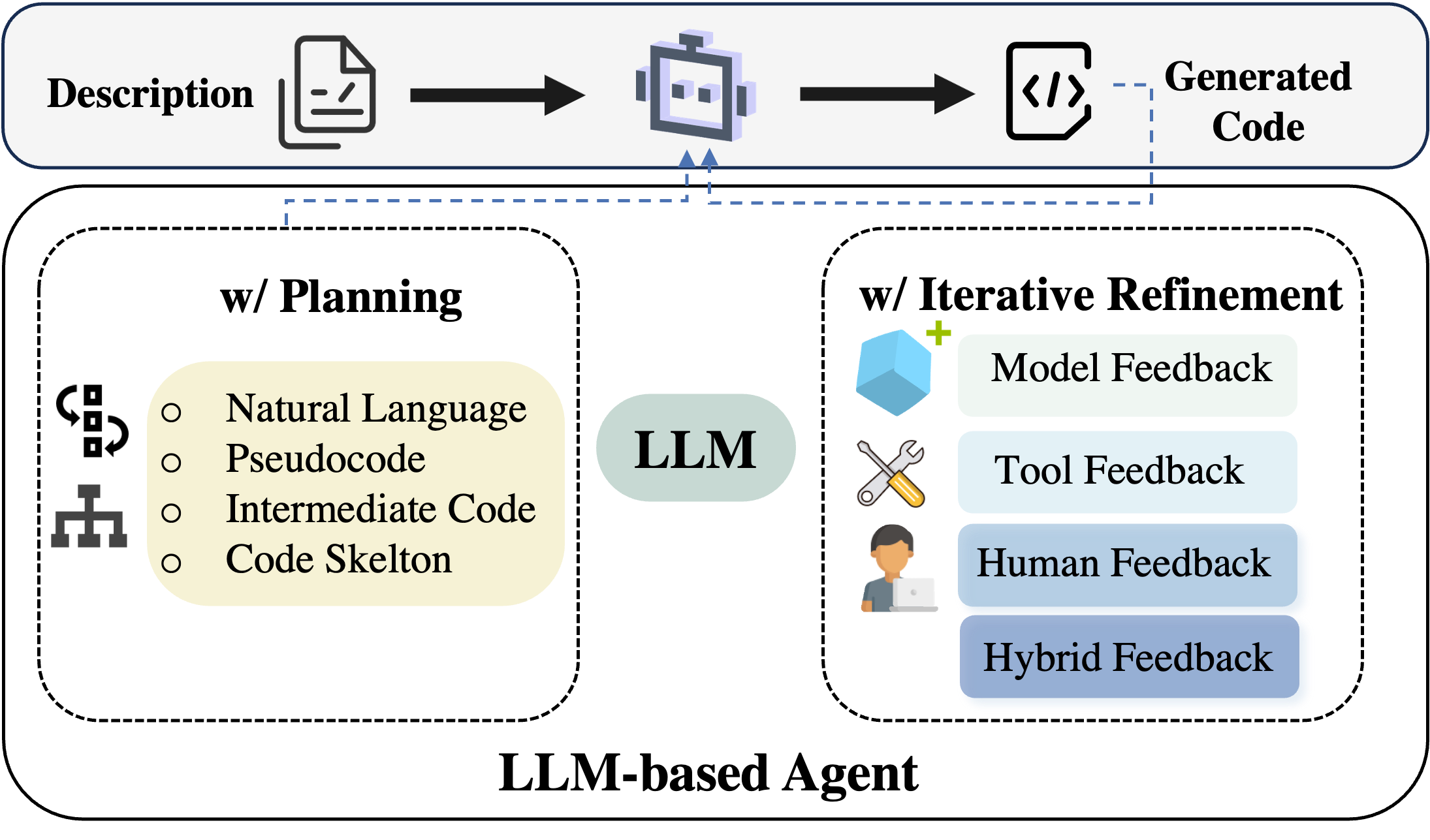}
    \caption{Pipeline of LLM-based Agents for Code Generation} 
    \label{fig:codegen}
\end{figure}
}
\subsubsection{Code Generation with Planning}
\junwei{LLM-based agents employ advanced strategies to extend the code generation capabilities, which can be basically divided into prompt engineering strategy and agentic strategy. }

\junwei{
\headt{Prompt engineering strategy}. Prompt engineering strategy refers to prompting an agent to break down the code generation task into step-by-step sub-tasks and achieve higher generation correctness.
Chain-of-thought (CoT)~\cite{cot} is the most popular prompt engineering strategy, which has two basic types: zero-shot CoT~\cite{zero-shot} and few-shot CoT~\cite{cot}. The difference between them is whether to provide completed task examples (shots) in the prompt as references. Among our collected papers, the number of studies using zero-shot CoT~\cite{parsel, he2024cocost, Flows, CodeChain} and few-shot CoT~\cite{liu2023dynamic, huang2024codecot, mint, AgentCoder} is roughly equal. Researchers who choose few-shot CoT often have additional formatting requirements for the output, necessitating at least one example as the format reference~\cite{mint, AgentCoder}.
For example, AgentCoder~\cite{AgentCoder} applies CoT on code generation with four predefined steps, \ie{} problem understanding
and clarification, algorithm and method selection, pseudocode creation, and code generation. Therefore, it provides an example to guide the model in following these steps during planning.
Moreover, these studies typically include merely one static pre-defined example rather than dynamically selecting examples based on relevance or similarity~\cite{zhang2023automatic, shi2023language}.
Despite the widespread application of CoT, the effectiveness of some other advanced prompt engineering strategies, such as self-consistency~\cite{wang2023selfconsistency} and least-to-most prompting~\cite{zhou2023leasttomost}, has not been explored in code generation tasks yet, which could be a potential direction for future research.}

\junwei{\headt{Agentic strategy.} On the other hand, \highlight{agentic strategy} refers to instructing agents to dynamically adapt the code generation plan based on historical thoughts, actions, and observations~\cite{shinn2023reflexion, CodeAct, bairi2023codeplan, yang2023intercode, CodeChain, Flows}.}
For example, CodePlan~\cite{bairi2023codeplan} employs an adaptive planning algorithm that dynamically detects the affected code snippets in the repository and adapts the modification plan accordingly. In addition, some works explore multi-path planning strategies. For example, LATS~\cite{LATs} simulates all possible generation paths as a tree and optimizes the plan with the Monte Carlo Tree Search algorithm. In MapCoder~\cite{MapCoder}, the planning agent generates multiple plans along with confidence scores for sorting. The highest-scoring plan is used to generate the target code. If the code is erroneous, the plan with the next highest confidence is selected to continue the iterative generation process. \junwei{Almost all of these agentic planning works assign only one agent as the planner; the only exception is Flows~\cite{Flows}, which compares the effectiveness of using a single planner versus a dual planner. However, the results demonstrate that the collaborative approach using two planners does not surpass the performance of using a single planner.}

\junwei{In addition to planning strategies, the planning representation also demonstrates significant diversity. Although text is still the most common form, some works propose to describe plans in several indirect code forms}, such as \textit{pseudocode}~\cite{AgentCoder}, \textit{intermediate code}~\cite{parsel}, and \textit{code skeleton}~\cite{SoA, CodeChain}. For example, AgentCoder~\cite{AgentCoder} prompts the agent to generate pseudocode after problem understanding and algorithm selection phases, which serves as a draft for the final code. \junwei{These code-based plans bridge the gap between the narrative-based steps and the final generated code, making them better suited for the code generation task.}

\junwei{
\headt{Comparison of Prompt Engineering Strategy and Agentic Strategy.}
Overall, code generation with planning is an important approach to decomposing programming steps and improving code generation accuracy. The two mainstream strategies, prompt engineering and agentic strategies, exhibit significant differences with respect to generalizability and planning iterations.
In terms of generalizability, prompt engineering strategies can be activated simply by incorporating straightforward instructions (\eg{} ``think step by step'') into the prompt, making them applicable to all instruction-following LLMs. In contrast, agentic strategies rely on environmental feedback and self-reflection mechanisms, making them suitable only for LLM-based agents. Regarding planning iterations, prompt engineering follows a one-time planning approach, where the plan is determined upfront and remains unchanged. In contrast, agentic strategies continuously refine the plan through iterative adjustments based on environmental feedback, allowing for more dynamic adaptability.
It is also worth noting that some studies have attempted to integrate traditional CoT strategies with agentic planning strategies. For example, RAT~\cite{RAT} proposes an iterative CoT optimization strategy. It uses the prefix steps along with the original prompt to retrieve information from the \textit{codeparrot/github-jupyter}~\cite{github_jupyter} dataset, which is then fed back to the agent for revising the next step in CoT iteratively. This combination makes RAT achieve better code generation accuracy than the basic CoT strategy.}

\subsubsection{Code Generation with Iterative Refinement}
One essential capability of agents is to act on the feedback from the environment. In the code generation scenario, agents also dynamically refine the previously-generated code based on the feedback via multiple iterations. We organize the relevant research based on the feedback sources, including model feedback, tool feedback, human feedback, and hybrid feedback. Table~\ref{tab:codegen} summarizes existing LLM-based agents for code generation with iterative refinement. Figure~\ref{fig:feedback} illustrates the four types of feedback.

\textbf{\begin{figure*}[htb]
    \centering
    \includegraphics[width=0.8\textwidth]{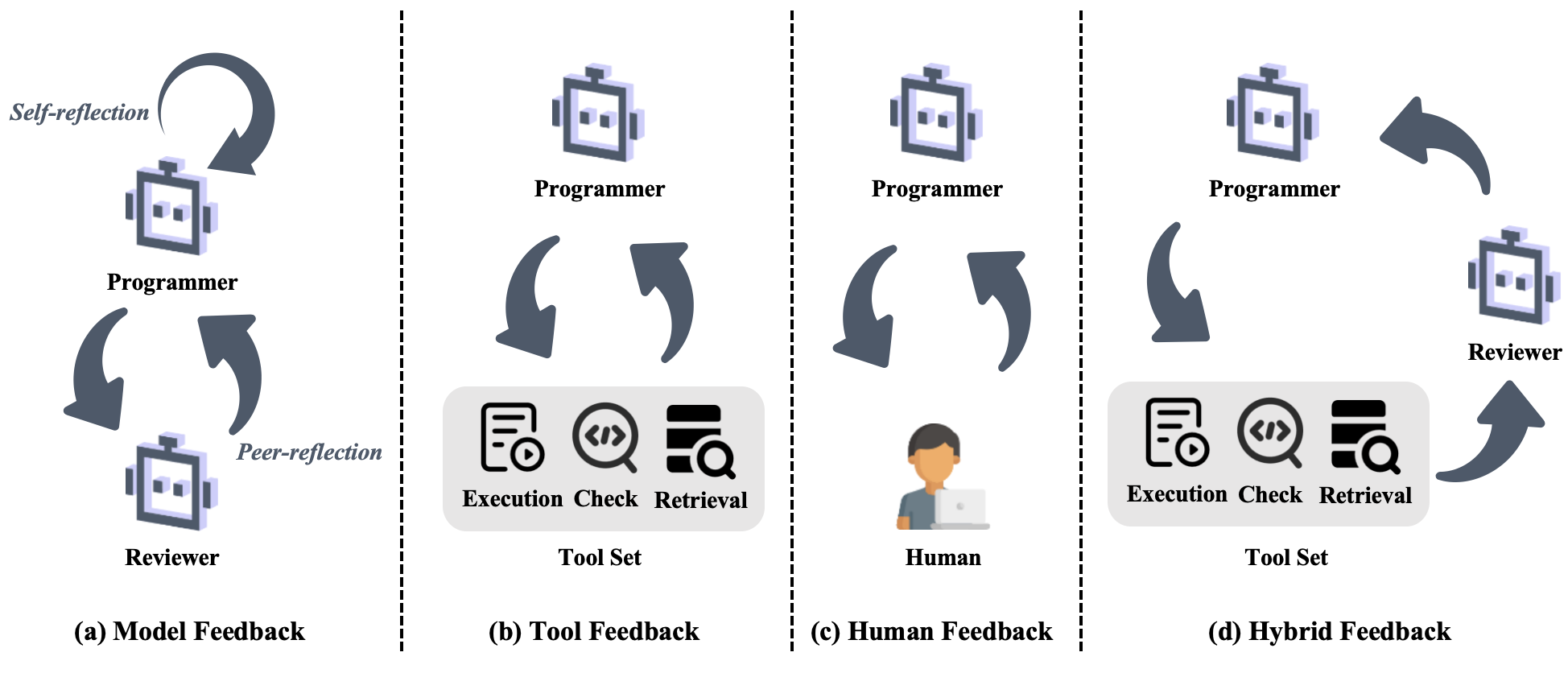}
    \caption{Four Types of Feedback in Code Generation} 
    \label{fig:feedback}
\end{figure*}
}

\headt{Model Feedback.} Model feedback can be classified into peer-reflection and self-reflection. 

\highlight{Peer-reflection} refers to information exchange and interaction between models, \ie{} the feedback is provided by other agents. The most common approach to facilitating peer-reflection is through role specialization and structured communication~\cite{li2023camel, shinn2023reflexion, AutoGen, liventsev2023fully,li2024agents}, which underscores the specialized responsibilities of each role and how they exchange information based on their responsibilities. For example, in AutoGen~\cite{AutoGen}, the SafeGuard agent will check the code safety and provide debugging feedback for the Writer agent.
\junwei{Moreover, there is a modality that treats each agent equally in expressing their opinions or engaging in debates to generate code, in which a selection mechanism is employed to retain the most suitable result. For example, AgentForest~\cite{li2024agents} and DyLAN~\cite{liu2023dynamic} both use Bilingual Evaluation Understudy (BLEU)~\cite{papineni2002bleu} to compute the similarity scores for code produced by each agent, aggregate these scores, and keep the top-scoring result.}

Apart from the interaction between different models, some works conduct \highlight{self-reflection}, in which the model will iteratively optimize its generated code based on the previous output~\cite{tian2024testcasedriven, chen2023teachinglargelanguagemodels, self-refine}. Le \et~\cite{CodeChain} guide LLMs to generate modularized code, leveraging cluster representatives from previously generated sub-modules in each iteration. Self-Debugging~\cite{chen2023teachinglargelanguagemodels} draws inspiration from the rubber duck debugging method used by programmers. During the explanation phase, the model provides a line-by-line explanation of the generated initial code. As Wang \et~\cite{mint} mention in their work, all models benefit from natural language feedback, with absolute performance gains by 2–17\% for each additional turn of natural language feedback.

\junwei{In summary, model feedback harnesses the contextual understanding and reasoning capabilities of LLMs. By mimicking the real-world iterative code-refinement process, the model can analyze generated code from either a programmer's perspective (self-reflection) or that of other expert roles (peer-reflection). Through step-by-step code interpretation, error detection, and contextual enrichment, the programming agent progressively enhances its understanding of the target code, thus leading to more accurate code generation. The reasoning ability of agents plays a crucial role in the iterative code generation process, distinguishing them significantly from traditional approaches based on programming templates~\cite{syriani2018systematic, Luhunu2017SurveyOT} or simple machine/deep learning methods~\cite{allamanis2018survey, mastropaolo2021studying}.}

\headt{Tool Feedback.}  
The code generated by models can be of limited quality with numerous uncertainties. One solution to address this challenge is to equip LLM-based agents with tools that can collect informative feedback and assist the agents in generating and refining code. 
\begin{itemize}[leftmargin=*, label=-]
    \item \textit{Dynamic Execution Tools.} One common group is to invoke the compiler, interpreter, and execution engine to directly compile or execute the code. This approach leverages the outputs and run-time behaviors, such as test results or compilation errors, as feedback for code improvement~\cite{self-edit, CodeAct, zhang2024codeagent, AgentCoder, piya2023llm4tdd, xu2023lemur, CodeChain, xu2023gentopia, huang2024codecot, jiang2023selfevolve, yang2023intercode, he2024cocost, MapCoder, ClarifyGPT, mint,SoA, AutoDev, 3DGen}.
    \item \textit{}{Static Checking Tools.} Agents can get more restricted knowledge of code constraints by applying code analysis tools. For example, some agents apply static analysis tools to obtain syntactically-valid program symbols/tokens~\cite{TOOLGEN} or dependencies between code during code generation~\cite{zhang2024codeagent, bairi2023codeplan}. Including the analyzed information in the prompt can guide LLMs toward generating valid code. 
    \item \textit{Retrieval Tools.} 
    Agents can access rich external resources by applying retrieval or searching tools. For example, some agents retrieve local knowledge repositories~\cite{he2024cocost} such as private API documentations~\cite{zhang2023toolcoder, zhang2024codeagent} and code repository~\cite{AutoDev} to facilitate better code generation; in addition, some apply online search engines~\cite{zhang2024codeagent, zhang2023toolcoder, xu2023gentopia,he2024cocost} or web crawling~\cite{FromMisusetoMastery} to collect information such as content from relevant websites (\eg{} \textit{StackOverflow} and \textit{datagy.io})~\cite{he2024cocost, zhang2023toolcoder, zhang2024codeagent, xu2023gentopia} and official online documentations~\cite{he2024cocost, FromMisusetoMastery}. Including the retrieved resources in the prompt can provide additional knowledge for language models. For example, ToolCoder~\cite{zhang2023toolcoder} integrates the agent with online search and local documentation search tools that provide helpful information for both public and private APIs, alleviating the hallucination of LLMs.

\end{itemize}

\junwei{In summary, tool feedback enhances code generation by integrating a wealth of mature coding tools and resources from the software engineering domain into LLM-based agents. This integration provides essential knowledge and diagnostic support, including but not limited to static analysis, runtime behavior monitoring, and performance evaluation across multiple dimensions. By leveraging these capabilities, tool feedback helps mitigate the inherent quality issues in LLM-generated code caused by hallucinations and randomness~\cite{hallucination_survey, LLM4CodeGeneration_survey}. }

\headt{Human Feedback.} Another approach involves incorporating human feedback into the process, as humans play a critical role in clarifying ambiguous requirements. 
For instance, in software development, humans can check whether the generated code aligns with their initial intent. Discrepancies are often attributed to vagueness or incompleteness in the requirements, prompting a revision of the requirement documents~\cite{Flows, mint}. 
To minimize human involvement in detecting vagueness, some methods enable the agent to handle the task of observing execution results.
For example, ClarifyGPT~\cite{ClarifyGPT} automatically identifies potential ambiguities in the manually-given requirements and proactively poses relevant questions for humans; then the responses from humans are further used to refine the requirements. 

\junwei{To sum up, the remarkable natural language understanding capabilities of LLM-based agents enable direct interaction with humans. In the context of code generation, this human-agent interaction allows the agent to confirm the user intent. By incorporating human feedback at key decision-making stages, such as requirement analysis, the agent can obtain clarifications and confirmations from the user, effectively preventing deviations from the original user intent and enhancing the consistency between generated code and user expectations.}

\headt{Hybrid Feedback.}  
\junwei{
To leverage the advantages of different types of feedback, some studies have explored integrating multiple feedback types to provide agents with a hybrid feedback mechanism.
While the three types of feedback (\ie{} model feedback, tool feedback, and human feedback) could theoretically be paired in various ways, we have merely observed the combination of tool feedback and model feedback in existing studies, which might stem from efforts to minimize human involvement and enhance automation.
Typically, these approaches first obtain precise error feedback of the generated code from program execution or testing tools. Then, an agent will analyze the tool feedback and provide explanations, suggestions, or new instructions accordingly~\cite{LDB, olausson2024selfrepair, chen2023teachinglargelanguagemodels, TGen, deshpande2024classlevelcodegenerationnatural, SoA, shinn2023reflexion, liventsev2023fully, tian2024testcasedriven, Flows, AutoCoder2405, LATs, AlphaCodium, AutoGen, INTERVENOR}. 
}
For example, in INTERVENOR~\cite{INTERVENOR}, a teacher agent is designated to observe the program execution results and provide error explanations and bug-fixing plans for the student coder to review and regenerate the code. 
\junwei{Nevertheless, for programs with complex data structures and control flows, merely analyzing the final execution outputs might be sub-optimal. Therefore, LDB~\cite{LDB} proposes a novel hybrid feedback mechanism that collects and analyzes the intermediate execution status of the generated program. Specifically, it divides the program into different code blocks based on the control flow graph and uses a breakpoint tool to collect the runtime states of variables before and after each code block's execution. Subsequently, the agent analyzes the runtime execution information along with the task description to explain the execution flow and assess the correctness of each block. The feedback is then fed to the agent to debug and regenerate a refined program. Experimental results show that LDB achieves better performance in code generation tasks compared to methods that merely analyze the final execution output (\eg{} Self-Debugging~\cite{chen2023teachinglargelanguagemodels}), demonstrating the unique contribution of intermediate state analysis.
}
\junwei{Overall, the current hybrid feedback methods leverage the strengths of both tool feedback and model feedback to provide interpretable information based on environmental feedback, thereby significantly improving the accuracy of model-generated code.}

\junwei{\textbf{Comparison of Iterative Feedback Mechanisms.}
The four types of feedback contribute to code generation from different perspectives, but each also comes with its own challenges.
While model feedback provides interpretable feedback, the inherent randomness and hallucination issues~\cite{hallucination_survey, LLM4CodeGeneration_survey} of feedback models can lead to cascading errors. Tool feedback, on the other hand, offers multi-dimensional code evaluation information, but this information often lacks necessary explanations and contextual relevance. Human feedback helps align the agent with human intent, but it reduces the method’s autonomy and introduces additional human effort.
However, these feedback mechanisms are complementary. For example, the lack of explanations and contextual relevance in tool feedback can be addressed through model feedback, while errors in model feedback can be mitigated by the precision of tool feedback. By adopting a hybrid feedback approach, the complementary strengths of different feedback mechanisms can be leveraged to maximize the advantages of all approaches.
However, in the current hybrid feedback approaches, the final feedback is still provided by the agent itself. Therefore, it may remain susceptible to cascading errors, requiring more innovative designs to mitigate this issue. }

\junweim{
\subsubsection{Common Failure Causes of LLM-based Agents in Code Generation.}
LLM-based agents have shown strong potential in automating code generation tasks, but they still face a number of common failure cases that limit their reliability and robustness in practice. These failures span across interaction quality, testing reliability, feedback effectiveness, and context management.
}

\junweim{
\textbf{Coordination Failures in Agent Collaboration.}
For agentic systems with multiple agents, it is common to encounter breakdowns in collaborative interactions due to poor coordination and role management. In systems like CAMEL~\cite{li2023camel}, agents may repeat user instructions without contributing new information, generate vague promises like “I will fix it”, or even fall into infinite conversational loops such as repeatedly saying “thank you” or “goodbye.” These issues typically stem from the agent’s inability to maintain consistent roles and task focus. To address these failures, potential solutions include limiting dialogue turns, enforcing token usage thresholds, or introducing explicit task-completion signals to terminate unproductive conversations.
}

\junweim{
\textbf{Low-Quality Tests Undermine Code Generation Accuracy.}
In many LLM-based code generation agents, the correctness of the generated code is assessed using automatically generated or existing test suites~\cite{shinn2023reflexion, tian2024testcasedriven, chen2023teachinglargelanguagemodels, he2024cocost}. However, flaky, incomplete, or poorly designed tests can mislead the self-correction process~\cite{shinn2023reflexion, MapCoder}. False positives occur when faulty tests pass incorrect solutions, causing premature task completion, while false negatives happen when correct code fails unreliable tests, leading to unnecessary revisions. Besides, insufficient test coverage is also a common bottleneck~\cite{tian2024testcasedriven}, which limits the confidence in the correctness of the generated code. 
Enhancing the reliability of test suites remains an open research direction.
}

\junweim{
\textbf{Cascading Errors from Incorrect or Noisy Feedback.}
Another common failure mode is the cascading effect caused by faulty feedback during iterative refinement. For LLM-based agents relying on model-generated feedback, the model may provide incorrect debugging suggestions~\cite{self-refine, liu2023dynamic,li2024agents}. For instance, in Self-Refine~\cite{self-refine}, 33\% of failed cases stemmed from feedback inaccurately identifying the error location, while 61\% resulted from feedback proposing inappropriate fixes.
Similarly, for LLM-based agents that utilize tool feedback, misleading tool outputs can propagate errors in subsequent iterations~\cite{he2024cocost, TOOLGEN}. For example, CoCoST~\cite{he2024cocost} depends on an online search tool to retrieve relevant information, but the retrieved content may contain inaccuracies. These findings highlight the critical importance of maintaining high-quality, well-calibrated feedback mechanisms throughout the generation process.
}

\junweim{
\textbf{Degraded Long-context Reasoning Capability.}
As the interaction history grows, LLM-based agents struggle to retain relevant information and maintain reasoning consistency. For example, both AutoGen~\cite{AutoGen} and InterCode~\cite{yang2023intercode} show that large volumes of accumulated context in multi-step refinement make it harder for agents to extract useful information for future actions. This leads to degraded performance over time. Potential solutions include increasing context window size, integrating memory management or retrieval modules, and developing adaptive planning strategies to maintain relevance across turns.
}

\junweim{
\subsubsection{Challenges of LLM-based Agents in Code Generation.} LLM-based agents for code generation face several key challenges. First, they heavily rely on test feedback to improve code quality, but high-quality tests are often unavailable or difficult to generate reliably, leading to concerns about test coverage and correctness~\cite{shinn2023reflexion, olausson2024selfrepair,huang2024codecot,ClarifyGPT, AgentCoder,piya2023llm4tdd,AlphaCodium,TGen,liu-etal-2025-llm}. Second, these agents typically require iterative refinement of the generated code, which introduces significant overhead compared to standalone LLM methods and poses challenges in terms of efficiency and cost~\cite{tian2024testcasedriven,FromMisusetoMastery,LATs,MapCoder}. Third, they depend on external tools to obtain feedback from the environment, but the reliability of these tools can be problematic~\cite{bairi2023codeplan, zhang2024codeagent,TOOLGEN,he2024cocost,deshpande2024classlevelcodegenerationnatural, RAT}. For example, the relevance of retrieved content or the accuracy of static analysis results may not always be guaranteed, affecting the agent's overall performance.
}

\subsection{Static Code Checking}
\label{sec:se:static}

Static code checking refers to examining the quality of code without executing the code. In particular, static code checking has been essential in the modern continuous integration pipeline, as it can identify diverse categories of code quality issues (\eg{} different bugs, vulnerabilities, or code smells) before extensively executing the tests. In practice, it is common to adopt static analysis techniques to automatically detect bugs/vulnerabilities (\ie{} static bug detection) or involve peer reviews to check the quality of code (\ie{} code review).

\subsubsection{Static Bug Detection}\label{sec:se:static:bug_detection}
Preliminary studies~\cite{LLM4SE1,LLM4SE2} show that LLMs can help identify potential quality issues in the given code under inspection. For example, fine-tuning LLMs on existing buggy/correct code or simply prompting LLMs has demonstrated promising effectiveness in identifying bugs, vulnerabilities, or code smells in the given code snippets~\cite{yuan2023evaluating, AdbGPT}. However, given the diversity and complexity of the root causes of different code issues as well as the long code contexts under inspection, standalone LLMs exhibit limited accuracy and recall in the real-world static code checking scenario~\cite{du2024vul}. Recently, researchers have built LLM-based agents to enhance the capabilities of isolated LLMs in vulnerability detection. Table~\ref{tab:bugdetection} summarizes these agents.

\begin{figure}[htb]
    \centering
    \includegraphics[width=1.0\columnwidth]{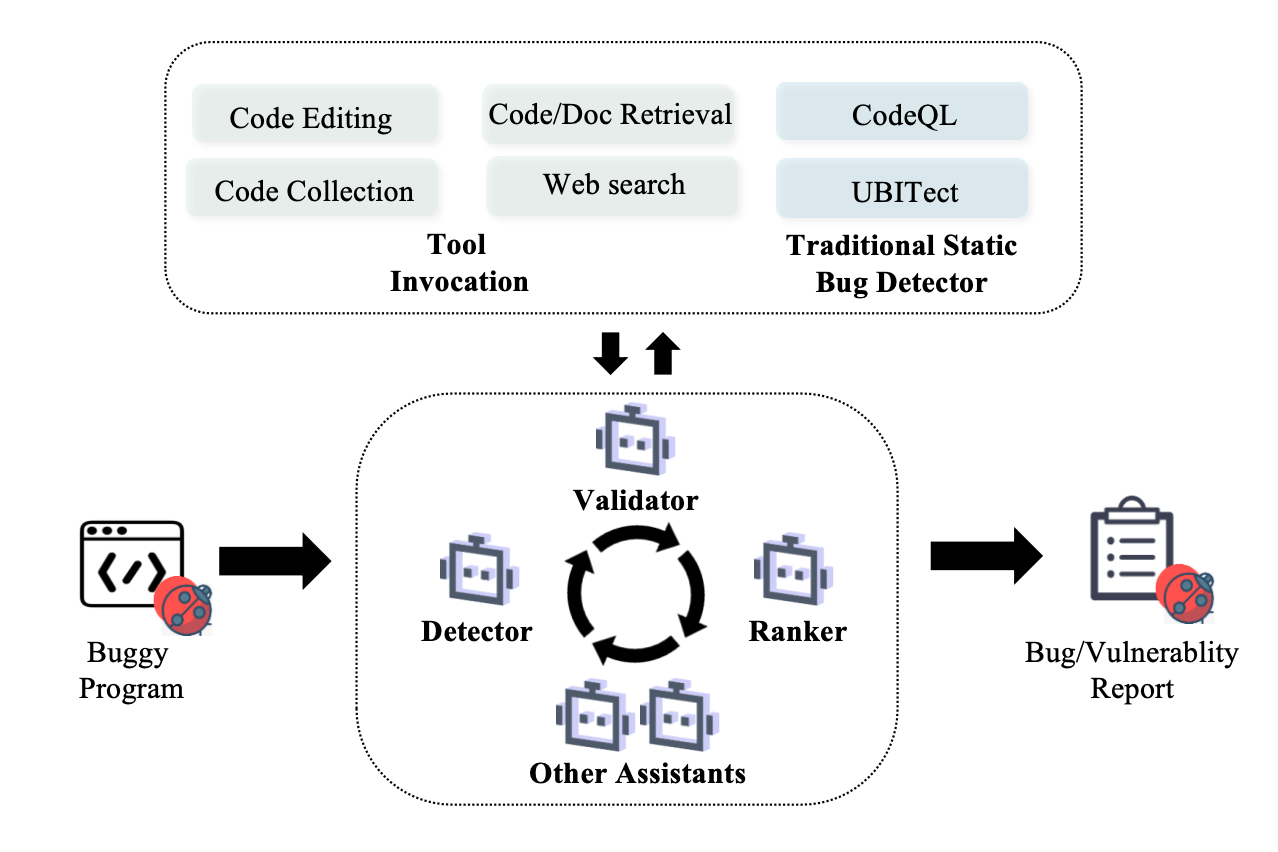}
    \caption{Pipeline of LLM-based Agents for Static Bug Detection} 
    \label{fig:bugdetect}
\end{figure}

\junwei{\textbf{Framework:} Figure~\ref{fig:bugdetect} illustrates the common framework of LLM-based agents for static bug detection. 
The agent systems will detect the input buggy programs and output a bug/vulnerability report. Based on responsibilities, there might be four types of roles in this process: the \textit{detector} is used to identify bugs in the code, the \textit{validator} is used to confirm and filter the detected bugs, the \textit{ranker} is used to rank the suspicious results, and different works might also involve some assistant roles (\eg{} planner and reporter).
These agents are equipped with a series of tools, including traditional bug detection tools such as CodeQL~\cite{avgustinov2016ql} and UBITect~\cite{zhai2020ubitect}.}

\begin{table*}[]
\centering
\caption{Existing LLM-based Agents for Static Bug Detection}
\renewcommand{\arraystretch}{1.4} 
\label{tab:bugdetection}
\begin{adjustbox}{width=\textwidth, keepaspectratio}

\begin{tabular}{lcccccc}
\hhline
\multirow{2}{*}{\textbf{Agents}} & \multirow{2}{*}{\textbf{Multi-Agent}} & \multicolumn{2}{c}{\textbf{Tool Utilization}} & \multirow{2}{*}{\textbf{Dataset}} & \multirow{2}{*}{\textbf{Target Program}} & \multicolumn{1}{c}{\multirow{2}{*}{\textbf{Bug Category}}} \\ \cline{3-4}
 &  & \multicolumn{1}{c}{\textbf{Tool Category}} & \textbf{Specific Tools} &  &  & \multicolumn{1}{c}{} \\ \hhline
ART~\cite{ART} & × & \multicolumn{1}{c}{Custom Toolkit} & \begin{tabular}[c]{@{}c@{}}Search Tool\\Code Generation Tool\\ Code Execution Tool\end{tabular} & BigBench~\cite{bigbench} & Python Program & Code Errors \\ \hline
\junweim{GPTLens}~\cite{GPTLENS} & \checkmark & \multicolumn{1}{c}{-} & - & Self-curated & Smart Contract & Smart Contract Vulnerability \\ \hline
ICAA~\cite{ICAA} & \checkmark & \multicolumn{1}{c}{Custom Toolkit} & \begin{tabular}[c]{@{}c@{}}Context  Splitting Tool\\ Code Retrieval Tool\\ Document Retrieval Tool\\ Web Search Tool\end{tabular} & \begin{tabular}[c]{@{}c@{}}NFBugs~\cite{non_fun_ds}\\ Self-curated\end{tabular} & \tabincell{c}{Python Program\\ Java Program} & \begin{tabular}[c]{@{}c@{}}Non-functional Bugs\\ API Misusage\end{tabular} \\ \hline
E\&V~\cite{EV} & × & \multicolumn{1}{c}{Static Analysis} & Clang~\cite{Clang} & Sampled syzbot~\cite{syzbot} & Linux Kernel & Kernel Address Sanitizer Bugs \\ \hline
LLM4Vuln~\cite{LLM4Vuln} & × & \multicolumn{1}{c}{Custom Toolkit} & \begin{tabular}[c]{@{}c@{}}Database Retrieval Tool\\ Context Collection Tool\end{tabular} & Self-curated & Smart Contract & Smart Contract Vulnerability \\ \hline
Mao \et{}~\cite{Multi-Role} & \checkmark & \multicolumn{1}{c}{-} & - & SySeVR~\cite{SySeVR} & C/C++ Program & \begin{tabular}[c]{@{}c@{}}Library/API Function Call\\ Arithmetic Expression\\ Array Usage\\ Pointer Usage\end{tabular} \\ \hline
IRIS~\cite{IRIS} & × & \multicolumn{1}{c}{Static Analysis} & CodeQL~\cite{avgustinov2016ql} & CWE-Bench-Java~\cite{IRIS} & Java Program & \begin{tabular}[c]{@{}c@{}}Path-Traversal\\ OS Command Injection\\ Cross-Site Scripting\\ Code Injection\end{tabular} \\ \hline
\junweim{LLift}~\cite{LLIFT} & × & \multicolumn{1}{c}{Static Analysis} & UBITect~\cite{zhai2020ubitect} & Rnd-300~\cite{LLIFT} & \tabincell{c}{Linux Kernel \\ C Program} & UBI Bugs \\ \hline
LLM4DFA~\cite{LLM4DFA}& \checkmark  & \multicolumn{1}{c}{Static Analysis} & \begin{tabular}[c]{@{}c@{}}tree-sitter~\cite{tree-sitter}\\ Z3 Solver~\cite{Z3SMT}\end{tabular} & Sampled Juliet Test Suite~\cite{Juliet} & Java Program & \begin{tabular}[c]{@{}c@{}}Divide-By-Zero (DBZ) bugs\\Cross-Site-Scripting (XSS) bugs\end{tabular}\\ \hline
PropertyGPT~\cite{PropertyGPT} & × & \multicolumn{1}{c}{Custom Toolkit} & Database Retrieval Tool & Self-curated & Smart Contract & Smart Contract Vulnerability \\ \hline
iAudit~\cite{iAudit} & \checkmark & \multicolumn{1}{c}{-} & - & Self-curated & Smart Contract & Smart Contract Vulnerability \\
\hhline
\end{tabular}

\end{adjustbox}
\end{table*}

\headt{Co-inspection with Multi-agent.} One effective vulnerability detection strategy focuses on the perspective of multi-agent collaboration. 
\junwei{Mao \et{}~\cite{Multi-Role} propose an approach for vulnerability detection through mutual discussion and consensus among the developer agent and the tester agent, which mimics the real-world code debugging process.
GPTLens~\cite{GPTLENS} is a two-stage framework for detecting vulnerabilities in smart contracts, where multiple auditor agents first generate potential vulnerabilities, and a critic agent then ranks the candidates to get the top-k vulnerabilities as the output. The evaluation on 13 real-world smart contracts demonstrates a successful vulnerability detection rate of 76.9\%.
Fan \et{}~\cite{ICAA} propose a code analysis framework named ICAA, with bug detection as a typical application. ICAA involves a linear pipeline for bug detection. After task planning and code preprocessing, ICAA assigns a react-based analysis agent which is equipped with a series of tools (\eg{} retrieval and static analysis tools) to identify bugs. The detected bugs will then be extracted by the report agent and filtered by the false pruner agent, with the refined report as the output.} 
The iAudit~\cite{iAudit} framework is a multi-agent system for smart contract auditing with justifications. It finetunes a Detector model alongside a Reasoner to determine whether the code is vulnerable and provide candidate explanations. Subsequently, an iterative debate ensues between the Ranker and the Critic to select the most compelling justification.

\headt{Additional Knowledge from Tool Execution.}
Another research direction is to enhance the knowledge of LLMs through tool invocation.
\junwei{ART~\cite{ART} is a framework designed to finish tasks by automatically generating multi-step reasoning decompositions and choosing appropriate tools (like search engines and code execution tools) with the help of retrieved task demonstrations. Employing ART for bug detection outperforms both few-shot prompting and the automated generation of CoT reasoning.}
ICAA~\cite{ICAA} also provides a document retrieval tool and a web search tool to enhance the bug analysis agent with both local and online knowledge.
LLM4Vuln~\cite{LLM4Vuln} enhances the vulnerability reasoning capabilities of LLMs by integrating various knowledge. First, it retrieves both the related vulnerability reports and the summarized vulnerability knowledge from self-constructed databases. Second, the agent invokes tools to obtain further context about the target code (\eg{} function or variable definitions). With the help of enriched knowledge, the agent identified 14 zero-day vulnerabilities in four pilot bug bounty programs.
PropertyGPT~\cite{PropertyGPT} is an agent designed to generate properties for the formal verification of smart contracts. When developing customized properties for the subject code, it first retrieves the knowledge base to find similar code and their reference properties. Based on this knowledge, PropertyGPT produces a set of candidates and refines them iteratively, resolving compilation issues through feedback from the compiler. These properties are then ranked and verified by the prover.

\headt{Combined with Traditional Static Bug Detection.}
Some researchers have combined LLM-based agents with traditional static checking techniques to improve their static bug detection capability. 
\junwei{\junweim{LLift}~\cite{LLIFT} is an agent framework for detecting Use-Before-Initialization (UBI) bug in Linux kernel, which is built on traditional UBI detection tool UBItect~\cite{zhai2020ubitect}. UBITect uses a two-stage pipeline: first, it performs flow-sensitive but path-insensitive static analysis to identify potential UBI bugs, which is fast but imprecise. In the second stage, symbolic execution filters out false positives by exploring feasible paths, but 40\% of the bugs are discarded due to time or memory limits. 
Based on these undecided bugs reported, \junweim{LLift} utilizes agents to identify potential initializers for a suspicious variable from a bug report, extract post-conditions for each initializer, and summarize the initialization status of the variable based on these initializers. Variables without any initializer that must initialize them are potential vulnerabilities.
E\&V~\cite{EV} is an agent designed to perform a static analysis of the Linux kernel code. The workflow of E\&V is a loop of employing an LLM-based agent for static analysis through pseudo-code execution, verifying the output of pseudo-code, and providing feedback for reanalysis. To mitigate hallucinations from missing necessary code (\eg{} inter-procedural call graphs), it retrieves required functions via traditional static analysis tools (\eg{} Clang~\cite{Clang}).}
IRIS~\cite{IRIS} is an agent augmented with CodeQL (a static analysis tool)~\cite{avgustinov2016ql} for vulnerability detection. IRIS first utilizes CodeQL to extract candidate APIs in the given repository. Then, it labels these APIs as potential sources or sinks of the given vulnerability via querying the LLM-based agent, which will be further handed over to CodeQL for detecting vulnerable paths. The final verdict is achieved by prompting the LLM agent to analyze the vulnerable paths and the surrounding code of the source and sink. 
\junwei{LLM4DFA~\cite{LLM4DFA} is a multi-agent system that employs data flow analysis to pinpoint dataflow-related bugs (\eg{} divide-by-zero bugs and cross-site-scripting bugs). The process unfolds in three stages: initially, it synthesizes scripts to extract sources and sinks from code using a parsing library (\ie{} tree-sitter~\cite {tree-sitter}). Then, the summarizer agent discerns dataflow facts within functions via few-shot learning and CoT prompting. Finally, the agent generates scripts to validate the dataflow facts against path conditions by invoking an SMT solver~\cite{Z3SMT}. The extraction and validation scripts are both refined by the agent through self-evaluation.}

\junwei{\textbf{Comparison of Bug Detection Enhancement Strategies.} In summary, the three most extensively employed approaches in static code checking agents are multi-agent collaboration, knowledge enhancement, and integration of static analysis tools. Multi-agent collaboration enhances detection effectiveness through the division of labor and task distribution across specialized agents. For single-agent approaches, it is more necessary to integrate additional knowledge and tools, which contribute to bug detection by providing the necessary vulnerability knowledge and code contexts. In particular, traditional static analysis tools can also be integrated into the agent-based methodology, which balances the semantic understanding ability of LLM with the precision of rule-based tools, thereby enhancing detection efficiency and accuracy. In addition, these three strategies can be combined to achieve better performance. For example, LLM4DFA~\cite{LLM4DFA} employs both multi-agent collaboration and static analysis tools.}

\subsubsection{Code Review} 
Developers review each other’s code changes to ensure and improve the code quality before merging the changes into the branch. To mitigate the manual efforts in code review, researchers leverage learning approaches to automate the code review procedure. In particular, code review is formulated as a binary classification problem (\ie{} code quality classification~\cite{khoshgoftaar2004comparative}) or a sequence-to-sequence generation problem (\ie{} review comment generation~\cite{li2022auger}), which are tackled by fine-tuning or prompting deep learning models (including LLMs). Different from these works, LLM-based agents mimic the real-world peer review procedure by including multiple agents as different code reviewers. Table~\ref{tab:code_review} summarizes existing agents for code review.
\begin{table*}[htb]
\centering
\caption{Existing LLM-based Agents for Code Review}

\label{tab:code_review}
\renewcommand{\arraystretch}{1.3}
\begin{adjustbox}{width=1.0\textwidth}
\begin{tabular}{
    l  
    >{\centering\arraybackslash}m{8cm}
    cccc        
}
\hhline

\multirow{2}{*}{\textbf{Agents}} & \multirow{2}{*}{\textbf{Multi-Agent Roles}} & \multicolumn{4}{c}{\textbf{Review Target}} \\ \cline{3-6} 
&  & \multicolumn{1}{c}{\textbf{Consistency}} & \multicolumn{1}{c}{\textbf{Vulnerability}} & \multicolumn{1}{c}{\textbf{Code Smell}} & \multicolumn{1}{c}{\textbf{Code Optimization}} \\ \hhline

CodeAgent~\cite{tang2024codeagent} &  User, CEO, CPO, CTO, Coder, Reviewer & \checkmark & \checkmark & \checkmark & \checkmark \\ \hline
Rasheed \et{}~\cite{AIPoweredCodeReview} & Code Review, Bug Report, Code Smell, Code Optimization Agent &  & \checkmark & \checkmark & \checkmark \\ \hline
ICAA~\cite{ICAA} & Context \& Prompt Incubation Agent, Consistency Checking Agent, Report Agent & \checkmark &  &  &  \\ 
\hline
CORE~\cite{CORE} & Proposer LLM, Ranker LLM &  &  &  & \checkmark \\ 
\hhline
\end{tabular}
\end{adjustbox}
\end{table*}

\junweim{
\textbf{Process-based Multi-agent Code Review.}
Process-based multi-agent systems organize agents to follow a structured, sequential workflow that mirrors traditional human-driven code review pipelines. By dividing the review process into distinct stages, these systems coordinate multiple agents with specialized roles to collaboratively complete tasks such as information gathering, code analysis, revision, and documentation in an orderly manner.
}

\junweim{
CodeAgent ~\cite{tang2024codeagent} is a multi-agent system that simulates a waterfall-like pipeline with four stages (\ie{} basic information synchronization, code review, code alignment, and document) and sets up a code review team with six agents of different characters (\ie{} user, CEO, CPO, CTO, coder, and reviewer). 
In the basic information synchronization phase, the CEO, CPO, and coder agents analyze the input modality and programming language. After that, the coder and reviewer agents collaborate to conduct a code review and produce the analysis report. In the code alignment phase, the coder and reviewer agents continue to revise the code based on the analysis reports. Finally, in the document phase, the CEO, CTO, and coder agents cooperate to document the holistic code review process. Experimental results demonstrate the effectiveness and efficiency of CodeAgent in various code review tasks, including consistency analysis, vulnerability analysis, format analysis, and code revision. 
}

\junweim{
ICAA~\cite{ICAA} designs a multi-agent system to identify code-intention inconsistencies. It first uses the Context \& Prompt Incubation Agent to collect necessary information from the code repository through a thinking-decision-action loop.  The Consistency Checking Agent will then analyze collected information and identify inconsistencies, which will be handed over to the Report Agent to form a final report.
}

\junweim{
CORE~\cite{CORE} designs a system with two agents, along with traditional static analysis tools to fix code quality issues automatically. 
Specifically, the Proposer agent takes the static analysis report, the suspicious file, and the issue documentation from language-specific static analysis tools (\eg{} CodeQL~\cite{avgustinov2016ql}) and the tool provider (\eg{} the QA team), and proposes candidate revisions for each suspicious file. After that, static analysis tools will prune revisions that still have issues, while the rest will be scored and re-ranked based on their likelihood of acceptance by the Ranker agent.
}

\junweim{
\textbf{Goal-based Multi-agent Code Review.}
Goal-based multi-agent systems focus on assigning agents to specialize in individual, well-defined tasks within the code review domain. Instead of following a fixed pipeline, each agent independently addresses a specific aspect of code quality, such as bug detection or code optimization, allowing modular development and targeted expertise for different review objectives. For example, Rasheed \et{} ~\cite{AIPoweredCodeReview} design an approach with each agent specialized for a single code review task. Notably, it proposes four agents, including the code review agent, bug report agent, code smell agent, and code optimization agent. Each agent is trained on relevant GitHub data and evaluated on 10 AI-based projects. The results demonstrate the potential of applying multi-agent systems in the code review task.
}

\junweim{
\textbf{Comparison of Different Multi-agent Code Review Strategies.}
Process-based multi-agent systems emphasize structured workflows where agents collaborate through well-defined sequential stages, enabling comprehensive and coordinated management of the code review lifecycle. This approach ensures clear role assignments and orderly progression of tasks. In contrast, goal-based systems prioritize specialization by assigning agents to independently tackle distinct code review subtasks. Such modularity allows focused expertise and flexible scalability but may lack the integrated coordination found in process-oriented designs. In summary, process-based frameworks offer end-to-end orchestration suitable for holistic review, while goal-based frameworks optimize for task-specific performance. Combining these paradigms may yield more effective multi-agent code review systems.
}

\junweim{
\subsubsection{Challenges of LLM-based Agents in Static Code Checking.} LLM-based agents face several challenges in static code checking tasks. First, these agents often incorporate traditional static bug detection tools such as UBITect and CodeQL. However, the integration remains relatively shallow. In many cases, the agents merely filter false positives based on the outputs of these tools, rather than enabling deeper collaboration. A tighter integration between the model and external tools could potentially improve performance and reduce noise~\cite{LLIFT, IRIS}. Second, static analysis tasks often require LLM-based agents to possess both a strong understanding of code and the capability to generate intermediate representations, such as pseudocode~\cite{EV}, execution specifications~\cite{EV}, or dataflow summaries~\cite{LLM4DFA}. These requirements place greater demands on the model's reasoning capabilities and often necessitate the use of high-performing proprietary models such as GPT-4~\cite{EV, LLIFT}.
Finally, LLM-based agents may still produce false positives, and there is a lack of effective automated mechanisms for verifying or filtering these incorrect results, which further limits their practical reliability~\cite{GPTLENS, ICAA, iAudit}.
}

\subsection{Testing}\label{sec:se:test}
Software testing is essential for software quality assurance. LLMs have demonstrated promising proficiency in test generation, including generating test code, test inputs, and test oracles. However, generating high-quality tests in practice can be challenging, as the generated tests should not only be syntactically and semantically correct (\ie{} both the inputs and oracles should satisfy the specification of the software under test) but also be sufficient (\ie{} the tests should cover as many states of the software under test as possible). As shown by previous work~\cite{Chattester}, the tests generated by standalone LLMs still exhibit correctness issues (\ie{} compilation errors, run-time errors, and oracle issues) and unsatisfactory coverage. Therefore, researchers build LLM-based agents to extend the capabilities of standalone LLMs in test generation.

\begin{table*}[htb]
\centering
\renewcommand{\arraystretch}{1.3}
\caption{Existing LLM-based Agents for Unit Testing}
\label{tab:unittest}
\begin{adjustbox}{width=1.0\textwidth}
\begin{tabular}{
    lc
    >{\centering\arraybackslash}m{5cm}
    cc
    }

\hhline
\textbf{Agents} & \textbf{Multi-Agent} & \textbf{Feedback Goal} & \textbf{Feedback Source} & \textbf{ Target Language} \\ \hhline

ChatTester~\cite{Chattester} &  $\times$  & Reduce compilation/execution errors & Error messages & Java, Python \\ \hline
TestPilot~\cite{TestPilot} &  $\times$  & Reduce compilation/execution errors & Error messages & JavaScript \\\hline
ChatUniTest~\cite{chatunitester} &  $\times$  & Reduce compilation/execution errors & Error messages & Java  \\ \hline
AgentCoder~\cite{AgentCoder} & $\times$ & Reduce compilation/execution errors & Error messages & Python \\\hline
TELPA~\cite{TELPA} &  $\times$ & Increase coverage &  Program analysis results & Python \\\hline
CoverUp~\cite{CoverUp} &  $\times$   & Increase coverage & Execution results \& Coverage & Python \\\hline
MuTAP~\cite{MuTAP} &  $\times$ &  Enhance fault detection & Surviving mutants & Python\\  \hline
AutoDev~\cite{AutoDev} &  $\checkmark$ & \tabincell{c}{Reduce compilation/execution errors\\Increase coverage} & Error messages & Java\\\hline
Mokav~\cite{Mokav} &  $\times$  & Enhance fault detection & Execution results & Python \\ 
\hhline

\end{tabular}
\end{adjustbox}
\end{table*}

\begin{figure}[htb]
    \centering
    \includegraphics[width=\columnwidth]{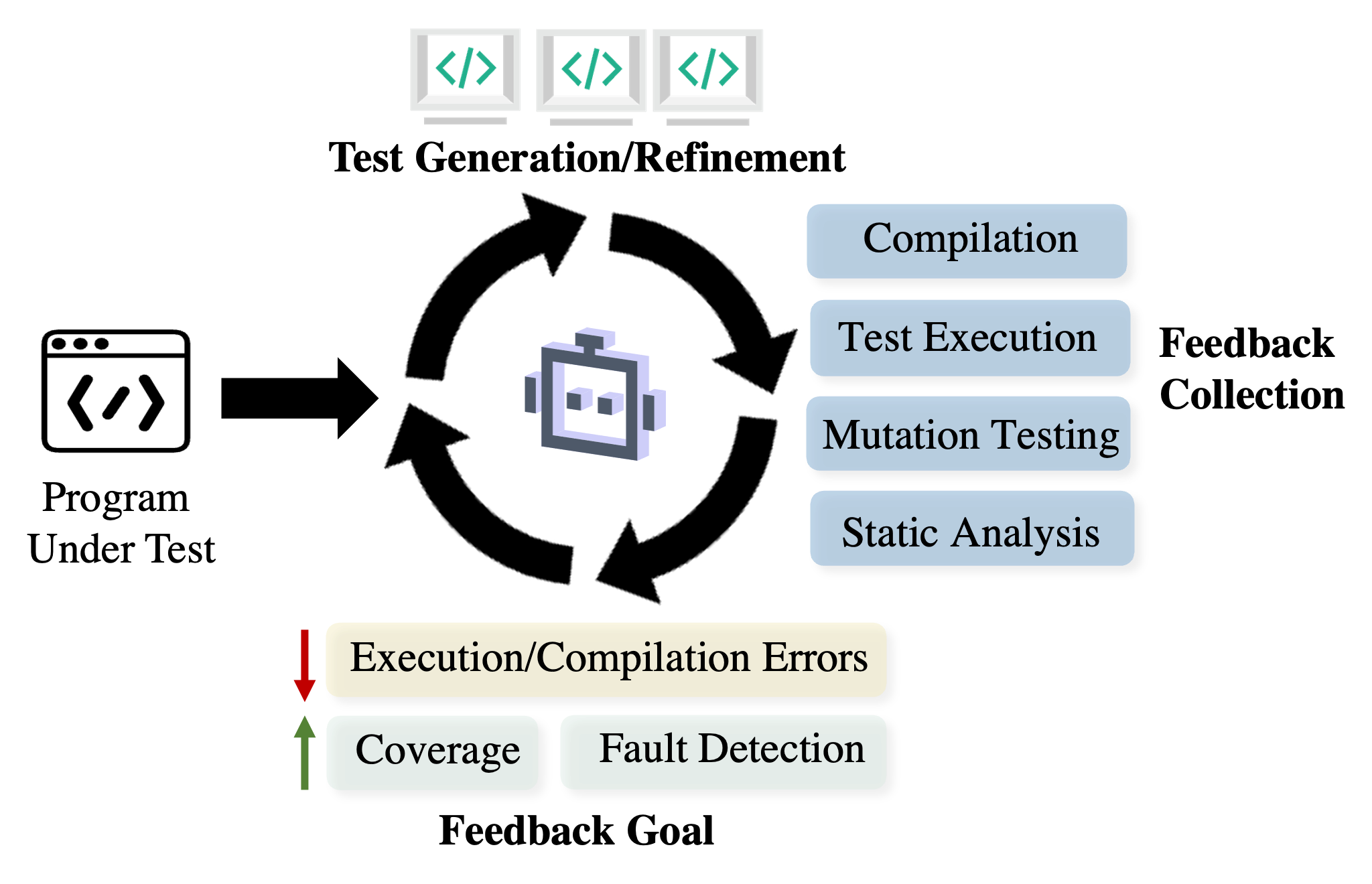}
    \caption{Pipeline of LLM-based Agents for Unit Testing} 
    \label{fig:ut}
\end{figure}

\subsubsection{Unit Testing}
Unit testing checks the isolated and small unit (\eg{} method or class) in the software under test, which helps quickly identify and locate bugs, especially for complicated software systems. Yuan~\et{}~\cite{Chattester} perform a study showing the potentials of LLMs (\eg{} ChatGPT) in generating unit tests with decent readability and usability. However, the unit tests generated by standalone LLMs still exhibit compilation/execution errors and limited coverage. 
\junwei{
Therefore, recent works have built LLM-based agents that extend standalone LLMs by iteratively refining the generated unit tests with distinct strategies and targets. We organize this section around the three primary enhancement directions, including reducing compilation/execution errors, improving test coverage, and enhancing the fault detection capability of tests.}
Table~\ref{tab:unittest} summarizes the existing LLM-based agents for unit test generation.

\junwei{\textbf{Framework:} Figure~\ref{fig:ut} illustrates the common framework of LLM-based agents for unit testing. Agent-based testing systems take the program under test (PUT) as input, generate initial test cases, and iteratively refine them by leveraging feedback from external tools such as compilation/testing outputs and static analysis results. The refinement target includes reducing execution/compilation errors, improving test coverage, and enhancing the fault detection capability.
}

\headt{Iterative Refinement to Fix Compilation/Execution Errors.}
The test cases directly generated by LLMs can exhibit compilation or execution errors. Therefore, inspired by program repair~\cite{ConversationalAPR}, LLM-based agents further eliminate such errors by iteratively collecting the error messages and fixing the buggy test code~\cite{TestPilot, Chattester, chatunitester, AutoDev, AgentCoder}.
\junwei{Existing LLM-based agents used for unit test generation all follow a generate-verify-fix pipeline, where error messages from test execution are fed back to refine the test script. The primary differences among these agents lie in the context integrated into their prompts. ChatUniTester~\cite{chatunitester} integrates code context into the prompt, including the focal method and class, as well as dependent methods and classes. AgentCoder~\cite{AgentCoder} designs a test generation prompt with three clear objectives: (i) to generate basic test cases, (ii) to cover edge test cases, and (iii) to cover large-scale inputs. This prompt enhances the accuracy and adequacy of the generated test scripts. TestPilot~\cite{TestPilot} designs a complicated prompt, including the signature, definition, doc comment, and usage snippets extracted from the documentation of the focal function. 
ChatTester~\cite{Chattester} introduces an intention prompt before the test generation prompt, which instructs the LLM to understand the intention of the focal methods first. 
Different from the above approaches, AutoDev~\cite{AutoDev}, as a React-style agent, primarily relies on feedback to improve the accuracy of test generation instead of prompt construction. 
In summary, different approaches for unit testing generation tend to focus on different contexts, such as more detailed code and documentation context, more specific test generation goals, and clearer method intentions. However, due to the different evaluation datasets and criteria used by each method, direct cross-method comparisons of their effectiveness are not feasible, which poses a challenge for establishing a unified standard.}

\headt{Iterative Refinement to Increase Coverage.} \junwei{In addition to enhancing the success rate of test execution, the feedback mechanism in LLM-based agents can also be employed to improve the coverage of unit testing.}
\junwei{CoverUp~\cite{CoverUp} is an LLM-powered test generation system aimed at achieving higher coverage rates. It initially employs the coverage analysis tool SlipCover~\cite{SlipCover} to measure existing test suite coverage and identify uncovered code segments using abstract syntax trees. This information, along with the methods under test, is provided to the LLM to generate new tests. If coverage does not increase after test execution (verified by iteratively invoke SlipCover with near-zero overhead), error messages are fed back to the LLM for re-generation. In addition to iterative feedback, the LLM is equipped with a tool to query types or variables in the code segments, which facilitates test generation.
TELPA~\cite{TELPA} is an LLM-based agent for enhancing test generation for hard-to-cover branches through iterative feedback. Based on the tests generated by existing tools (\eg{} the search-based software test technique Pynguin~\cite{Pynguin} and the LLM-based technique \junweim{CodaMosa}~\cite{lemieux2023codamosa}) to cover easy-to-reach branches, it performs backward and forward method invocation analyses to extract relevant information for constructing complex objects and understanding inter-procedural dependencies. Finally, TELPA employs a feedback-based process with the LLM, using counter-examples to iteratively refine and generate tests that improve coverage of difficult branches.
}
AutoDev~\cite{AutoDev} iteratively generates, executes, and revises tests, achieving 99.3\% test coverage on the HumanEval~\cite{Humaneval} dataset,  which is comparable to the human-written tests’ coverage.

\headt{Iterative Refinement to Increase Fault Detection Capabilities.} 
\junwei{Besides execution success rate and test coverage, the quality of test cases has also garnered research attention, particularly in generating test cases with enhanced fault detection capabilities.}
MuTAP~\cite{MuTAP} is a single LLM-based agent system that aims at generating unit tests of better bug detection capabilities with the feedback of mutation testing. It employs prompt augmentation with surviving mutants and refining steps to correct syntax and intended behavior. During each iteration, the LLM first generates initial test cases and self-refines their syntax errors and wrong behaviors, with the help of the Python parser to locate the erroneous line. Then the tests run against the mutated programs, while the surviving mutants serve as feedback to direct the LLM in improving the test cases. Mokav~\cite{Mokav} is an agent to generate difference-exposing tests. It first summarizes the descriptions of the two programs under test. Subsequently, Mokav engages in an iterative process, crafting tests that are refined based on the feedback from execution outcomes, until the tests are capable of exposing the distinctions between the two programs.

\subsubsection{System Testing}~\label{sec:systest}
System testing is a comprehensive process that assesses an integrated software system/component to guarantee that it fulfills its specification and operates as intended across diverse settings. For example, fuzzing testing and GUI (Graphical User Interface) testing are common testing paradigms at the system level. 
\junwei{Leveraging LLMs for system testing can be challenging, as generating valid and effective system-level test cases should satisfy the constraints that are contained implicitly and explicitly in the specifications or domain knowledge of the software system under test. Besides, the entire system not only involves multiple interacting components and modules but also contains different execution paths and behaviors, resulting in numerous dependencies, interactions and scenarios that standalone LLMs may struggle to fully obtain and account for when generating test cases. LLM-based agents are designed to better incorporate the domain knowledge and dynamically explore the software system under test compared to generating system-level tests via standalone LLMs.
}
We then organize these works according to the software systems under test.
Table~\ref{tab:systemtest} summarizes the existing agents for different software systems.

\begin{table*}[htb]
\centering
\caption{Existing LLM-based Agents for System Testing}
\label{tab:systemtest}
\begin{adjustbox}{width=1.0\textwidth}
\renewcommand{\arraystretch}{1.3} 
\begin{tabular}{
  cccccc
}
\hhline
\multirow{2}{*}{\textbf{Software System}} & \multirow{2}{*}{\textbf{Agents}} & \multirow{2}{*}{\textbf{Multi-Agent}} & \multicolumn{2}{c }{\textbf{Tool}} & \multirow{2}{*}{\textbf{Output}} \\ \cline{4-5}
 &  &  & \multicolumn{1}{c }{\textbf{Tool Category}} & \textbf{Specific Tools} &  \\ \hhline
OS Kernel & KernelGPT~\cite{KernelGPT} & × & \multicolumn{1}{c }{Static Analysis} & \begin{tabular}[c]{@{}c@{}}syz-extract~\cite{Syzkaller}\\ LLVM Toolchain~\cite{LLVMtoolchain}\end{tabular} & Syzkaller Specifications \\ \hline
\multirow{2}{*}{Compiler} & WhiteFox~\cite{WhiteFox} & \checkmark & \multicolumn{1}{c }{-} & - & Test Cases \\ \cline{2-6} 
 & LLM4CBI~\cite{LLM4CBI} & × & \multicolumn{1}{c }{Static Analysis} & \begin{tabular}[c]{@{}c@{}}OClint~\cite{OCLint}\\ srcSlice~\cite{srcSlice}\\ Gcov~\cite{Gcov}\\ Frama-C~\cite{Frama-C}\end{tabular} & Mutated Programs \\ \hline
\multirow{6}{*}{Mobile App} & GPTDroid~\cite{GPTDroid} & × & \multicolumn{1}{c }{Execution Environment} & \begin{tabular}[c]{@{}c@{}}VirtualBox~\cite{virtualbox}\\ pyvbox~\cite{pyvbox}\\ Android UIAutomator~\cite{UIAutomator}\\ Android Debug Bridge~\cite{adb}\end{tabular} & Test Scripts \\ \cline{2-6} 
 & DroidAgent~\cite{DroidAgent} & \checkmark & \multicolumn{1}{c }{Custom Toolkit} & Navigation Action Toolkit & Test Scripts \\ \cline{2-6} 
  & InputBlaster~\cite{InputBlaster} & × & \multicolumn{1}{c }{Execution Environment} & Android UIAutomator~\cite{UIAutomator} & Unusual Text Inputs \\ \cline{2-6} 
 & AXNav~\cite{AXNav} & \checkmark & \multicolumn{1}{c }{Custom Toolkit} & Navigation Action Toolkit & Bug Replay Video \\ \cline{2-6} 
 & AdbGPT~\cite{AdbGPT} & × & \multicolumn{1}{c }{Execution Environment} & \begin{tabular}[c]{@{}c@{}}Genymotion~\cite{Genymotion}\\ Android UIAutomator2~\cite{uiautomator2}\\ Android Debug Bridge~\cite{adb}\end{tabular} & Bug Replay Steps \\ \cline{2-6} 
  & VisionDroid~\cite{VisionDroid} & \checkmark & \multicolumn{1}{c }{Execution Environment} & \begin{tabular}[c]{@{}c@{}}VirtualBox~\cite{virtualbox}\\ pyvbox~\cite{pyvbox}\\ Android UIAutomator~\cite{UIAutomator}\\ Android Debug Bridge~\cite{adb}\end{tabular} & Detected Bugs \\ \cline{2-6} 
 & XUAT-Copilot~\cite{XUAT-Copilot} & \checkmark & \multicolumn{1}{c }{-} & - & Test Scripts \\ \hline
Web App & RESTSpecIT~\cite{RESTSpecIT} & × & \multicolumn{1}{c }{-} & - & OpenAPI Specification \\ \hline
\multirow{4}{*}{Universal} & Fuzz4All~\cite{fuzz4all} & \checkmark & \multicolumn{1}{c }{-} & - & Test Cases \\ \cline{2-6} 
 & PentestGPT~\cite{PentestGPT} & \checkmark & \multicolumn{1}{c }{Testing Tool} & Metasploit~\cite{Metasploit} & Test Operations \\ \cline{2-6} 
 & \multirow{2}{*}{Fang~\et{}~\cite{fang2024llmagentsautonomouslyexploit}} & \multirow{2}{*}{×} & \multicolumn{1}{c }{Custom Toolkit} & \begin{tabular}[c]{@{}c@{}}Web Browsering Tool\\ File Creation and Editing Tool\end{tabular} & \multirow{2}{*}{Exploit Actions} \\ \cline{4-5}
 &  &  & \multicolumn{1}{c }{Execution Environment} & \begin{tabular}[c]{@{}c@{}}Terminal\\ Code Interpreter\end{tabular} & \\ \hhline
\end{tabular}
\end{adjustbox}
\end{table*}

\headt{OS Kernel.} 
KernelGPT~\cite{KernelGPT} is an LLM-based agent for kernel fuzzing. 
Initially, KernelGPT uses a code extractor and analysis LLM to identify device operation handlers and infer device names and initialization specifications. It then iteratively analyzes the source code to generate syscall specifications, including command values, argument types, and type definitions. 
\junwei{Finally, it invokes the Syzkaller tool~\cite{Syzkaller} (\eg{} \textit{syz-extract} and \textit{syz-generate}, which can detect errors in the generated specifications), and repairs any invalid specifications by consulting the LLM with error messages iteratively.} 

\headt{Compiler.}
WhiteFox~\cite{WhiteFox} encompasses two LLM-based agents, an analysis agent and a generation agent. While the former examines the low-level optimization source code and produces requirements on the high-level test programs that can trigger the optimizations, the latter crafts test programs based on summarized requirements. The generation agent further incorporates tests that have successfully triggered optimizations as feedback during the iterative process, thereby producing more satisfactory tests. 
LLM4CBI~\cite{LLM4CBI} is a single agent that aims at isolating compiler bugs by generating test cases with better fault detection capabilities. The agent utilizes tools to collect static information about the program (\eg{} srcSlice~\cite{srcSlice} for data flow) to construct precise prompts to guide the LLM for program mutation. The memorized component records meaningful prompts and selects better ones to instruct LLMs to generate variants. The generated programs undergo validation by a static analysis tool (\eg{} the Frama-C~\cite{Frama-C}, \junwei{which is an open-source static analysis toolset for C language}), and the feedback helps LLMs to avoid the same mistakes. The final test cases are used to identify suspicious files with spectrum-based fault localization techniques.

\headt{Mobile Applications.}
LLM-based agents are proposed to automate the testing process of mobile applications, including \textit{GUI testing}, \textit{bug replay}, and \textit{user acceptance testing}.  

Some agents are developed to execute \highlight{GUI testing} for mobile applications. GUI testing is a commonly used software testing method aimed at verifying whether the user interface meets service specifications and user requirements. Previous LLM-based GUI testing approaches lack adequate autonomy, long-term planning, and coherence~\cite{QTypist, DroidBot-GPT}. The emergence of LLM-based agents enables GUI testing to focus more on higher-level test objectives~\cite{GPTDroid, DroidAgent, AXNav, InputBlaster, VisionDroid}, such as clear task objectives, without relying on specific GUI states.
Liu \et~\cite{GPTDroid} propose a framework called GPTDroid, where the LLM iterates the entire process by perceiving GUI page information, generating test scripts in the form of Q\&A, executing these scripts through tools, and receiving feedback from the application. GPTDroid keeps a long-term memory to retain testing knowledge, which would help to improve the reasoning process.
The DroidAgent~\cite{DroidAgent} framework employs multiple LLM-based agents coordinating through different memory modules and can set its own tasks according to the functionalities of the apps under test. It is composed of four LLM-based agents: planner, actor, observer, and reflector, each with specific roles and supported by memory modules that enable long-term planning and interaction with external tools.
AXNav~\cite{AXNav} is another multi-agent system designed for replaying accessibility tests on mobile apps. It includes the planner agent, the action agent, and the evaluation agent, which together form the LLM-based UI navigation system. These agents translate test instructions into executable steps, conduct tests on a cloud-based iOS device, and summarize the test results in a chaptered video annotated with potential issues in the application, respectively.
InputBlaster~\cite{InputBlaster} is an agent designed to generate unusual text inputs for mobile app crash detection. Initially, it infers the input constraints and generates a valid text input, based on the GUI page information. Building on this valid input and constraints, it then generates appropriate mutation rules with corresponding test generators, in the form of natural language and code snippets, respectively. Each test generator produces a batch of test inputs, and the test execution feedback will help the agent to produce more diversified outcomes. Additionally, the agent can retrieve relevant examples of buggy input for a better understanding of the task.
VisionDroid~\cite{VisionDroid} is a multi-agent system designed to detect non-crash functional bugs via multimodal LLM. It is composed of the function-aware explorer and the logic-aware bug detector. The explorer takes both the image and text information to comprehend the GUI page, generating actions to explore the functionality of the app and memorize the testing history. The intra-page bugs can be found by the explorer. The detector then segments the exploration history to check the inconsistency between the process logic and the GUI change history, which leads to the detection of inter-page bugs.

For automating Android \highlight{bug replay}, Feng \et{}~\cite{AdbGPT} introduce AdbGPT. Equipped with the knowledge of Step-to-Reproduce (S2R) entity specifications (\ie{} predefined actions and action primitives), AdbGPT analyzes bug reports to translate identified entities into a sequence of actions for bug reproduction using the CoT strategy. It then perceives GUI states dynamically and maps the S2R entities to actual GUI events to replicate the reported bug.

To increase the automation of the \highlight{user acceptance testing} process, Wang \et~\cite{XUAT-Copilot} propose XUAT-Copilot. The system is primarily comprised of three LLM-based agents responsible for action planning, state checking, and parameter selection, as well as two additional modules for state awareness and case rewriting. These agents interact with the testing equipment collaboratively, making human-like decisions and generating action commands.

\headt{Web Applications.}
RESTful APIs are popular among web applications as they provide a standardized, stateless, and easily integrable means of communication that enhances scalability and performance through a resource-oriented approach. RESTSpecIT \cite{RESTSpecIT} leverages LLMs to automatically infer RESTful API specifications and conduct black-box testing. Given an API name, RESTSpecIT generates and mutates HTTP requests through a reflection loop. By sending these requests to the API endpoint, it analyzes the HTTP responses for inference and testing. The LLM uses valid requests as feedback to refine the mutations in each iteration. Requests are validated based on the status code and message of the returned response.

\headt{Universal Software Categories.}  Some agent systems are not designed with a task-specific workflow, enabling them to be universally applicable across various target software systems. Xia \et\cite{fuzz4all} present Fuzz4All, the first universal LLM-based fuzzer for general and targeted fuzzing across multiple programming languages. For a higher cost-effectiveness ratio, Fuzz4All consists of two agents: (i) the distillation LLM for user input distillation and initial prompt generation, and (ii) the generation LLM for fuzzing input generation. They are powered by LLMs with different capabilities. In the fuzzing loop, the generation LLM refers to the previously generated samples and dynamically adjusts its strategy, thereby producing diverse fuzzing inputs.  
Deng \et\cite{PentestGPT} design a modular framework, PentestGPT, to conduct Penetration Testing. The system includes inference, generation, and parsing modules. With the planning strategy of Pentesting Task Tree (which is based on the cybersecurity attack tree~\cite{attack_tree}) and CoT methods, PentestGPT solves the problems of context loss and inaccurate instruction generation that may be encountered during automated penetration testing.
Fang \et{}~\cite{fang2024llmagentsautonomouslyexploit} develop a benchmark consisting of 15 one-day vulnerabilities to assess the efficacy of their agent framework in exploiting such weaknesses, utilizing various LLM backbones. Their agents are imbued with an understanding of the Common Vulnerabilities and Exposures (CVE) descriptions and are capable of harnessing a suite of tools to facilitate the exploitation process. These tools include web browsing capabilities for navigation, web search functionalities for traversing web pages, as well as terminal and code interpreter access for the generation and execution of scripts.

\junweim{
\subsubsection{Challenges of LLM-based Agents in software testing.}
LLM-based agents face several unique challenges in software testing tasks. First, the complexity of test environments often requires analyzing testing targets that are embedded within class-level or even project-level contexts. This issue arises not only in system-level testing but also in unit testing, which can depend on subtle interactions across multiple components. As a result, accurate test generation frequently demands sophisticated context augmentation mechanisms that can incorporate broader structural and semantic information from the codebases and documentations~\cite{AXNav, CoverUp, TELPA, chatunitester}. Second, despite the availability of mature traditional testing tools, effectively integrating them into the execution workflow of LLM-based agents remains a significant challenge. For example, WhiteFox~\cite{WhiteFox} can guide input generation to support traditional fuzzing methods, such as NNSmith~\cite{nnsmith}, and TestPilot~\cite{TestPilot} can produce initial tests to support traditional feedback-directed techniques, such as Nessie~\cite{Nessie}. However, these tools are typically used alongside agents rather than being tightly integrated into a unified testing strategy. Finally, current LLM-based testing agents tend to focus on narrow, task-specific goals. For example, InputBlaster~\cite{InputBlaster} targets the generation of unusual text inputs for mobile app crash detection. However, testing is inherently multidimensional, involving aspects such as test quality, coverage, security, and assertion generation. Multi-agent architectures hold promise for coordinating these diverse concerns within a more complete testing workflow, but this direction remains underexplored.
}
\subsection{Debugging}~\label{sec:se:debugging}
Software debugging typically includes two phases: \textit{fault localization}~\cite{wong2016survey} and \textit{program repair}~\cite{gazzola2018automatic}. In particular, fault localization techniques aim at identifying buggy elements (\eg{} buggy statements or methods) of the program based on the buggy symptoms (\eg{} test failure information); then, based on the buggy elements identified in the fault localization phase, program repair techniques generate patches to fix the buggy code. In addition, recent works also propose \textit{unified debugging} to bridge fault localization and program repair in a bidirectional way~\cite{FixAgent}. We then organize the works in LLM-based agents for debugging into three parts, \ie{} fault localization, program repair, and unified debugging. 

\begin{figure}[htb]
    \centering
    \includegraphics[width=1.0\columnwidth]{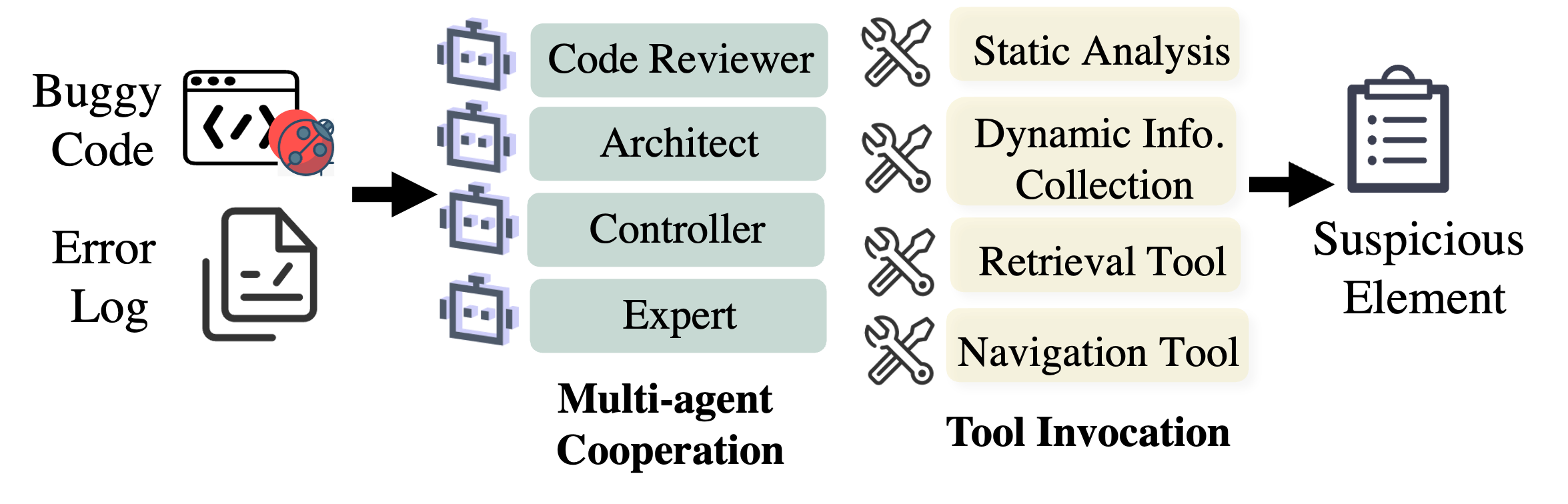}
    \caption{Pipeline of LLM-based Agents for Fault Localization} 
    \label{fig:fl}
\end{figure}

\begin{table*}[htb]
\centering
\renewcommand{\arraystretch}{1.2}
\caption{Existing LLM-based Agents for Fault Localization}
\label{tab:fl}
\begin{adjustbox}{width=\textwidth}
\begin{tabular}{ccccccc}
\hhline
\multirow{2}{*}{\textbf{Agents}} & \multirow{2}{*}{\textbf{Multi-Agent}} & \multicolumn{2}{c}{\textbf{Tools}} & \multirow{2}{*}{\textbf{Input Context}} & \multirow{2}{*}{\textbf{FL Granularity}} & \multirow{2}{*}{\textbf{Target Language}} \\ \cline{3-4}
 &  & \multicolumn{1}{c}{\textbf{Tool Category}} & \textbf{Specific Tools} &  &  &  \\ \hhline
AgentFL~\cite{AgentFL} & \checkmark & \multicolumn{1}{c}{Static Analysis} & Tree-sitter~\cite{tree-sitter} & Project Level & Method & Java \\ 
\junweim{AutoFL}~\cite{AUTOFL} & × & \multicolumn{1}{c}{Custom Toolkit} & Repository Retrieval Tools & Project Level & Method & Java \\ \hhline
\end{tabular}
\end{adjustbox}
\end{table*}

\subsubsection{Fault Localization}~\label{sec:se:debugging:FL}
Learning-based fault localization has been widely studied before the era of LLM, which typically trains deep learning models to predict the probability that each code element is buggy or not~\cite{li2019deepfl}. However, precisely identifying the buggy elements of software is challenging, given the scale of software systems and the massive, diverse error messages, which are often beyond the capabilities of standalone learning models, including LLMs. Therefore, recent works build LLM-based agents, which incorporate multi-agents and tool usage to help LLMs tackle these challenges. Table~\ref{tab:fl} summarizes the existing LLM-based agents for fault localization.

\junwei{\textbf{Framework:} Figure~\ref{fig:fl} illustrates the common framework of LLM-based agents for fault localization. Generally, the buggy code and error log will be fed into the agent systems, in which common roles like code reviewer, architect, and other experts will cooperate to explore the suspicious elements in the buggy code. In this process, these agents can invoke pre-defined tools, such as static analysis and code retrieval tools, to collect relevant code segments within the repository.}

\headt{Multi-agent Synergy.} 
AgentFL~\cite{AgentFL} is a multi-agent system for project-level fault localization. The main insight of AgentFL is to scale up LLM-based fault localization to project-level code context via the synergy of multiple agents. The system consists of four distinct LLM-driven agents: test code reviewer, source code reviewer, software architect, and software test engineer. Each agent is customized with specialized tools and a unique set of expertise. With the four agents, AgentFL streamlines the project-level fault localization process by breaking it down into three phases: fault comprehension, codebase navigation, and fault confirmation.

\headt{Tool Invocation.} 
\junweim{AutoFL}~\cite{AUTOFL} is a single-agent system, which enhances standalone LLMs with tool invocation (\ie{} four specialized function calls) to better explore the repository. It first performs root cause explanation, invoking tools to oversee the source code repository for pertinent information, requiring only a single failing test and its failure stack. During this stage, it autonomously decides whether to continue function calling or to terminate with the production of a root cause explanation. Subsequently, a post-processing step is used to correlate the outputs with exact code elements, aiming at bug localization. In addition, AgentFL~\cite{AgentFL} also incorporates tool invocation (\eg{} static analysis, dynamic instrumentation, and code base navigation) into its framework.

\subsubsection{Program Repair}
Fine-tuning and fixed prompting are the most widely adopted paradigms for program repair techniques based on standalone LLMs. In particular, program repair is formulated as a translation problem~\cite{sequenceR} (\ie{} translating the buggy code to correct code) or a generation problem~\cite{xia2022less} (\eg{} infilling the correct code in the buggy code context). However, patches generated by LLMs in a single iteration are not always correct; they may fail to pass all tests or overfit to the test cases. Therefore, existing LLM-based agents follow an iterative paradigm to refine patch generation based on the tool or model feedback in each iteration.
Table~\ref{tab:apr} summarizes the existing LLM-based agents for program repair. 

\begin{figure}[htb]
    \centering
    \includegraphics[width=1.0\columnwidth]{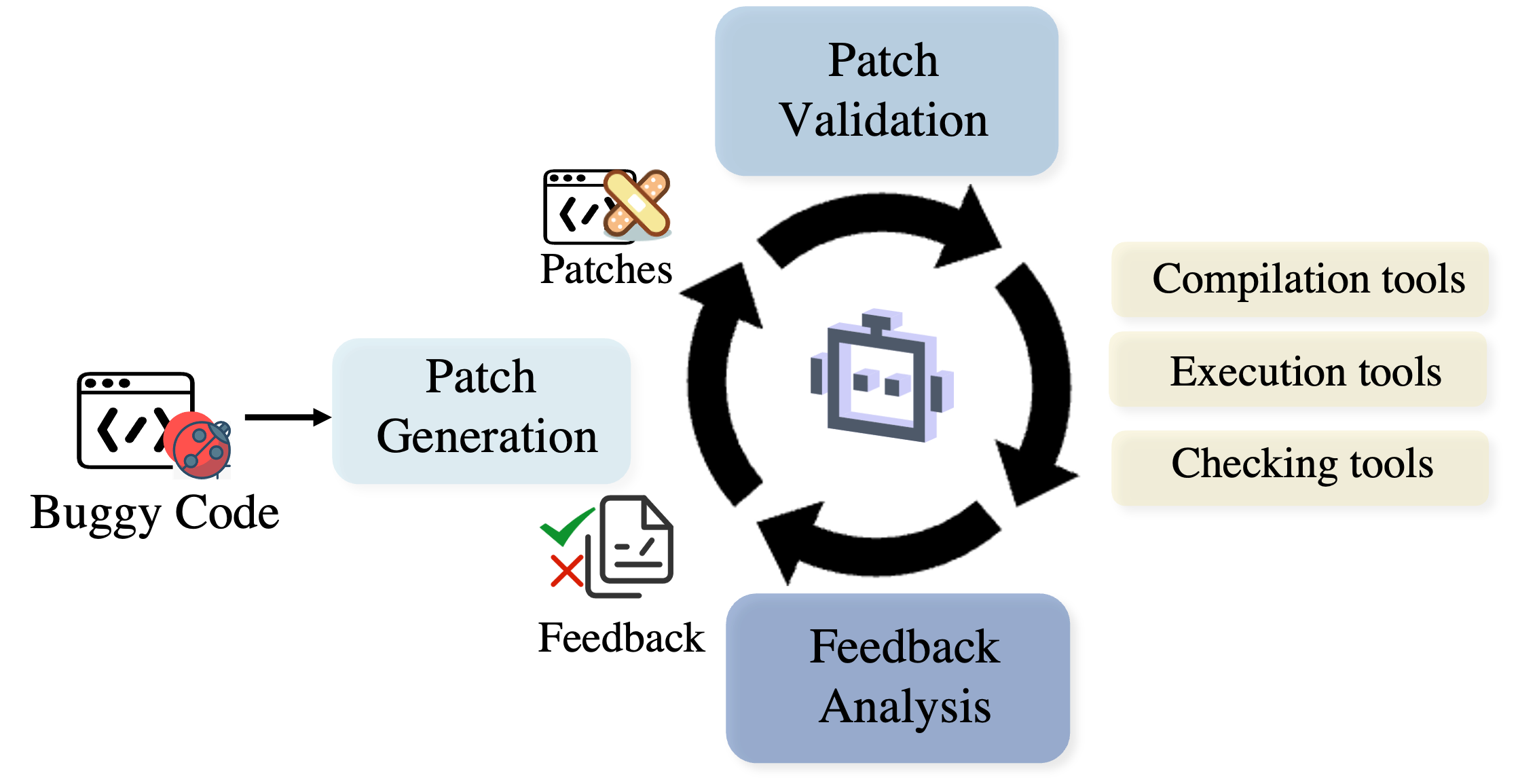}
    \caption{Pipeline of LLM-based Agents for Program Repair} 
    \label{fig:apr}
\end{figure}

\junwei{\textbf{Framework:} Figure~\ref{fig:apr} illustrates the common framework of LLM-based agents for program repair. First, they generate an initial candidate patch for the buggy code.  The patch is then validated against predefined test cases through compilation and execution.  Based on the feedback from these validation steps, including compilation errors, runtime failures, or test case outcomes, the patch undergoes iterative refinement.  This cycle continues until the patch meets the acceptance criteria, such as achieving full test suite pass coverage.}

\begin{table*}[htb]
\centering
\renewcommand{\arraystretch}{1.2}
\caption{Existing LLM-based Agents for Program Repair}
\label{tab:apr}
\begin{adjustbox}{width=1.0\textwidth, keepaspectratio}
\renewcommand{\arraystretch}{1.1} 
\begin{tabular}{ccccccc}
\hhline
\multirow{2}{*}{\textbf{Agents}} & \multirow{2}{*}{\textbf{Multi-Agent}} & \multirow{2}{*}{\textbf{Feedback Source}} & \multirow{2}{*}{\textbf{Target Software}} & \multirow{2}{*}{\textbf{Benchmark}} & \multicolumn{2}{c}{\textbf{Correct Fix}} \\ \cline{6-7} 
 &  &  &  &  & \multicolumn{1}{c }{\textbf{Rate}} & \textbf{Metric} \\ \hhline
ChatRepair~\cite{ChatRepair} & × & Execution/Compilation & Java/Python & \begin{tabular}[c]{@{}c@{}}Sampled Defects4J~\cite{defects4j}\\ QuixBugs~\cite{QuixBugs}\end{tabular} & \multicolumn{1}{c }{\begin{tabular}[c]{@{}c@{}}162/337 (Defects4J)\\ 80/80 (QuixBugs)\end{tabular}} & \begin{tabular}[c]{@{}c@{}}Pass Tests\\ Semantic Equivalence (Manual)\end{tabular} \\ \hline
CigaR~\cite{CigaR} & × & Execution/Compilation & Java & \begin{tabular}[c]{@{}c@{}}Sampled Defects4J\\ HumanEval-Java~\cite{Humaneval}\end{tabular} & \multicolumn{1}{c }{\begin{tabular}[c]{@{}c@{}}69/267 (Defects4J)\\ 102/162 (HumanEval)\end{tabular}} & \begin{tabular}[c]{@{}c@{}}Pass Tests\\ AST Match\end{tabular} \\ \hline
RepairAgent~\cite{RepairAgent} & × & Execution/Compilation & Java & Defects4J & \multicolumn{1}{c }{164/835} & \begin{tabular}[c]{@{}c@{}}Pass Tests\\ Syntax Match (Automatic)\\ Semantic Equivalence (Manual)\end{tabular} \\ \hline
AutoSD~\cite{AUTOSD} & \checkmark & Execution & Java/Python & \begin{tabular}[c]{@{}c@{}}Defects4J\\ BugsInPy~\cite{BugsInPy}\\ Almost-Right HumanEval\end{tabular} & \multicolumn{1}{c }{\begin{tabular}[c]{@{}c@{}}189/835 (Defects4J)\\ 187/200 (HumanEval)\end{tabular}} & \begin{tabular}[c]{@{}c@{}}Pass Tests\\ Semantic Equivalence (Manual)\end{tabular} \\ \hline
\junweim{ACFix}~\cite{ACFIX} & \checkmark & \begin{tabular}[c]{@{}c@{}}Static Checking\\ Model Debate\end{tabular} & Smart Contract & Self-curated Dataset & \multicolumn{1}{c }{112/118} & \begin{tabular}[c]{@{}c@{}}Comparison with Author Fixes\\ Execute Exploit Scripts\\ Manual Inspection\end{tabular} \\ \hline
FlakyDoctor~\cite{FlakyDoctor} & \checkmark & \begin{tabular}[c]{@{}c@{}}Static Checking\\ Execution\end{tabular} & Java & \begin{tabular}[c]{@{}c@{}}Sampled IDoFT~\cite{idoft}\\ DexFix dataset~\cite{dexfix}\\ Sampled ODRepair dataset~\cite{odrepair}\end{tabular} & \multicolumn{1}{c }{\begin{tabular}[c]{@{}c@{}}311/541 (Implementation-Dependent Flakiness)\\ 189/332 (Order-Dependent Flakiness)\end{tabular}} & Manual Inspection \\ \hline
SRepair~\cite{SRepair} & \checkmark & - & Java & \begin{tabular}[c]{@{}c@{}}Sampled Defects4J\\ QuixBugs\end{tabular} & \multicolumn{1}{c }{\begin{tabular}[c]{@{}c@{}}332/665 (Defects4J)\\ 80/80 (QuixBugs)\end{tabular}} & \begin{tabular}[c]{@{}c@{}}Pass Tests\\ Semantic Equivalence (Manual)\end{tabular} \\ \hhline
\end{tabular}
\end{adjustbox}
\end{table*}

\junwei{ChatRepair~\cite{ChatRepair, ConversationalAPR} is the first automated approach to refine patch or program generation based on environmental feedback iteratively. Specifically, it incorporates previously generated patches and test execution feedback into the prompt and feeds it into the LLM to generate new patches, thereby learning from both incorrect and plausible patches to ultimately achieve correct repairs. Ultimately, considering that the test suite may be incomplete, patches generated based on it may not always cover all intended uses of the underlying code. Therefore, ChatRepair continues the iteration with only plausible patches provided to the LLM to generate more plausible patches for manual inspection. 
CigaR~\cite{CigaR} adopts a similar approach, with the generated patches and test failure messages serving as the feedback for the LLM to generate new patches. It also employs a plausible patch multiplication stage to iteratively produce more plausible patches. A notable difference is that CigaR employs a reboot mechanism. If the LLM fails to generate a plausible patch within the maximum number of invocations, the entire repair process is restarted. Experimental results show that this reboot strategy enables the LLM to efficiently explore different parts of the search space, avoiding wasting tokens on dead ends.}

\junwei{RepairAgent~\cite{RepairAgent} is a highly autonomous LLM-based agent. It consists of three main components: an LLM agent, a set of tools for interacting with the codebase (\eg{} reading code, 
searching code base, and running tests), and a middleware, which is a state machine containing four states: \textit{understand the bug}, \textit{collect information to fix the bug}, \textit{try to fix the bug} and \textit{done}. The middleware accepts the action request from the LLM agent and invokes tools, with the results dynamically updating the prompt and then being fed into the LLM agent. Experimental results demonstrate that RepairAgent repairs 164 bugs in Defects4j~\cite{defects4j}, including 39 bugs not fixed by prior techniques.}

AutoSD~\cite{AUTOSD} is a multi-agent system that iteratively fixes the buggy program via simulating the scientific debugging~\cite{ScientificDebugging}. AutoSD includes four components: LLM-based hypothesis generator, execution-based validator, LLM-based conclusion maker, and LLM-based fixer. In each iteration, the generator first generates a hypothesis about the bug, then invokes the debugger tool for hypothesis validation; the conclusion maker further identifies whether the hypothesis is rejected or not; and the fixer finally returns potential patches with explanations.

\junweim{ACFix}~\cite{ACFIX} is a multi-agent system for fixing the access control vulnerabilities in smart contracts. By specializing LLMs with different roles, \junweim{ACFix} includes a Role-based Access Control (RBAC) mechanism identifier, a role-permission pair identifier, a patch generator, and a validator. In particular, the validator checks the validity of the generated patches with both tool feedback (static grammar rule checking) and model feedback (multi-agent debate process). The feedback is further provided to iteratively refine the patch.

FlakyDoctor~\cite{FlakyDoctor} is an agent to repair flaky tests. It takes the test execution results and the location of test failures into consideration. Following this, the agent generates targeted repairs and tests them for validation. This process is iterative, with the aim of continually refining the repairs until the issue of test flakiness is resolved.

SRepair~\cite{SRepair} is a dual-agent system for function-level program repair. One agent first analyzes the auxiliary repair-relevant information (\eg{} the buggy code, error messages) to produce fix suggestions via the CoT technique, while the other generates fixed functions with the help of the suggestions.

\subsubsection{Unified Debugging}
Instead of tackling fault localization or program repair as isolated phases, unified debugging techniques treat them as a unified procedure, which leverages the outputs of each phase to refine the other. In particular, traditional unified debugging techniques~\cite{unifieddebuggingissta, unifieddebuggingtse, unifieddebuggingase} primarily pre-define heuristic rules to refine fault localization based on the patch validation results during program repair. Recently, LLM-based agents have enhanced traditional unified debugging techniques with more flexibility by leveraging LLMs to comprehend, utilize, and unify the outputs of both fault localization and program repair.

\junwei{FixAgent~\cite{FixAgent} is a multi-agent framework for automated debugging, 
with each agent simulating the ``rubber duck debugging'' strategies to explain their detailed work. It includes an LLM Localizer for identifying bugs, an LLM Repairer for generating patches, an LLM Crafter for creating additional test inputs to ensure patch generalizability, and an LLM Revisitor for analyzing symptoms of the buggy code and the rationale of the patch. The agents cooperate in an iterative manner, in which the downstream agent depends on the results from the upstream agent but can also influence the upstream agent in the next turn. For instance, if the Repairer modifies code segments different from those identified by the Localizer, the Localizer will adjust its localization results accordingly.
LDB~\cite{LDB} also adopts an iterative approach. In each iteration, it divides the current program into different blocks and precisely analyzes the changes in variables before and after each block during test execution. Based on this information, it queries LLMs to verify each block according to the task, thereby locating and repairing bugs.}

\junwei{
In summary, to coordinate fault localization and program repair and achieve unified debugging, these agents employ a division of labor, with two different agents responsible for localization and repair separately. Moreover, these approaches involve fine-grained analysis. For example, FixAgent requires each agent to explain its work to a ``rubber duck'', while LDB incorporates the runtime analysis. As a key distinction, FixAgent considers the interplay between the fault localization and program repair phases, which leverages the output of each phase to refine the other, distinguishing it from traditional single-stage fault localization and program repair works.
}

\junweim{
\subsubsection{Common Failure Causes of LLM-Based Agents in Debugging.}
Based on the failure analyses of existing LLM-based agents in debugging, we summarize several common failure causes that reflect the current limitations of LLM-based debugging agents in real-world scenarios.
}

\junweim{
\textbf{Lack of Understanding of Complex Project Context.} Effective debugging often entails a comprehensive traversal of the software project to grasp its class hierarchies, inter-method dependencies, and underlying test architecture. This exploration is necessary to identify the root cause of bugs, but it can also introduce significant noise. For example, in AutoFL~\cite{AUTOFL}, most failures stemmed from the agent spending excessive rounds trying to understand project-specific structures such as custom classes, helper functions, and test frameworks, leaving insufficient budget for inspecting potentially buggy code. This suggests that current agents need more effective strategies for reducing the search space and selectively incorporating project-specific context.
}

\junweim{
\textbf{Lack of coherence across multi-step debugging workflows.} A successful debugging process typically involves a sequence of reasoning, hypothesis testing, and verification. However, these steps are not always well-aligned in current agents. For example, in AutoSD~\cite{AUTOSD}, the agent proposed a breakpoint that was never hit during test execution. Instead of revisiting the hypothesis or suggesting alternative breakpoints, the agent incorrectly suggested that the test was flawed. This disjointed reasoning between the analysis and verification steps led to an incorrect debugging trajectory. Such inconsistencies highlight the need for tighter integration between reasoning and feedback in iterative debugging loops.
}

\junweim{
\textbf{Other Failures.}
There are several lower-frequency but still impactful failure types. For instance, in some cases, buggy methods are too long to fit within the agent's context window~\cite{AUTOFL}, resulting in truncated analysis or context overflow. In others, the agent introduces logical errors or edits only a subset of the necessary locations in its fix proposals~\cite{AUTOFL, RepairAgent}. There are also errors caused by the agent generating tool-incompatible invocations. For example, AutoSD~\cite{AUTOSD} might insert multiple print statements in contexts that only allowed one. 
While less prevalent, these issues expose limitations in the agent’s handling of edge cases, context constraints, and system-level instructions.
}

\junweim{
\subsubsection{Challenges of LLM-based Agents in Debugging.}
LLM-based agents face several distinctive challenges in debugging tasks. Many current approaches rely heavily on failing test cases, but using intermediate execution signals alone to assess program correctness remains an open research direction~\cite{LDB, AUTOFL, RepairAgent}. In addition, LLM-based agent systems often incorporate static analysis or runtime tools to enhance fault localization. A key challenge lies in designing effective integration strategies that allow LLM-based agents to leverage these tools' strengths while mitigating issues such as performance overhead, data latency, and inconsistency in results~\cite{LDB, SRepair}. Coordination mechanisms that enable tighter coupling between LLM reasoning and tool feedback are essential to improve debugging efficiency and accuracy. Furthermore, although multi-agent collaboration can potentially enhance the reasoning capabilities of debugging systems, it also introduces additional system complexity and significant performance bottlenecks. For example, AutoSD takes approximately five times longer to generate a patch compared to a standalone LLM~\cite{AUTOSD, FixAgent}.
Finally, most existing debugging agents generate only plausible patches, and verifying whether these patches are semantically equivalent to the intended golden patch still requires manual effort. This limits both the level of automation and the scalability of current LLM-based debugging agents~\cite{SRepair}.
}
\junwei{
\subsection{IT Operations}\label{sec:se:operations}
IT operations (Ops) involve managing and maintaining the technology infrastructure of an organization, ensuring systems run smoothly, and quickly addressing any issues that arise.  With the rise of AI, Ops are becoming increasingly automated and intelligent~\cite{AIOpsSurvey}.
However, standalone LLMs struggle with this task, as they lack the ability to directly interact with systems or perform real-time actions. In contrast, LLM-based agents can integrate with IT environments, enabling them to analyze data, automate processes, and respond to incidents, making them highly effective in IT operations. However, different LLM-based agents might adopt various collaborative diagnosis strategies, which we categorize into three types: self-consistency with embedding voting, agent chain with blockchain-inspired voting, and tree search with majority voting.
}

\textbf{\begin{figure}[h]
    \centering
    \includegraphics[width=1.0\columnwidth]{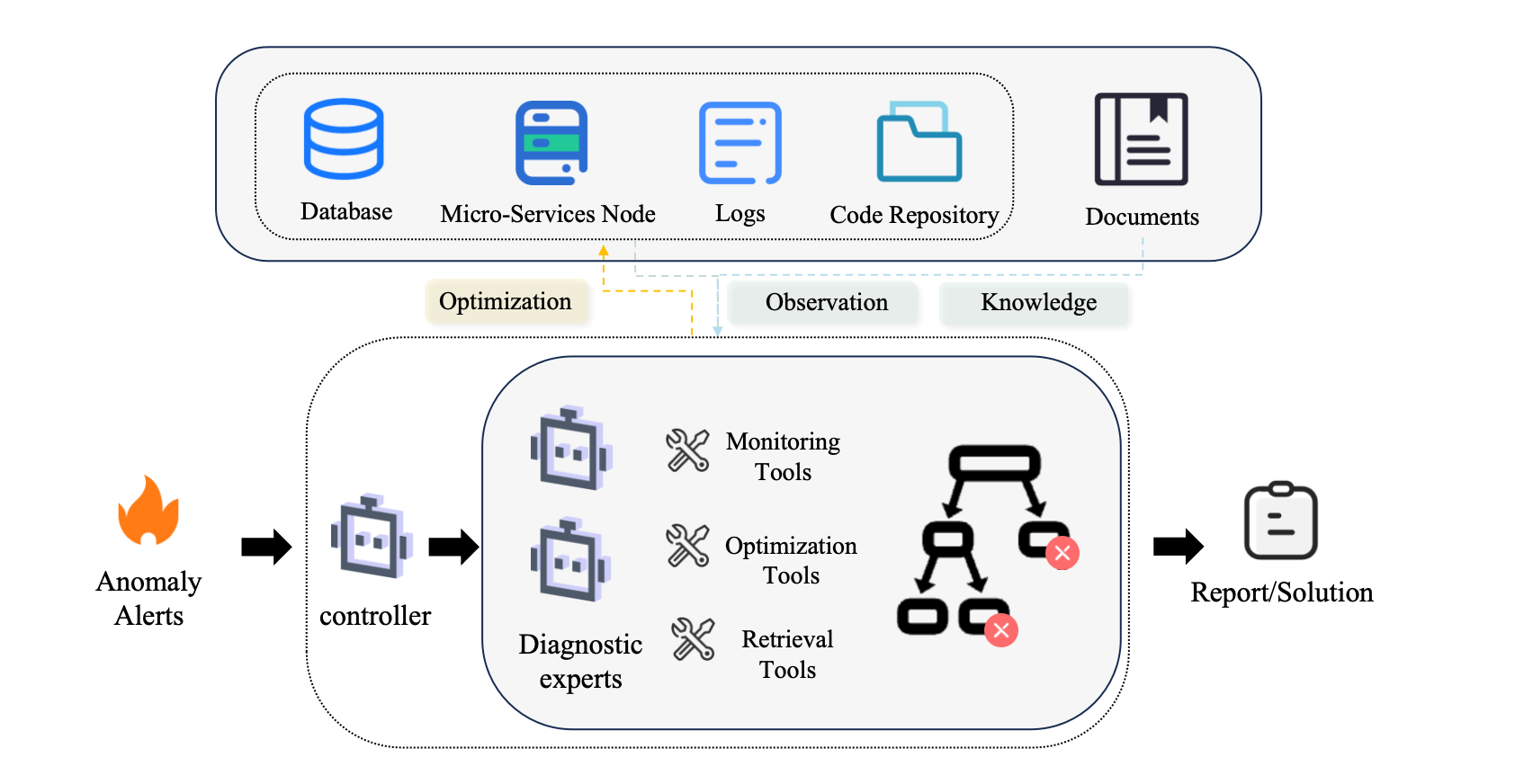}
    \caption{Pipeline of LLM-based Agents for IT Operations} 
    \label{fig:aiops}
\end{figure}
}

\junwei{\textbf{Framework:} The overall pipeline for LLM-based agents in IT operations is illustrated in Figure~\ref{fig:aiops}. Upon receiving anomaly alerts, the controller will dispatch the diagnosis tasks to diagnostic expert agents, who will then collaborate to analyze the root cause through interaction with the target system. Each expert agent is equipped with predefined toolkits to interact with the environment or access domain-specific reference documents. During the diagnostic process, they will try different exploration paths, ultimately producing a diagnostic report or solution.}

\junweim{
\subsubsection{Self-Consistency with Embedding Voting.}
RCAgent~\cite{RCAgent} proposes a self-consistency mechanism that shares preliminary steps across multiple reasoning trajectories, and uses vector similarity in embedding space to aggregate and weigh outcomes.
As a multi-agent system for root cause analysis in industrial cloud settings, RCAgent includes two components: the controller agent and the expert agents. The controller agent oversees the comprehensive thought-action-observation cycle, while the expert agents act for specialized tasks(\ie{} code and log analysis).
Through the self-consistency mechanism, RCAgent mitigates computational overhead during inference, thereby enhancing both efficiency and robustness. 
}

\junweim{
\subsubsection{Agent Chain with Blockchain-Inspired Voting.}
mABC~\cite{mABC} proposes a blockchain-inspired voting mechanism to ensure agreement and reliability among different diagnostic experts. The main workflow of mABC is as follows: when an alert arises, the Alert Receiver agent will prioritize incoming alerts and select the most critical one for investigation. The Process Scheduler then breaks down the root cause analysis task into smaller subtasks and assigns them to specialized agents, including the Data Detective, Dependency Explorer, Probability Oracle, and Fault Mapper agents. Finally, the Solution Engineer agent proposes remediation strategies based on historical knowledge.
Throughout this process, all agents compose a decentralized structure, Agent Chain, and participate in a blockchain-inspired voting mechanism to ensure reliable consensus on the root cause and solution.
}

\junweim{
\subsubsection{Tree Search with Majority Voting.}
D‑Bot~\cite{D-Bot} leverages LLM-based agents to diagnose database anomalies through several key steps. First, it extracts diagnostic knowledge from manuals and documentation offline. Then, it generates prompts automatically by matching relevant knowledge and diagnostic tools. For complex anomalies involving multiple root causes, D‑Bot employs a collaborative multi-agent mechanism where multiple LLM agents (\eg{} CPU Expert and I/O Expert) work together. The core is an LLM-powered root-cause analysis using a tree-search approach that allows multi-step reasoning and exploration of alternative hypotheses. When the search tree branches into multiple candidate paths, the system leverages LLM-based majority voting to evaluate the plausibility of each branch. Multiple reasoning agents independently assess the intermediate evidence and potential outcomes of every path, casting votes for the most promising direction. The path with the highest consensus score is then selected for further exploration, ensuring that the reasoning process remains both diverse in exploration and robust against individual model errors.
}

\junweim{
\subsubsection{Challenges of LLM-based Agents in IT Operations.}
LLM-based agents face several challenges in IT operations tasks. First, these agents are often required to monitor and analyze information from multiple system dimensions. For example, in D-Bot~\cite{D-Bot}, upon receiving an alert, the assigner agent dispatches specialized expert agents to investigate potential root causes from various perspectives, such as CPU, memory, workload, and I/O behavior. This demands strong capabilities in both agent scheduling and the handling of heterogeneous data sources. Additionally, IT operations typically involve complex, iterative processes with multiple rounds of exploration. Designing more efficient planning algorithms to guide the agent through these exploratory steps is crucial to reducing computational overhead while improving diagnostic accuracy~\cite{D-Bot, mABC}. Finally, most LLM-based agents focus primarily on root cause analysis, whereas other important operational tasks, such as system performance analysis and optimization, remain largely unexplored. Addressing these broader challenges will require new strategies for integrating LLM-based agents with existing operations tools, such as logging systems and performance profilers.}
\begin{table*}[]
\centering
\caption{Existing LLM-based Agents for End-to-end Software Development}
\label{tab:endtoendsd}
\begin{adjustbox}{width=1.0\textwidth}
\renewcommand{\arraystretch}{1.3}

\begin{tabular}{lccccc}
\hhline
\textbf{Agents} & \textbf{Multi-Agent} & \textbf{Process Model} & \textbf{Roles Creation} & \textbf{Collaboration Mode} & \textbf{Communication Protocal} \\ \hhline
Self-Collaboration~\cite{self-collaboration} & \checkmark & Waterfall & Pre-defined & Vertical & Memory \\
Low-code LLM~\cite{low-code-llm} & \checkmark & - & Pre-defined & Vertical & Direct Communication \\
Prompt Sapper~\cite{sapper} & $\times$ & - & Pre-defined & Vertical & Direct Communication \\
Talebirad \et{}~\cite{multi-agent-collaboration} & \checkmark & - & Task-Adaptive & Vertical & Direct Communication \\
ChatDev~\cite{ChatDev} & \checkmark & Waterfall & Pre-defined & Vertical + Horizontal & Direct Communication + Memory \\
MetaGPT~\cite{MetaGPT} & \checkmark & Waterfall & Pre-defined & Vertical & Direct Communication + Memory\\
AgentVerse~\cite{AgentVerse} & \checkmark & - & Task-Adaptive & Vertical & Direct Communication \\
AutoAgents~\cite{AutoAgents} & \checkmark & - & Task-Adaptive & Vertical & Direct Communication + Memory \\
Rasheed \et{}~\cite{rasheed} & \checkmark & Waterfall & Pre-defined & Vertical + Horizontal & Direct Communication \\ 
Co-Learning~\cite{Co-Learning} & \checkmark & - & Pre-defined & Vertical + Horizontal & Direct Communication \\
AISD~\cite{aisd} & \checkmark & Waterfall & Pre-defined & Vertical & Direct Communication \\
LLM4PLC~\cite{LLM4PLC} & $\times$ & - & Pre-defined & Vertical & Direct Communication \\
CodePori~\cite{CodePori} & \checkmark & Waterfall & Pre-defined & Vertical & Direct Communication \\
\junweim{FlowGen}\textsubscript{Waterfall}~\cite{lcg} & \checkmark & Waterfall & Pre-defined & Vertical + Horizontal & Direct Communication \\
\junweim{FlowGen}\textsubscript{TDD}~\cite{lcg} & \checkmark & Agile & Pre-defined & Vertical & Direct Communication \\
\junweim{FlowGen}\textsubscript{Scrum}~\cite{lcg} & \checkmark & Agile & Pre-defined & Vertical + Horizontal & Direct Communication \\
CodeS~\cite{CodeS} & \checkmark & - & Pre-defined & Vertical & Direct Communication \\
Qian \et{}~\cite{IterativeExperienceRefinement} & \checkmark & - & Pre-defined & Vertical + Horizontal & Direct Communication \\
CTC~\cite{CTC} & \checkmark & Waterfall & Pre-defined & Vertical + Horizontal & Direct Communication + Memory \\
AgileCoder~\cite{AgileCoder} & \checkmark & Agile & Pre-defined & Vertical & Direct Communication + Memory \\ 
\junweim{MacNet}~\cite{MACNET} & \checkmark & - & Pre-defined & Vertical + Horizontal & Direct Communication + Memory \\ 
Sami \et{}~\cite{sami} & \checkmark & Waterfall & Pre-defined & Vertical & Direct Communication \\ 

\hhline
\end{tabular}
\end{adjustbox}
\end{table*}
\subsection{End-to-end Software Development}~\label{sec:se:development}
Given the high autonomy and the flexibility of multi-agent synergy, LLM-based agent systems can further tackle the end-to-end procedure of software development (\eg{} developing a Snake Game application from scratch) beyond an individual phase of software development. In particular, like the real-world software development team,  these agent systems can cover the entire software development life cycle (\ie{} requirements engineering, architecture design, code generation, and software quality assurance) by incorporating the synergy between multiple agents that are specialized with different roles and relevant expertise.  Table~\ref{tab:endtoendsd} summarizes the existing LLM-based agents for end-to-end software development.

\subsubsection{Software Development Process Model}
\junwei{
End-to-end software development requires a well-structured workflow. In the field of software engineering, software development process models (\eg{} waterfall~\cite{waterfall}, incremental model~\cite{thakur2016survey}, unified process model~\cite{jacobson1999unified}, and agile development~\cite{aoyama1997agile}) are used to describe methodologies for organizing software development processes.  
Inspired by these classic methods, the pipelines of some end-to-end software development agents are adapted from traditional process models, primarily Waterfall and Agile. Figure~\ref{fig:processmodel} illustrates the adapted development pipelines used by these approaches. For agents that do not adopt classic software development process models, we will discuss their collaboration mechanisms in Section~\ref{e2edevelopment_collaboration}.
}

\textbf{\begin{figure*}[htb]
    \centering
    \includegraphics[width=0.9\textwidth]{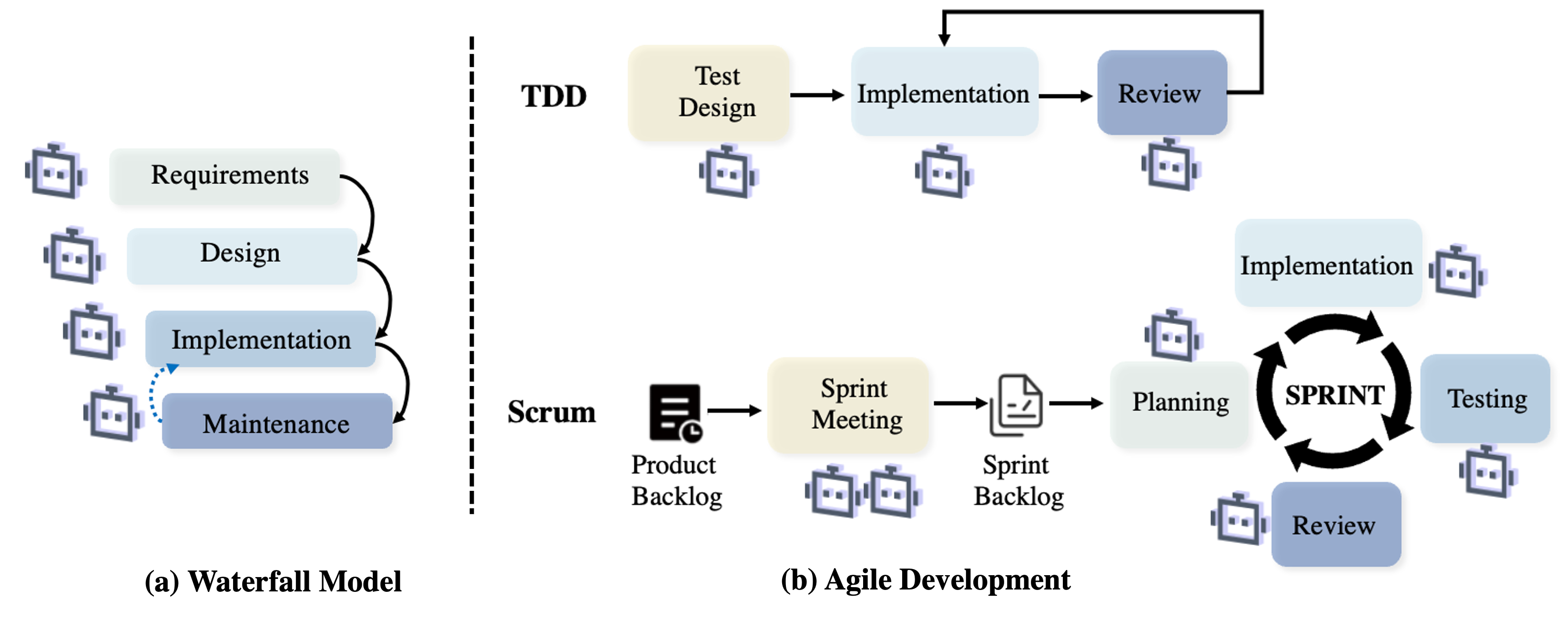}
    \caption{Adapted Process Models Adopted by LLM-based Agents for End-to-end Software Development} 
    \label{fig:processmodel}
\end{figure*}
}

\headt{Waterfall Process Model.} 
The Waterfall model is the most popular approach among current LLM-based agent methods for end-to-end software development.
The traditional waterfall process model~\cite{waterfall} is a linear and sequential software development workflow that divides the project into distinct phases, \ie{} requirements engineering, design, code implementation, testing, deployment, and maintenance. Once a phase is finished, the project moves forward to the next phase without iteration.
\junwei{
However, due to the randomness and hallucination of LLMs~\cite{hallucination_survey, LLM4CodeGeneration_survey}, feedback mechanisms are often introduced to improve accuracy. Therefore,} these end-to-end software development agents~\cite{lcg, aisd, MetaGPT,self-collaboration} further extend the traditional waterfall process by including iterations in specific phases to ensure the high quality of the generated content. For example, the results of the testing phase might be fed back to the developer agent to revise the generated code. 

\headt{Agile Development.} Some works explore the potential of LLM-based agents with agile development, including Test-Driven-Development (TDD)~\cite{lcg} and Scrum~\cite{lcg, AgileCoder}. TDD prioritizes writing tests before the actual coding and fosters a cycle of writing test suites, implementing the code to pass the test suites, and concluding with a reflective phase of refinement.
Scrum is an agile software development process model that breaks down software development into several sprints, achieving complex software systems through iterative updates.
\junwei{However, as shown in Figure~\ref{fig:processmodel}, the actual Scrum model adopted in current LLM-based agents omits the ``Daily Scrum'', which serves as a short meeting for team members to sync progress and discuss issues. This may be due to the fact that agents can share information through the memory mechanism. }
Experiments on function-level code generation benchmarks show that the Scrum model can achieve the best and most stable performance, followed by the TDD model~\cite{lcg}.

\subsubsection{Role Specialization of Software Development Team} 
Imitating real-world software development teams, multi-agent systems for end-to-end software development often assign different roles to tackle specialized sub-tasks and collaborate throughout the software development life cycle.

\headt{Role Categories.} 
\junwei{Most end-to-end frameworks emulate real-world software development teams by assigning key roles as follows:
\begin{itemize}[leftmargin=*,label=-]
\item \textit{Managers.} In virtual software teams, managers serve as the team leaders and have diverse responsibilities. 
One of the main responsibilities is to analyze and extract user requirements, such as the Product Manager~\cite{multi-agent-collaboration, MetaGPT, aisd, AgileCoder} and the Manager in CodePori~\cite{CodePori}.
On the other hand, some manager roles are responsible for \textit{task decomposition and allocation}~\cite{MetaGPT, rasheed, low-code-llm}. For example, Rasheed \et{}~\cite{rasheed} proposed a framework in which the Project Planning Agent defines the scope, objectives, and development plan. In Low-code LLM~\cite{low-code-llm}, the Planning LLM is tasked with generating a flowchart that systematically breaks down the final task into small steps. Some studies also introduce a CEO role to assist in \textit{software design}~\cite{ChatDev, CTC}. 
Finally, the Scrum Master is a role specific to the Scrum agile model. In FlowGen\textsubscript{Scrum}~\cite{lcg}, the Scrum Master is responsible for summarizing sprint meetings and extracting user story lists. In AgileCoder~\cite{AgileCoder}, the Scrum Master provides feedback to the Product Manager to optimize the requirements list. 
\item \textit{Requirements Engineers.} Requirements engineers are responsible for understanding user inputs and generating requirement documents~\cite{rasheed, lcg, self-collaboration, sami}. Typically, if managers are already assigned to decompose requirements within the workflow, the requirements engineers will not be introduced.
\item \textit{Designers.} Designers are responsible for various aspects of system design. Architectural design is the most common one, where a role such as architect or system Designer is introduced to perform high-level system design based on requirement documents~\cite{multi-agent-collaboration, aisd, lcg, MetaGPT, sami, rasheed}. Prior work also introduces specialized designers for User Experience (UX) and User Interface (UI) design~\cite{multi-agent-collaboration}.
\item \textit{Developers.} Developers are responsible for writing code based on requirements and design specifications. In the collected works, we identify two types of developer agents. The first type is the basic developer, who is responsible for the actual coding process~\cite{multi-agent-collaboration, rasheed, CodePori, lcg, AgileCoder, self-collaboration, ChatDev, aisd, MetaGPT, sami, CTC}. The second type is the senior developer, who provides feedback to basic developers to ensure code quality. Examples include the Senior Developer in AgileCoder~\cite{AgileCoder} and the CTO in ChatDev~\cite{ChatDev, CTC}.
\item \textit{Quality Assurance Experts.} QA Experts can be categorized into three main types: (i) Software Testers, who are responsible for generating test code, executing tests, and providing test feedback~\cite{self-collaboration, multi-agent-collaboration, ChatDev, rasheed, MetaGPT, aisd, lcg, AgileCoder, sami, CTC}. (ii) Debuggers, tasked with identifying and resolving software defects~\cite{multi-agent-collaboration}. (iii) Reviewers, who enhance code quality by identifying issues through code review~\cite{ChatDev, CTC}.
\item \textit{Deployment Engineers.} They are responsible for formulating software release strategies, such as the Deployment Plan Agent proposed in ~\cite{rasheed}.
\item \textit{Assistants.} Assistants primarily support the output of the aforementioned roles by providing feedback, summarization, and criticism. Examples include the Oracle Agent\cite{multi-agent-collaboration}, the QA Agent\cite{rasheed}, and the action observer~\cite{AutoAgents}.
\end{itemize}
}
The detailed categories of roles in existing SE agents are discussed in Section~\ref{sec:agent:role}.

Instead of simulating the real-world development teams, some agents design their specialized agent workflow and break down roles accordingly. For example, CodeS~\cite{CodeS} decomposes the complex code generation task into the implementation of repository, file, and method layers, and sets up the roles of RepoSketcher, FileSketcher, and SketchFiller. Co-Learning~\cite{Co-Learning} and its subsequent work~\cite{IterativeExperienceRefinement} abstract the code generation process into instruction-response pairs, thus only setting up the roles of instructor and assistant.

\headt{Role Creation.} 
\junwei{ The roles in multi-agent end-to-end software development systems are either created in a predefined way or in a task-adaptive way. Most approaches predefine fixed roles and workflows through manual design~\cite{self-collaboration, low-code-llm, ChatDev, MetaGPT, Co-Learning, aisd, CodePori, lcg, CodeS, IterativeExperienceRefinement, CTC, AgileCoder, MACNET, sami, rasheed}. In contrast, some agents only predefine a few meta-roles, which discuss and derive the actual agent roles to solve the specific problem. This role-creation approach is typically applied to general-purpose agents to handle various types of tasks, or in scenarios involving multi-turn operations where different roles are utilized in each round. }
For example, AutoAgents~\cite{AutoAgents} designs a drafting stage that aims at determining the roles of the multi-agent group via the communication between two meta agents: the planner and the agent observer. In AgentVerse~\cite{AgentVerse}, a group of different roles is established through an expert recruitment stage, and the roles recruited in each round may vary.
Talebirad \et{}~\cite{multi-agent-collaboration} propose a novel framework and enable an agent to spawn additional agents to the system. Such dynamic strategies aim at creating roles in a more diverse and flexible way. 

\subsubsection{Collaboration Mechanism in Multi-agent}\label{e2edevelopment_collaboration}
Within the multi-agent systems for end-to-end software development, it is essential to schedule how each agent coordinates with the other. We then discuss the collaboration mode and the communication protocol adopted in existing agents. 

\textbf{\begin{figure}[htb]
    \centering
    \includegraphics[width=1.0\columnwidth]{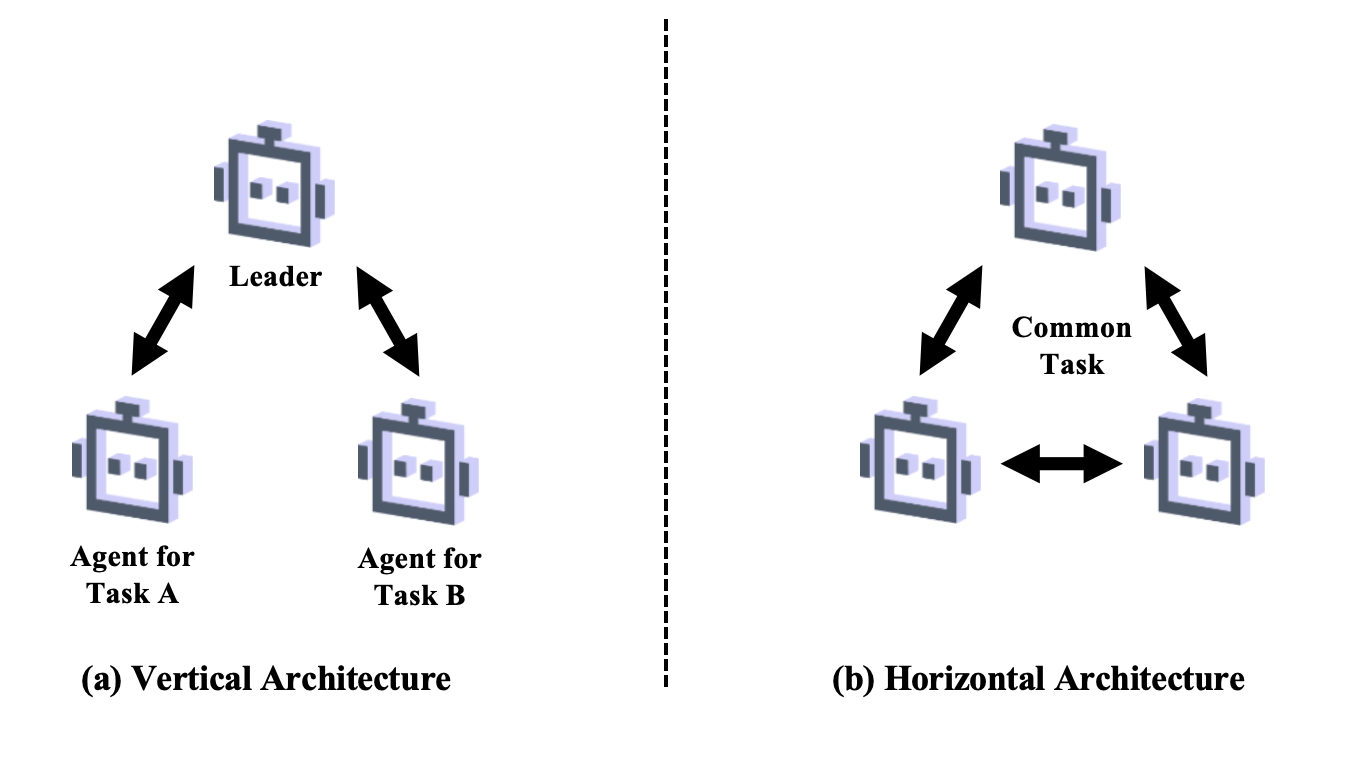}
    \caption{Vertical and Horizontal Collaboration Architectures} 
    \label{fig:e2esoftware_collaboration}
\end{figure}
}

\headt{Collaboration Mode.}
\junwei{ The collaboration mechanism of multiple agents can generally be divided into two types: vertical architecture and horizontal architecture. As illustrated in Figure~\ref{fig:e2esoftware_collaboration}, in the vertical architecture, a leader assigns tasks to different agents for execution and integrates the final results. In the horizontal architecture, the agents are treated as equals and collaborate through discussion to jointly advance the execution of the same task~\cite{DBLP:journals/corr/abs-2404-11584}.
All the end-to-end software development agents choose to use vertical architecture in the overall pipeline. However, some of these works incorporate horizontal collaboration in some phases.}

\junwei{\headfi{Vertical Architecture.} 
Previous research~\cite{AgentVerse, AutoAgents} suggests that vertical collaboration is preferable for tasks like software development, which produces only the final refined decision. For example, in AgentVerse~\cite{AgentVerse}, the Solver agent plays the role of the leader, deciding the final solution by integrating the feedback from other agents. In AutoAgents~\cite{AutoAgents}, the Action Observer is predefined as the leader, responsible for assigning tasks, validating the outputs of different agents, and dynamically adjusting the plan. Except for these two agents, all other end-to-end software development agents omit the role of leader, with the output of each phase directly serving as the input for the next, thus forming a linear workflow~\cite{CodePori, aisd, CodeS, Co-Learning, MetaGPT, ChatDev, self-collaboration, IterativeExperienceRefinement, CTC, AgileCoder, sami, rasheed}.}

\junwei{ \headfi{Vertical + Horizontal Architecture.}
Some agents in end-to-end software development introduce horizontal collaboration into certain phases.
A common approach is to arrange multiple agents at each phase, instructing them to collaboratively solve the tasks through discussion.
Depending on the configuration, the collaborating roles may include two~\cite{ChatDev, CTC, rasheed, Co-Learning, IterativeExperienceRefinement, MACNET} or more agents~\cite{lcg}.
For example, in ChatDev~\cite{ChatDev}, each task is participated in by two agents, who reach a consensus through dialogue. In \junweim{FlowGen}\textsubscript{scrum}~\cite{lcg}, agents with different roles will equally present their opinions in the sprint meeting, and in the end, the Scrum Master will summarize and extract a list of user stories.
Another approach to integrating horizontal collaboration is to run multiple pipelines that use vertical collaboration. For instance, CTC~\cite{CTC} sets up different teams of agents to develop the same software in parallel. Teams that contribute low-quality information will be eliminated, and in the end, the software development will be completed based on the high-quality information contributed by different teams.
}
    
\headt{Communication Protocol.} 
Within the end-to-end software development systems, agents communicate with other agents to exchange information. 
For those adjacent agents, the most common communication protocol is direct dialogue~\cite{self-collaboration, lcg, CodePori, aisd, AutoAgents, AgentVerse, ChatDev, MACNET, rasheed}, which leverages natural language to exchange information. This approach allows for flexible expression of intent and is close to human communication. In contrast, some agents (\eg{} MetaGPT~\cite{MetaGPT}) structure communication by having agents exchange documents and diagrams instead of relying solely on dialogue, as pure natural language may be insufficient for solving complex tasks due to distortion in multi-turn communication. Likewise, Sami \et{}~\cite{sami} proposed a framework that defines the output format of different agents. For example, the requirements engineering agent transcribes user inputs into structured software requirements, which can be downloaded in CSV format, and the architecting agent parses the structured requirements to generate PlantUML~\cite{PlantUML}.
\junwei{Except for direct communication, some end-to-end software development agents integrate the memory mechanism to establish communication between non-adjacent agents in the pipeline~\cite{self-collaboration, ChatDev, MetaGPT, AutoAgents, CTC, AgileCoder, MACNET}. For example, in MetaGPT~\cite{MetaGPT}, all agents have a shared information pool from which they can obtain the required information.}

\begin{table*}[]
\centering
\caption{Benchmarks for End-to-end Software Development}
\label{tab:endtoenddata}
\begin{adjustbox}{width=1.0\textwidth}
\renewcommand{\arraystretch}{1.3}

\begin{tabular}{
    l l c 
    >{\centering\arraybackslash}m{4cm}
    c
    c 
    >{\centering\arraybackslash}m{4cm}
}

\hhline
\textbf{Granularity} & \textbf{Benchmarks} & \textbf{\# Tasks} & \textbf{Input Scale} & \textbf{Output Scale} & \textbf{Language} & \textbf{Evaluated Agents} \\ 
\hhline

\multirow{5}{*}{\textbf{Project}} 
& SRDD~\cite{ChatDev} & 1,200 & \tabincell{c}{Software Description \\ (55 words)} & Multiple Files & Python & \cite{ChatDev, Co-Learning, IterativeExperienceRefinement, CTC, MACNET}  \\ \cline{2-7}
& CAASD~\cite{aisd} & 72 & \tabincell{c}{Software Description \\ (50 words)}  & Multiple Files & Python  & \cite{aisd, ChatDev, MetaGPT} \\ \cline{2-7}
& SoftwareDev~\cite{MetaGPT} & 70 & \tabincell{c}{Software Description \\ (30 words)} & Multiple Files & Python & \cite{MetaGPT, ChatDev, AgentVerse} \\ \cline{2-7}
& SketchEval~\cite{CodeS} & 19 & \tabincell{c}{README \\ (421 words)} & Structured Multiple Files & Python & \cite{CodeS, ChatDev} \\ \cline{2-7}
& ProjectDev~\cite{AgileCoder} & 14 & \tabincell{c}{Software Description \\ (262 words)} & Multiple Files & Python & \cite{AgileCoder, ChatDev, MetaGPT} \\ \cline{2-7}
\hhline

\multirow{2}{*}{\textbf{Method}} 
& \tabincell{l}{HumanEval~\cite{Humaneval} \\ HumanEval-ET~\cite{CodeScore}}  
& 164 & \tabincell{c}{Function Description \\ (68 words)} & Single Function & Python & \cite{lcg, CodePori, AutoAgents, AgentVerse, MetaGPT, self-collaboration, AgileCoder, MACNET} \\ \cline{2-7}
& \tabincell{l}{MBPP~\cite{MBPP} \\ MBPP-ET~\cite{CodeScore}}  
& 974 & \tabincell{c}{Function Description \\ (15 words)} & Single Function & Python & \cite{lcg, CodePori, MetaGPT, self-collaboration, AgileCoder} \\
\hhline

\end{tabular}
\end{adjustbox}
\end{table*}

\begin{table}
\small
\caption{Metrics Used in Evaluating Agents for End-to-end Software Development}
\label{tab:endtoendmetric}
\begin{adjustbox}{width=1.0\columnwidth}
\begin{tabular}{
    >{\centering\arraybackslash}m{3cm} 
    >{\centering\arraybackslash}m{3cm}
    c
}
\hhline 
\textbf{Category} & \textbf{Metrics} & \textbf{Used Agents} \\ 
\hhline 
Execution Validation & \tabincell{c}{Pass Rate \\ Pass\@K \\ Executability \\ \# Errors} & \tabincell{c}{\cite{ChatDev, MetaGPT, self-collaboration} \\ \cite{AgentVerse, AutoAgents, Co-Learning} \\ \cite{IterativeExperienceRefinement, lcg, CodePori} \\ \cite{CTC, AgileCoder, MACNET}} \\
\hline
Similarity & \tabincell{c}{SketchBLEU~\cite{CodeS} \\ Cosine Distance} & \tabincell{c}{\cite{CodeS, ChatDev, Co-Learning} \\ \cite{IterativeExperienceRefinement, CTC, MACNET}} \\
\hline
Costs & \tabincell{c}{Running Time \\ Token Usage \\ Expenses \\ \#Sprints}  & \cite{MetaGPT, AgileCoder} \\
\hline
Manual Efforts & Human Revision Costs  & \cite{MetaGPT} \\
\hline
Generated Code Scale & \tabincell{c}{Line of Code \\ Code Files \\ Completeness} & \tabincell{c}{\cite{MetaGPT, ChatDev, Co-Learning} \\ \cite{IterativeExperienceRefinement, CTC, MACNET}} \\ \hhline 

\end{tabular}
\end{adjustbox}
\end{table}
\subsubsection{Agent Evaluation}
Given the complexity of end-to-end software development, researchers further build diverse benchmarks and metrics for a comprehensive evaluation. 

\headt{Benchmarks.}
Table~\ref{tab:endtoenddata} summarizes the benchmarks used for evaluating existing LLM-based agents for end-to-end software development. In particular, we can observe that there are still a large number (\ie{} 8) of papers using the function-level code generation benchmarks (\eg{} HumanEval~\cite{Humaneval} or MBPP~\cite{MBPP}) for evaluating end-to-end software development. Although these traditional code benchmarks can represent end-to-end software development to some extent, they still involve simplified, small-scale development tasks (\ie{} input of short function descriptions and output of a single function, as shown in Table~\ref{tab:endtoenddata}). In addition, there are five more complicated project-level benchmarks that aim at simulating the end-to-end software development, \ie{} SRDD~\cite{ChatDev, Co-Learning, IterativeExperienceRefinement, CTC, MACNET}, CAASD~\cite{aisd}, SoftwareDev~\cite{MetaGPT}, SketchEval~\cite{CodeS}, and ProjectDev~\cite{AgileCoder}. The tasks in these benchmarks include more complicated and longer requirement descriptions (\eg{} the average length of software description in ProjectDev is 262 words), and their expected outputs are supposed to contain multiple files. In particular, the benchmark SketchEval is built upon the real-world GitHub repositories, and its input descriptions are extracted from the README file of the software. 
\junwei{However, although these project-level benchmarks are more complex than function-level benchmarks, their overall complexity remains limited (\eg{} most project inputs contain no more than 60 words), failing to capture the intricacies of real-world SE challenges. Moreover, most methods are evaluated on self-constructed benchmarks, which limits the comparability between different approaches. Drawing inspiration from recent high-quality software engineering datasets~\cite{SWE-bench, yu2024codereval}, future research can focus on constructing a high-quality benchmark to advance end-to-end software development techniques, potentially by mining existing GitHub repositories and incorporating manual annotation and filtering.}

\headt{Metrics.} 
Table~\ref{tab:endtoendmetric} summarizes the metrics used for evaluating existing LLM-based agents for end-to-end software development. In fact, given the difficulty of generating a complicated program, it is possible that the generated program cannot perfectly pass the tests. Therefore, in addition to the common metrics (\eg{} Pass Rate or Pass@K) that execute the generated program for validation, there are multiple dimensions for assessing how existing agents perform in end-to-end software development. In particular, there are (i) the similarity metrics between the generated program and the ground truth (\eg{} SketchBLEU~\cite{CodeS} measures the structure similarity), (ii) the costs of executing or generating the program, (iii) the manual efforts to further refine the generated program, and (iv) the scale of the generated program. 
\junwei{However, existing evaluation metrics are insufficient for end-to-end software development tasks. For instance, software accuracy is merely assessed based on manually scored executability or code similarity, which may not fully capture the quality of the generated software. Furthermore, some critical evaluation dimensions, such as robustness, security, and cost, remain underexplored, limiting the impact and reliability of these approaches. Future research could focus on these fine-grained metrics. Moreover, key metrics in the agent workflow (\eg{} crash rate, valid iteration rounds, \etc{}) could also be considered to form a more systematic and comprehensive evaluation.}

\junweim{
\subsubsection{Challenges of LLM-based Agents in End-to-end Software Development.} LLM-based agents face several challenges in the end-to-end software development task. First, most existing LLM-based agents still follow a linear, waterfall-style workflow without true iterative or evolutionary mechanisms~\cite{self-collaboration, ChatDev, MetaGPT}. This workflow makes it difficult to handle the complexity and dynamism of real-world software development, where requirements often evolve and feedback loops are essential. Second, current agents are predominantly designed for Python or basic Web application development~\cite{ChatDev, MetaGPT, Co-Learning,CTC}, limiting their applicability to a narrow subset of software types. In practice, software systems span a wide range of languages, frameworks, and architectures, many of which involve complex dependencies and integration constraints. As a result, the generalizability of current approaches remains limited. Third, this field lacks standardized benchmarks and evaluation metrics for assessing end-to-end software generation. This is partly due to the open-ended nature of the task, \ie{} when the target software is totally unknown in advance, it is difficult to define test cases upfront. Furthermore, important aspects of software quality, such as component reusability or architectural soundness, are challenging to quantify and are currently not well-captured by existing evaluation methods. These limitations highlight the need for more adaptive development workflows, broader language and domain support, and better-defined evaluation standards.
}

\begin{table*}[]
\centering
\caption{Existing LLM-based Agents for End-to-end Software Maintenance}

\label{tab:issue}
\begin{adjustbox}{width=1.0\textwidth}
\renewcommand{\arraystretch}{1.3}

\begin{tabular}{
    c|c|c
    >{\centering\arraybackslash}m{1cm}
    >{\centering\arraybackslash}m{4cm}
    >{\centering\arraybackslash}m{1cm}
    >{\centering\arraybackslash}m{3cm}
    >{\centering\arraybackslash}m{3cm}
    >{\centering\arraybackslash}m{1cm}
}
\hhline
\multirow{2}{*}{\textbf{Agents}} & \multirow{2}{*}{\textbf{Multi-Agent}} & \multicolumn{6}{c}{\textbf{Phases}} &  \\ \cline{3-9} 
 &  & \textbf{Preprocessing} & \textbf{\begin{tabular}[c]{@{}c@{}}Issue \\ Reprod.\end{tabular}} & \textbf{\begin{tabular}[c]{@{}c@{}}Issue \\ Localization\end{tabular}} & \textbf{\begin{tabular}[c]{@{}c@{}}Task \\ Decomp. \end{tabular}} & \textbf{\begin{tabular}[c]{@{}c@{}}Patch \\ Generation\end{tabular}} & \textbf{\begin{tabular}[c]{@{}c@{}}Patch \\ Verification\end{tabular}} & \textbf{\begin{tabular}[c]{@{}c@{}}Patch \\ Ranking\end{tabular}} \\ \hhline
MAGIS~\cite{MAGIS} & \checkmark & $\times$ & $\times$ & Retrieval-based & $\times$ & w/ local context & Code Review & $\times$ \\
AutoCodeRover~\cite{AutoCodeRover} & \checkmark & $\times$ & $\times$ & Navigation/Spectrum-based & $\times$ & w/ cross-file context & Static Check & $\times$ \\
SWE-agent~\cite{yang2024swe} & $\times$ & $\times$ & \checkmark & Navigation-based & $\times$ & w/ local context & Static Check & $\times$ \\
CodeR~\cite{CodeR} & \checkmark & Plan Selection & \checkmark & Spectrum-based & $\times$ & w/ cross-file context & Dynamic Check & $\times$ \\
\junweim{LingmaAgent}~\cite{RepoUnderstander} & \checkmark & Knowledge Graph Const. & $\times$ & Simulation & \checkmark & w/ cross-file context & Static Check & $\times$ \\
MASAI~\cite{MASAI} & \checkmark & Test Template Generation & \checkmark & Navigation-based & \checkmark & w/ local context & Static/Dynamic Check & \checkmark \\
Agentless~\cite{Agentless} & $\times$ & Repository Tree Const. & $\times$ & Navigation-based & $\times$ & w/ local context & Static/Dynamic Check & \checkmark \\ 
SpecRover~\cite{SpecRover} & \checkmark & $\times$ & \checkmark & Navigation-based & $\times$ & w/ cross-file context & Static/Dynamic Check & \checkmark \\
\junweim{DEIBase}~\cite{DEIBASE} & \checkmark & $\times$ & $\times$ & $\times$ & $\times$ & $\times$ & Static/Dynamic Check & \checkmark \\
\hhline
\end{tabular}

\end{adjustbox}
\end{table*}
\subsection{End-to-end Software Maintenance}~\label{sec:se:improvement}
Software systems undergo maintenance as requirements continuously change (\ie{} adding, deleting, or modifying features) or unexpected software behaviors arise. In practice, users report unsatisfactory behaviors that they encounter; developers then diagnose the reported issues and modify the software to fix them. Such an end-to-end software maintenance process can be time-consuming and labor-intensive in practice, as it involves multiple phases, including understanding user-reported issues, localizing code for maintenance, and precisely editing code to address issues. Recently, there has been an increasing number of multi-agent systems aiming at automatically solving issues of real-world software projects. Table~\ref{tab:issue} summarizes the characteristics of these agents.

\begin{figure*}[htb]
    \centering
    \includegraphics[width=1.0\textwidth]{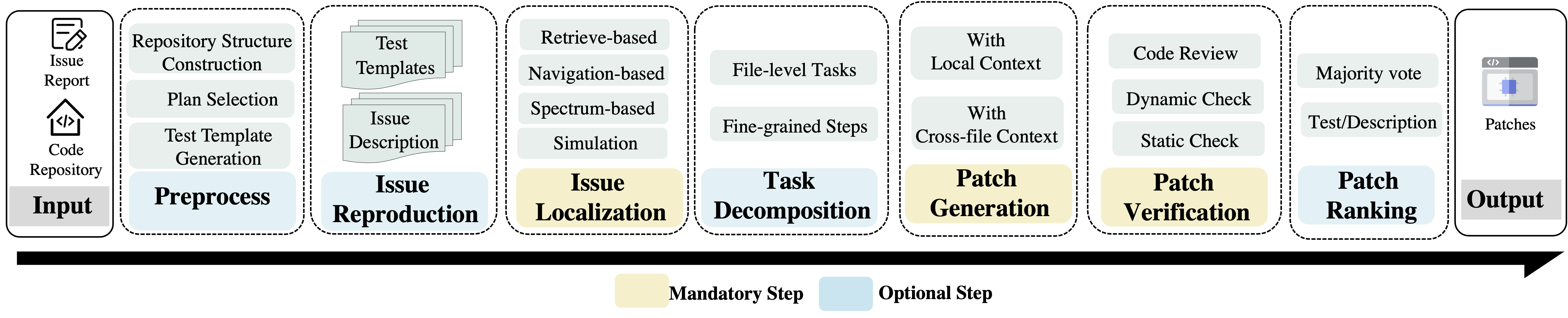}
    \caption{Pipeline of LLM-based Agents for End-to-end Software Maintenance}  
    \label{fig:issue_resolution}
\end{figure*}

\junwei{\textbf{Framework:} Figure~\ref{fig:issue_resolution} illustrates the pipeline of existing LLM-based agent systems for end-to-end software maintenance. All of the existing agents follow a common pipeline of three mandatory phases, \ie{} \textit{issue localization}, \textit{patch generation}, and \textit{patch verification}, where different agents incorporate various strategies to tackle each phase. In addition, some optional phases can be included in the pipeline to improve performance, \ie{} \textit{preprocessing}, \textit{issue reproduction}, \textit{issue localization}, \textit{task decomposition}, \textit{patch generation}, \textit{patch verification}, and \textit{patch ranking}. }

\subsubsection{Preprocessing.}
To better understand the whole repository, some agents first perform preprocessing to prepare prior knowledge before the entire procedure. The agent system \junweim{LingmaAgent}~\cite{RepoUnderstander} constructs a knowledge graph of the entire code repository to facilitate the subsequent process of issue localization. Meanwhile, Agentless~\cite{Agentless}, which is simplistic and less agentic, simply turns the whole project into a tree-like structure that demonstrates all directories and files of the repository in a hierarchical format, which facilitates the issue localization phase.
In CODER~\cite{CodeR}, a manager agent first chooses a plan from several workflows pre-defined by human experts.
In MASAI~\cite{MASAI}, the test template generator is used to analyze the testing setup of the repository and generate a test template with the running command, which further serves as an example for the following issue reproduction phase.

\subsubsection{Issue Reproduction.}
A test script that triggers the unexpected behaviors users encounter is essential for issue resolution. It not only helps with issue localization but also serves as the verification criterion for patch correctness. However, in practice, users do not always provide such reproduction tests when they report issues, and such reproduction tests are often added by developers after they fix the buggy software. Therefore, some agents design the issue reproduction phase that aims at generating the test script that can trigger the unexpected behaviors encountered by users. For example, SWE-agent~\cite{yang2024swe}, CodeR~\cite{CodeR}, and SpecRover~\cite{SpecRover} all directly leverage the agent to generate reproduction tests based on issue descriptions when there is no existing reproduction script in issue descriptions. 
However, generating reproduction tests can be challenging, as the tests must be executable and ideally should fail on the buggy software version while passing on the fixed version. Therefore, to increase the success rate of issue reproduction,  the multi-agent system MASAI~\cite{MASAI} includes a two-stage approach for issue reproduction, which first investigates the test framework and existing tests for generating a sample test template (generated in the preprocessing phase) and then uses the template as a demonstration to create the reproduction script.

\subsubsection{Issue Localization.}
Issue localization is one of the most important phases where the agents are supposed to precisely identify the code elements (\eg{} classes, methods, or code blocks) that are related to issues and should be edited. 
\junwei{
Unlike fault localization discussed in Section~\ref{sec:se:debugging:FL}, which is a task triggered by the developers during the debugging stage and involves test files that reproduce the bug, issue localization is a user-triggered process, typically only accompanied by a natural language description of the issue from the user's perspective and lacking reproducible test scripts. This characteristic poses a unique challenge for issue localization and makes it heavily dependent on the natural language understanding capabilities of LLMs.
}
We then summarize the common localization strategies used in existing LLM-based agents. 

\headt{Retrieval-based Localization.} 
Retrieval-based localization is the most fundamental approach, where agents identify suspicious code elements based on their similarity to issue descriptions. For example, MAGIS~\cite{MAGIS} ranks all code files using BM25~\cite{BM25} and selects the Top-K most relevant ones as potential issue locations. However, retrieval-based methods typically operate at a coarse granularity (\eg{} file level) and can not pinpoint the exact functions or lines that need modification. To improve localization accuracy, agents often integrate additional strategies such as navigation, spectrum analysis, or simulation. 
However, the similarity with the issue description still plays a crucial role in these strategies, for example, by filtering out irrelevant files to narrow the search space~\cite{Agentless}, assessing whether the localization path deviates from the description~\cite{RepoUnderstander}, or combining it with other strategies~\cite{CodeR}.

\headt{Navigation-based Localization.} 
\junwei{
The navigation-based localization approach generally equips agents with a series of actions for browsing the directory structure (\eg{} listing files within a directory, searching directories or files, scrolling within files, \etc{}), allowing agents to autonomously explore the project directory and locate specific code snippets~\cite{yang2024swe, SpecRover, MASAI, AutoCodeRover}. The only exception is Agentless~\cite{Agentless}, which directly presents the entire repository's file and directory structure in the prompt to locate the relevant files.
Another distinction between different approaches lies in their navigation strategies. For instance, SWE-agent~\cite{yang2024swe} and MASAI~\cite{MASAI} both allow the agents to autonomously determine the exploration paths. In contrast, the navigation path in AutoCodeRover~\cite{AutoCodeRover} and SpecRover~\cite{SpecRover} is influenced by previously identified relevant code snippets, which are inserted into the prompt as context to guide further exploratory navigation. Agentless~\cite{Agentless} employs a hierarchical approach and instructs the LLM to gradually localize files, classes, functions, and concrete edit locations.}

\headt{Spectrum-based Localization.}
\junwei{Some agents integrate spectrum-based fault localization (SBFL)~\cite{wong2016survey}, which assigns a suspiciousness score to code elements based on their coverage in passing and failing tests~\cite{AutoCodeRover, CodeR}. The difference between these approaches lies in the source of the test suite.
AutoCodeRover~\cite{AutoCodeRover} explores spectrum-based fault localization under ideal conditions, utilizing the developer-written test cases from SWE-bench Lite as the test suite. 
In contrast, CODER~\cite{CodeR} adopts a more practical approach by first generating reproduction tests and then calculating a score based on both the suspiciousness scores and the BM25 scores.
The experimental results show that the issue resolution rate increases from 17.00\% to 20.33\% in AutoCodeRover. However, in CodeR, the localization accuracy of using SBFL alone is lower than that of combining BM25 scores, which might be attributed to differences in test script quality (human-written vs. model-generated).}

\headt{Simulation-based Localization.}
Simulation is a special technique adopted by \junweim{LingmaAgent}~\cite{RepoUnderstander} for issue localization. It applies the classic Monte Carlo Tree Search (MCTS)~\cite{browne2012survey} algorithm. By recursively incorporating nodes with higher similarity with the issue, it evaluates and ranks the most relevant paths in the repository knowledge graph. The collected code is then summarized for issue localization.

\subsubsection{Task Decomposition.}
Before generating patches, some agents decompose the task into more fine-grained sub-tasks. For instance, in MAGIS~\cite{MAGIS}, its manager agent breaks down the issue into file-level tasks and delegates them to a newly-formed development team; similarly, in \junweim{LingmaAgent}~\cite{RepoUnderstander}, its summary agent summarizes the collected code and issue description, and then outlines the fine-grained steps for issue resolution.
 
\subsubsection{Patch Generation.}
In this phase, the agents generate patches for the localized suspicious code elements. The input context for this phase typically includes the issue/task description and the suspicious code elements to be modified~\cite{MAGIS, yang2024swe, MASAI, Agentless}. In addition, some agents (\eg{} AutoCodeRover~\cite{AutoCodeRover}, CodeR~\cite{CodeR}, and \junweim{LingmaAgent}~\cite{RepoUnderstander}) further refine the input contexts by including relevant cross-file code contexts that are collected by retrieval APIs. Notably, SpecRover~\cite{SpecRover} provides ancillary function summaries for all collected code snippets, which reflect the high-level intent of related functions and can further assist patch generation.

\subsubsection{Patch Verification.}
Agents further verify the correctness of the generated patches, which is challenging as the reproduction tests are not always available in practice. Therefore, agents incorporate different verification strategies. 

\headt{Code Review.}
Some agents design a quality assurance agent to review the quality of generated patches. 
For example, in MAGIS~\cite{MAGIS}, each developer agent is paired with a QA engineer agent to review the code change and provide timely feedback.
SpecRover~\cite{SpecRover} assigns a reviewer agent to check the correctness of both the patch and reproducer test, which can not only mitigate misjudgments caused by errors in either but also provide comprehensive feedback to assist with iterative modifications.

\headt{Static Checking.}
Some agents (\eg{} AutoCodeRover~\cite{AutoCodeRover}, \junweim{LingmaAgent}~\cite{RepoUnderstander}, MASAI~\cite{MASAI}, Agentless~\cite{Agentless}, and SWE-agent \cite{yang2024swe}) use static checking approaches to assess the syntactic correctness, indentation, and compatibility of the generated patch with the repository environment. 

\headt{Dynamic Checking.}
Since the static checking cannot find the semantic violation of the patches, some agents (\eg{} CodeR~\cite{CodeR} and MASAI~\cite{MASAI}) further perform dynamic checking by executing the reproduction test on the patch. The patch that passes the reproduction test can be considered as effectively resolving the issue. In particular, existing reproduction tests are reused (if available); otherwise, reproduction tests generated during the issue reproduction phase are used. Agentless~\cite{Agentless} also implements a dynamic checking approach by conducting regression testing to filter out incorrect candidate patches. Notably, SpecRover~\cite{SpecRover} performs both reproduction test and regression test to guarantee the correctness of the final selected patch.

\subsubsection{Patch Ranking.}
In the patch verification phase, code review and static checking strategies are not sufficient for checking the correctness of generated patches. The dynamic checking methods require reproduction or regression test scripts, which are not always available, and may not be able to decide whether the generated patches are correct or not. For example, the generated reproduction test may be incorrect itself, and the correct patch may edit some code snippets that can resolve the issue, but fail the existing regression tests. 
To tackle these problems, some agents are designed to generate multiple patches and further include a patch ranking phase to identify the optimal patch. For example, in MASAI~\cite{MASAI}, a ranker agent is responsible for ranking all potential patches based on the issue description and reproduction tests; In Agentless~\cite{Agentless}, all patches are normalized and re-ranked based on the number of occurrences with the majority voting strategy. In SpecRover~\cite{SpecRover}, patches failing some tests, together with the issue descriptions, are provided to the selection agent, which then deeply analyzes the cause of the issue and chooses the best patch.
\junweim{DEIBase}~\cite{DEIBASE} is a framework that integrates multiple expert agents (\eg{} Agentless~\cite{Agentless}, Moatless~\cite{Moatless}, and Aider~\cite{Aider}). It chooses the optimal patch from candidates generated by existing expert agents to achieve a higher resolve rate. To rerank candidate patches, it assigns an LLM-based code review committee, which takes the issue description, the relevant context, the original code, and the patched code as input and scores each candidate patch based on its analysis and explanation.

\begin{figure}[htb]
    \centering
    \includegraphics[width=1.0\columnwidth]{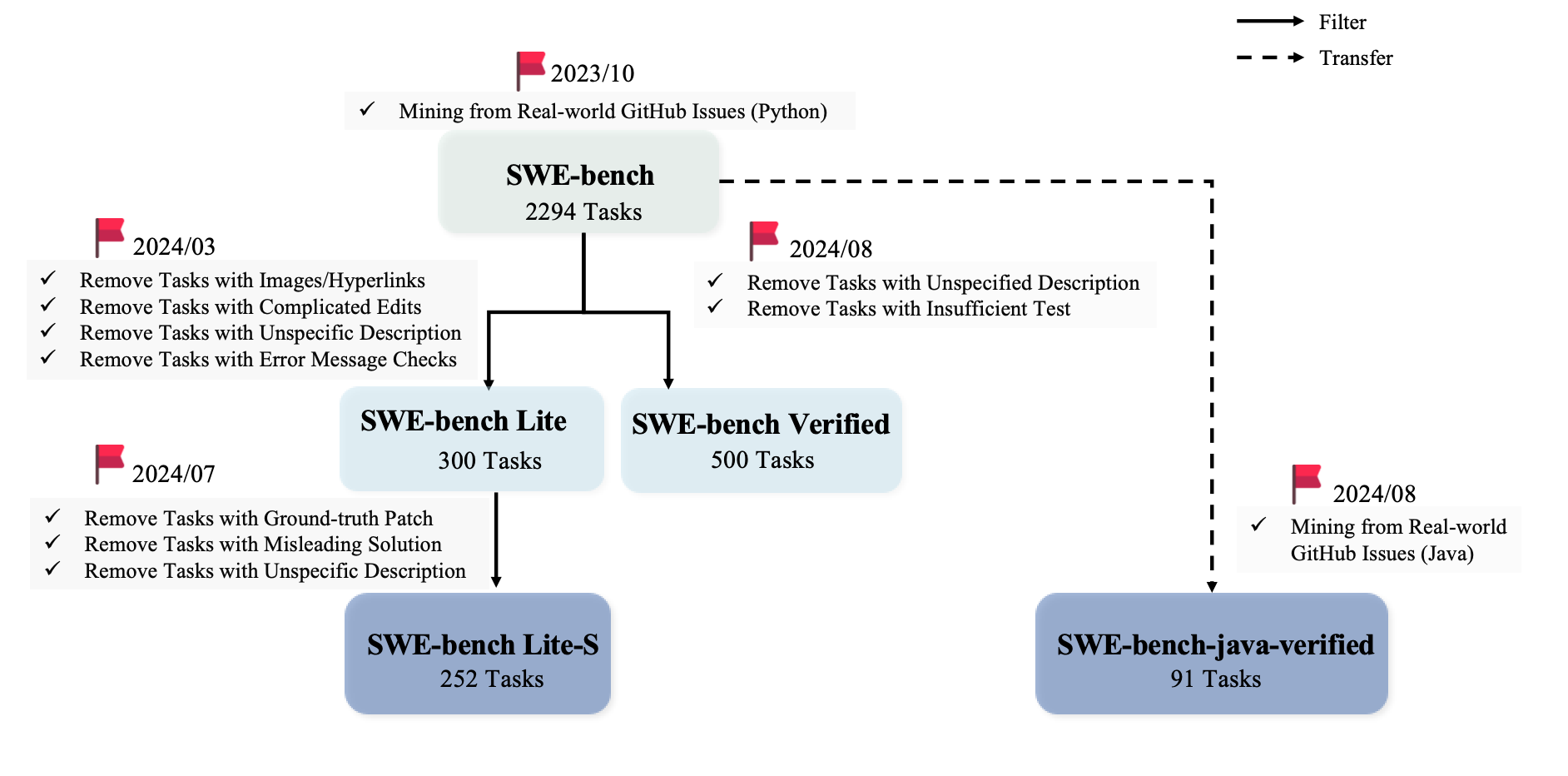}
    \caption{Benchmark Evolution in Software Maintenance} 
    \label{fig:swebench}
\end{figure}

\subsubsection{Agent Evaluation}
In this section, we will discuss the benchmarks used for end-to-end software maintenance tasks and analyze the performance of existing approaches.

\headt{Benchmarks.}
\junwei{Traditional fault localization often relies on Defects4J~\cite{defects4j}, a dataset that extracts real-world code bugs from software projects, providing pre-fix and post-fix repositories, repair patches, and bug-exposing tests. While Defects4J includes real bugs, it also provides well-filtered test cases with test names, root causes, and stack traces. However, in real-world maintenance scenarios, issues are typically reported by users, who usually provide only descriptions of the problem without reproducible test scripts. Additionally, Defects4J contains only 357 bugs from five open-source projects, with 92.41\% of patches modifying only a single file~\cite{sobreira2018dissection}. These limitations create a notable gap between Defects4J and real-world software maintenance in both scale and complexity.}

Therefore, to evaluate how LLM-based agents tackle real-world end-to-end software maintenance, researchers build benchmarks by mining user-reported issues from GitHub, including SWE-bench~\cite{SWE-bench}, SWE-bench Lite~\cite{swe-bench-lite}, SWE-bench Lite-S~\cite{Agentless}, SWE-bench Verified~\cite{swe-bench-verified}, and SWE-bench-java-verified~\cite{swe-bench-java-verified}. 
\junwei{Figure~\ref{fig:swebench} summarizes the evolution timeline and relationships of existing benchmarks for end-to-end software maintenance, with the ``Filter'' relationship referring to extracting a subset and the ``Transfer'' relationship referring to adopting a similar construction strategy.}

SWE-bench~\cite{SWE-bench} is the first benchmark for end-to-end software maintenance, which consists of 2,294 real-world GitHub issues across 12 popular Python repositories. Each task in SWE-bench includes an original text from a GitHub issue (\ie{} the issue description or problem statement), the entire code repository, the execution environment (\ie{} Docker environment), and validation tests (\ie{} tests that are hidden from the evaluated agents).  
\junwei{A typical evaluation process is as follows: the agents under evaluation will take the issue description and the buggy code repository as input and attempt to locate and fix the issue, with passing the validation tests serving as a success indicator.
The \textit{resolve rate}, which is defined as the ratio of resolved issues to total issues, is the common metric used to evaluate the agents' performance on these end-to-end software maintenance datasets.
}

However, the full SWE-bench benchmark can take too much evaluation costs, and it contains particularly difficult or problematic tasks~\cite{Agentless}, which can underestimate the evaluation of LLM-based agents. Therefore, researchers have dedicated considerable manual efforts to extract subsets of SWE-bench that feature high-quality tasks with acceptable cost, reasonable difficulty, self-contained information, informative issue descriptions, and sufficient evaluation tests.
For example, SWE-bench Lite~\cite{swe-bench-lite} is a subset of SWE-bench that manually removes tasks requiring complicated edits (\eg{} editing more than one file), and the tasks including images or hyperlinks; SWE-bench Lite-S~\cite{Agentless} further removes tasks that contain exact patches, misleading solutions, or insufficient information in the issue descriptions; similarly, SWE-bench Verified~\cite{swe-bench-verified} removes cases with unspecified descriptions or insufficient tests.  

In addition to issues in Python projects, some researchers transfer the build process of SWE-bench and construct benchmarks in other popular languages (\eg{} Java).  SWE-bench-java-verified~\cite{swe-bench-java-verified} is constructed by collecting Java projects from GitHub and the Defects4j dataset. Through rigorous validation and filtering, it eventually includes 91 issues across 6 Java projects. Experiments based on the SWE-agent and DeepSeek-Coder achieve the best resolve rate of 9.89\%, fixing 9/91 issues.

\junwei{
\headt{Performance Comparison.} 
SWE-bench Lite is the most widely used evaluation benchmark in practice. The reported resolve rate data on SWE-bench Lite are illustrated in Figure~\ref{fig:swebench_lite_comprison}, which are collected from the original papers.
It is worth noting that most of these agents are based on the GPT-4 series models, such as GPT-4~\cite{yang2024swe, MAGIS, AutoCodeRover, CodeR, RepoUnderstander}, and GPT-4o~\cite{MASAI, Agentless, DEIBASE}. Therefore, the model capabilities do not significantly contribute to the differences in performance. The only exception is SpecRover~\cite{SpecRover}, which primarily uses the Claude-3.5-Sonnet model~\cite{claude}.
}

\junwei{In conjunction with the phases and techniques presented in Table~\ref{tab:issue}, we have drawn some interesting observations:
\begin{itemize}[leftmargin=*, label=-]
    \item Pure autonomous localization methods do not necessarily lead to better performance. A typical example is SWE-agent~\cite{yang2024swe}, which allows for fully autonomous issue localization by designing an interaction interface between the agent and the computer. However, it performed the worst among all the agents on SWE-bench Lite.
    \item Agents that employ dynamic patch verification and patch ranking strategies generally achieve higher resolve rates. Among the top five agents with the highest repair rates, Agentless~\cite{Agentless}, MASAI~\cite{MASAI}, SpecRover~\cite{SpecRover}, and \junweim{DEIBase}~\cite{DEIBASE} all adopt both dynamic patch verification (\ie{} test execution) and patch ranking strategies. CodeR~\cite{CodeR} does not use patch ranking, but it still adopts dynamic code checking.
    \item Approaches adopting traditional fault localization strategies, such as Agentless~\cite{Agentless}, have surpassed many agentic approaches, which place higher demands on evaluating the effectiveness of complex agent designs.
\end{itemize}}

\textbf{\begin{figure}[h]
    \centering
    \includegraphics[width=1.0\columnwidth]{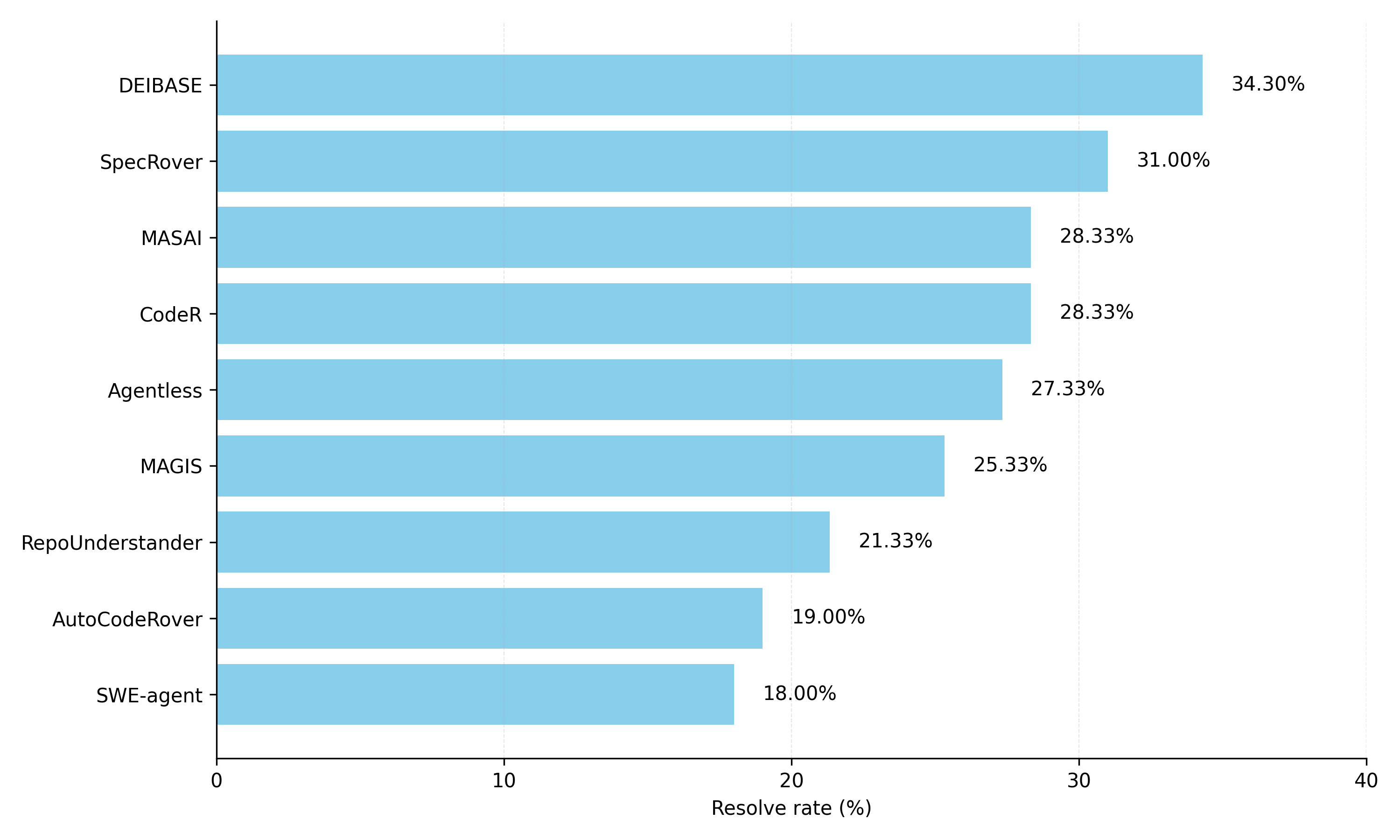}
    \caption{Resolve Rate of LLM-based Agents on SWE-bench Lite} 
    \label{fig:swebench_lite_comprison}
\end{figure}
}

\junweim{
\subsubsection{Challenges of LLM-based Agents in End-to-end Software Maintenance.}
LLM-based agents encounter several key challenges in the end-to-end software maintenance task. First, issue descriptions submitted by users are often unstructured and may include multimodal content such as links and screenshots. Effectively interpreting such inputs requires multimodal understanding and the ability to extract relevant information from external sources, which are absent in current LLM-based systems~\cite{SWE-bench}. Second, in end-to-end maintenance workflows, the ability to reproduce user-reported issues is critical for both issue localization and patch verification. However, existing approaches lack reliable mechanisms for ensuring issue reproducibility~\cite{MAGIS, AutoCodeRover, Agentless}. Third, current LLM-based agents often rely on limited strategies for patch verification, such as static checking or code review heuristics~\cite{MAGIS, AutoCodeRover, yang2024swe}, which are insufficient for guaranteeing correctness. While some agents execute tests to validate patches, they still face ambiguity in determining correctness. In some cases, the agent-generated tests may be incorrect; in others, the agent may fail to generate any test~\cite{yang2024swe, CodeR}. Some LLM-based agents rely on existing regression tests, but they may break due to unrelated code changes, leading to false negatives even when the patch is semantically correct~\cite{SpecRover, Agentless}. These challenges highlight the need for improved capabilities in understanding, reproducing, and verifying real-world software issues.
}

\section{Analysis from Agent Perspective}\label{sec:agent}
This section organizes the collected papers from the perspective of agents. 
Specifically, Section~\ref{sec:agent:framework} summarizes the components of existing LLM-based agents for SE; Section~\ref{sec:agent:multiagent} focuses on existing multi-agent systems for SE by summarizing their roles, collaboration mechanisms, information flows, and real-world applications; and Section~\ref{sec:agent:human} summarizes how humans coordinate with agents for SE. 

\subsection{Agent Framework}~\label{sec:agent:framework}
Based on the common framework of LLM-based agents~\cite{agents_fudannlp, agents_NotreDame, agents_Renmin}, this section first summarizes the common paradigms of the planning, perception, memory, and action components in existing LLM-based agents for SE, then discusses the impact of different foundation models on the effectiveness of agent systems.
\subsubsection{Planning}
\label{sec:agent:planning}
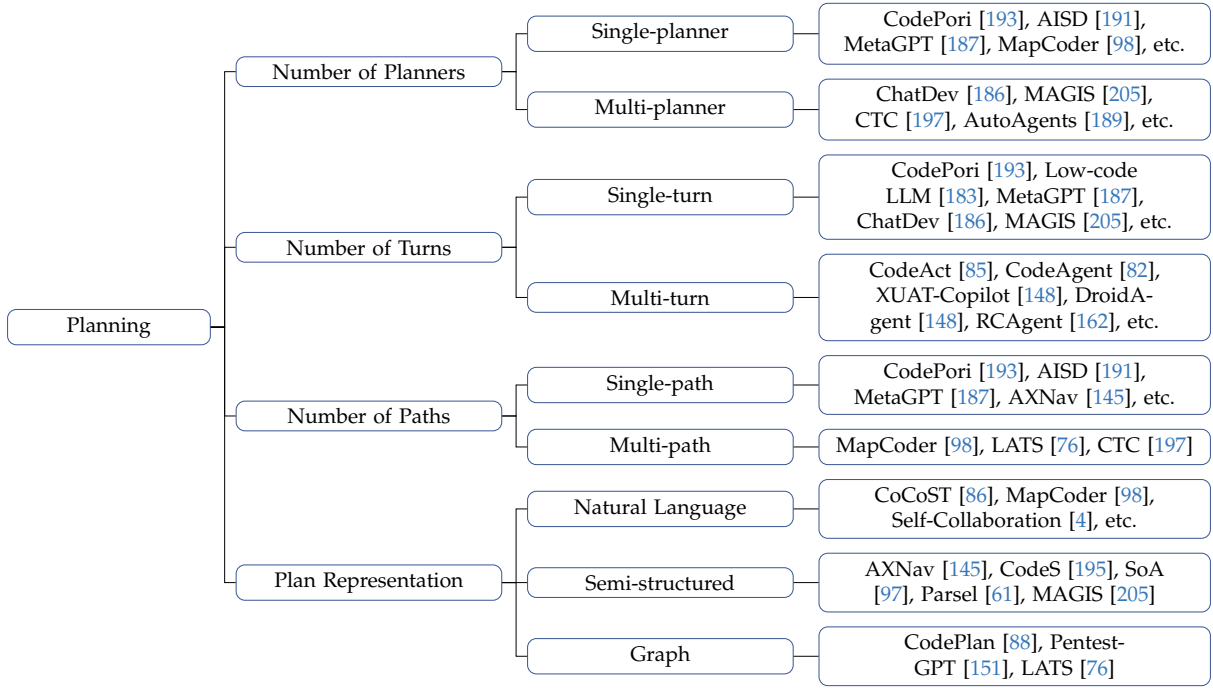
\begin{figure*}[htbp]
  \centering
  \begin{adjustbox}{width=0.9\textwidth}
    \begin{forest}
for tree={
    rounded corners,
    child anchor=west,
    parent anchor=east,
    grow'=east,
    text width=4cm,
    draw=darkblue,
    anchor=west,
    node options={align=center},
    edge path={
      \noexpand\path[\forestoption{edge}]
        (.child anchor) -| +(-5pt,0) -- +(-5pt,0) |-
        (!u.parent anchor)\forestoption{edge label};
    },
    where n children=0{text width=6cm}{}
  },
[Planning, fill=white,text width=3cm
    [Number of Planners
        [Single-planner
            [{CodePori~\cite{CodePori}, AISD~\cite{aisd}, MetaGPT~\cite{MetaGPT}, MapCoder~\cite{MapCoder}, \etc{}}]
        ]
        [Multi-planner
            [{ChatDev~\cite{ChatDev}, MAGIS~\cite{MAGIS}, CTC~\cite{CTC}, AutoAgents~\cite{AutoAgents}, \etc{}}]
        ]
    ]    
    [Number of Turns
        [Single-turn
            [{CodePori~\cite{CodePori}, Low-code LLM~\cite{low-code-llm}, MetaGPT~\cite{MetaGPT}, ChatDev~\cite{ChatDev}, MAGIS~\cite{MAGIS}, \etc{}}]
        ]        
        [Multi-turn
            [{RAT~\cite{RAT}, CodeAct~\cite{CodeAct}, CodeAgent~\cite{zhang2024codeagent}, XUAT-Copilot~\cite{XUAT-Copilot}, DroidAgent~\cite{XUAT-Copilot}, \etc{}}]
        ]
    ],
    [Number of Paths
        [Single-path
            [{CodePori~\cite{CodePori}, AISD~\cite{aisd}, MetaGPT~\cite{MetaGPT}, AXNav~\cite{AXNav}, \etc{}}]
        ],
        [Multi-path
            [{MapCoder~\cite{MapCoder}, LATS~\cite{LATs}, CTC~\cite{CTC}}]
        ]
    ]
    [Plan Representation
        [Natural Language
            [{CoCoST~\cite{he2024cocost}, MapCoder~\cite{MapCoder}, Self-Collaboration~\cite{self-collaboration}, \etc{}}]
        ]
        [Semi-structured
        [{AXNav~\cite{AXNav}, CodeS~\cite{CodeS}, SoA\cite{SoA}, Parsel~\cite{parsel}, MAGIS~\cite{MAGIS}}]
        ]
        [Graph
            [{CodePlan~\cite{bairi2023codeplan}, PentestGPT~\cite{PentestGPT}, LATS~\cite{LATs}}]
        ]
    ]  
  ]
\end{forest}
\end{adjustbox}
  \caption{Taxonomy of Planning Strategies in LLM-based Agents for Software Engineering}
  \label{fig:planning-tree}
\end{figure*}

In SE, intricate tasks such as development and maintenance activities necessitate the orchestrated efforts of various agents through multiple iterative cycles. Therefore, planning is an essential component for agent systems by meticulously delineating task sequences and strategically scheduling agents to ensure the seamless progression of the SE process. Figure~\ref{fig:planning-tree} presents the taxonomy of the planning components in existing LLM-based agents for SE.

\headt{Single Planner vs. Multiple Planners.}
In LLM-based agent systems, planning is typically handled by a specialized agent~\cite{CodePori,aisd, DroidAgent,ICAA,MetaGPT,Flows,self-collaboration,low-code-llm,MapCoder,parsel,AXNav} or as a core responsibility of the sole single agent~\cite{CodeAct, zhang2024codeagent, XUAT-Copilot, LLM4PLC,RCAgent,LATs,he2024cocost, RAT}. Some works use the function-calling interface~\cite{FunctionCalling} provided by GPT-3.5~\cite{GPT3.5} or GPT-4~\cite{GPT4}, handing over the planning task to high-performance models~\cite{zhang2024codeagent}.
However, given the pivotal role that planning plays in influencing subsequent action steps, some works incorporate a collaborative approach among several agents to further enhance the accuracy and practicality of the plans formulated~\cite{AutoAgents, Flows, ChatDev, MAGIS, CTC, AgileCoder, rasheed}. 
\junwei{Single-planner strategies incur lower overhead and demonstrate broader applicability, but are more susceptible to the risks of hallucination. In contrast, multi-planner approaches benefit from mutual correction among different agents and the integration of specialized knowledge from diverse perspectives, thus reducing factual errors~\cite{du2024improving, wang-etal-2024-unleashing}. However, they also introduce additional token consumption and time overhead~\cite{ChatDev}, making them more suitable for complex tasks such as competitive coding~\cite{Flows}, end-to-end software development~\cite{AutoAgents, ChatDev, CTC, AgileCoder, rasheed}, and end-to-end code maintenance~\cite{MAGIS}.}

\headt{Single-turn Planning vs. Multi-turn Planning.}
The fundamental planning strategy is to craft a holistic plan from the very beginning, and then proceed to implement it in successive rounds~\cite{CodePori,zhang2024codeagent,aisd, LLM4PLC,ICAA, MetaGPT,Flows,ChatDev,self-collaboration,MAGIS,CodeS,he2024cocost,MapCoder,parsel,CTC,AgileCoder, low-code-llm, rasheed, RAT}. 
In contrast, some SE agents have adopted a ReAct-like~\cite{React} architecture, which implements a multi-turn planning mechanism wherein the next-round actions will not be determined until receiving the environmental feedback from the previous round. This form allows for dynamic revision and expansion of the plan, enabling it to adapt to more flexible task scenarios, such as issue resolution~\cite{MASAI, LATs}, iterative code generation~\cite{CodeAct, zhang2024codeagent}, mobile app testing~\cite{XUAT-Copilot, DroidAgent, AXNav}, among others~\cite{RCAgent}.
\junwei{Single-turn planning requires as much knowledge as possible prior to planning to alleviate hallucinations. In addition, it faces the challenge of performing fine-grained task decomposition for complex tasks. On the other hand, multi-turn planning allows for plan adjustments based on environmental feedback, thereby improving fault tolerance. However, for complex tasks, the gradually accumulating trajectories may exceed the model’s limited context window, leading to hallucinations and forgetting~\cite{huang2024understanding}. Consequently, some memory mechanisms are necessary to prevent the agent from deviating from its initial objectives.}

\headt{Single-path Planning vs. Multi-path Planning.}
Most LLM-based agents use single-path planning strategies, \ie{} they plan and execute tasks in a linear manner~\cite{CodePori, aisd, LLM4PLC, ICAA, MetaGPT, PentestGPT, ChatDev, self-collaboration, low-code-llm, MAGIS, CodeS, he2024cocost, parsel, AgileCoder, lcg, RAT, rasheed, CodeAct, zhang2024codeagent, XUAT-Copilot, DroidAgent, RCAgent, AXNav, bairi2023codeplan, SoA, MASAI}. 
However, agents inherit the randomness from the backbone LLMs, leading to fluctuations in task decomposition. 
One approach to address this issue is to design a multi-path planning strategy, which instructs the agents to generate or simulate multiple plans, and select~\cite{LATs}, switch~\cite{MapCoder}, or aggregate~\cite{CTC} the optimal paths for execution. 
\junwei{The single-path strategy is constrained by the inherent stochasticity of LLMs~\cite{ozkara2025stochastic}, undermining its reliability. In contrast, the multi-path strategy endows the agent with a certain degree of trial-and-error capability, thus increasing the probability of completing the task~\cite{MapCoder}. However, exploring multiple paths also introduces additional time and token costs~\cite{LATs}, limiting its practicality in certain situations.}

\headt{Plan Representation.}
The plan can be exhibited in different forms, including natural language descriptions, semi-structured representations, or graphs. 
Most agents describe the plan in natural language, especially as a list of procedural steps~\cite{ICAA, Flows, self-collaboration, low-code-llm, he2024cocost, MapCoder, RAT} or features to be implemented~\cite{aisd, lcg, MetaGPT, AgileCoder}.
Furthermore, some agents implement semi-structured plans~\cite{AXNav, CodeS, SoA, parsel, LLM4PLC}. For example, AXNav~\cite{AXNav} represents the action list in JSON format. Some code-generating agent systems directly output the code skeleton~\cite{CodeS, SoA, parsel, LLM4PLC} or present the plan as executable code~\cite{MAGIS}, which can be seen as a special plan form in SE tasks. Graphs can also be implemented as a special plan form, which facilitates the expansion and traceability of execution paths~\cite{bairi2023codeplan, PentestGPT, LATs}.

\junwei{\headt{Challenges in Planning.}
Although agents applied in software engineering  have explored various planning strategies and representations, planning still faces numerous challenges in practical applications:
\begin{itemize}[leftmargin=*, label=-]
    \item \textit{Hallucination}. LLM-based models can suffer from hallucination issues when generating plans, especially in an insufficient context. For example, agents may generate inaccurate plan steps operating non-existent methods or variables. Furthermore, during iterative planning, there is a risk of redundant steps or repetitive sequences, leading to loops in the paths planned by LLMs.
    \item \textit{Limited reliability in complex tasks}. Current planning primarily relies on the inherent reasoning capability of LLMs, which results in limited reliability in complex tasks. For example, experiments of CoCoST~\cite{he2024cocost} have shown that in scenarios with higher complexity of individual function codes, the contribution of planning significantly decreases. In addition, there is still a gap between agent-generated plans and human-provided plans for complex tasks. For instance, experiments of Flows~\cite{Flows} have demonstrated that providing just a small segment of a human-designed plan can lead to a substantial performance increase in competitive programming tasks (from 26.9\% to 74.5\% and from 47.5\% to 80.8\% on novel problems).
    \item \textit{Lack of fine-grained evaluation}. Although planning plays an important role in agents, its effectiveness is primarily reflected through performance on the task results. The validity of planning itself has not been sufficiently assessed, which is often crucial for task success. Future work should focus on fine-grained evaluation of planning, such as cost and the effectiveness of planning steps.
\end{itemize}} 
\subsubsection{Memory}
\label{sec:agent:memory}
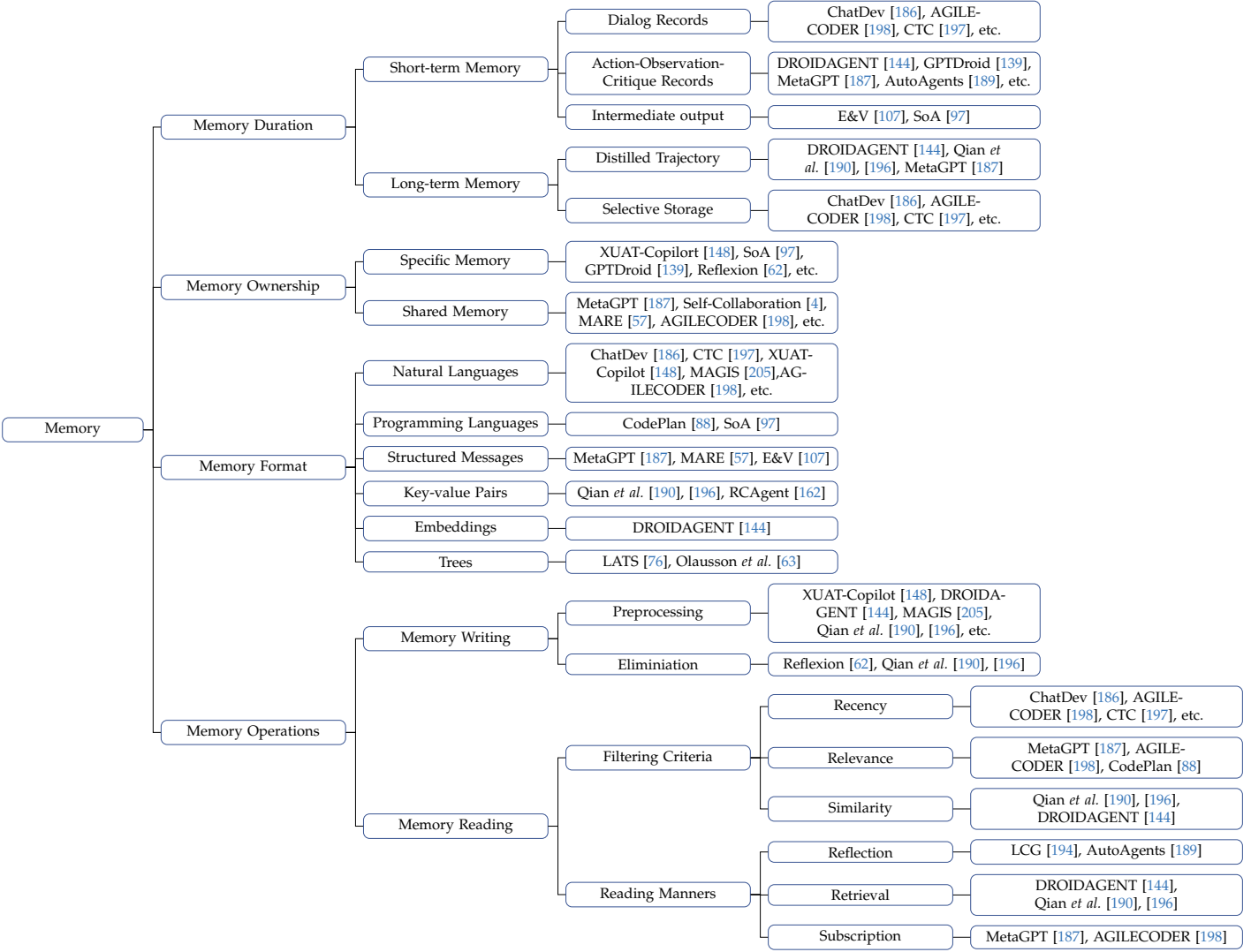
\begin{figure*}[htbp]
  \centering
  \begin{adjustbox}{width=1.0\textwidth}
    \begin{forest}
for tree={
    rounded corners,
    child anchor=west,
    parent anchor=east,
    grow'=east,
    text width=4cm,
    draw=darkblue,
    anchor=west,
    node options={align=center},
    edge path={
      \noexpand\path[\forestoption{edge}]
        (.child anchor) -| +(-5pt,0) -- +(-5pt,0) |-
        (!u.parent anchor)\forestoption{edge label};
    },
    where n children=0{text width=6cm}{}
  },
[Memory, fill=white, text width=3cm
    [Memory Duration
        [Short-term Memory
            [Dialog Records
                [{ChatDev\cite{ChatDev}, \junweim{AgileCoder}\cite{AgileCoder}, CTC~\cite{CTC}, \etc{}} ]
            ]
            [Action-Observation-Critique Records
                [{\junweim{DroidAgent}~\cite{DroidAgent}, GPTDroid~\cite{GPTDroid}, MetaGPT~\cite{MetaGPT}, AutoAgents~\cite{AutoAgents}, \etc{}}]
            ]
            [Intermediate output
                [{E\&V~\cite{EV}, SoA~\cite{SoA}}]
            ]
        ]
        [Long-term Memory
            [Distilled Trajectory                [{\junweim{DroidAgent}~\cite{DroidAgent}, Qian \et{}~\cite{Co-Learning, IterativeExperienceRefinement}, MetaGPT~\cite{MetaGPT}}]
            ]
            [Selective Storage
                [{ChatDev~\cite{ChatDev}, \junweim{AgileCoder}~\cite{AgileCoder}, CTC~\cite{CTC}, \etc{}}]
            ]
        ]            
    ]
    [Memory Ownership
        [Specific Memory
            [{XUAT-Copilort~\cite{XUAT-Copilot}, SoA~\cite{SoA}, GPTDroid~\cite{GPTDroid}, Reflexion~\cite{shinn2023reflexion}, \etc{}}]
        ]
        [Shared Memory
            [{MetaGPT~\cite{MetaGPT}, Self-Collaboration~\cite{self-collaboration}, MARE~\cite{MARE}, \junweim{AgileCoder}\cite{AgileCoder}, \etc{}}]
        ]
    ]
    [Memory Format
        [Natural Languages
            [{ChatDev~\cite{ChatDev}, CTC~\cite{CTC}, XUAT-Copilot~\cite{XUAT-Copilot}, MAGIS~\cite{MAGIS},\junweim{AgileCoder}~\cite{AgileCoder}, \etc{}}]
        ]
        [Programming Languages
            [{CodePlan~\cite{bairi2023codeplan}, SoA~\cite{SoA}}]
        ]
        [Structured Messages
            [{MetaGPT~\cite{MetaGPT}, MARE~\cite{MARE}, E\&V~\cite{EV}}]
        ]
        [Key-value Pairs
            [{Qian \et{} \cite{Co-Learning, IterativeExperienceRefinement}, RCAgent~\cite{RCAgent}}]
        ]
        [Embeddings
            [{\junweim{DroidAgent}~\cite{DroidAgent}}]
        ]
        [Trees
            [{LATS~\cite{LATs}, Self-Repair~\cite{olausson2024selfrepair}}]
        ]
        [Images
            [{VisionDroid~\cite{VisionDroid}}]
        ]
    ]
    [Memory Operations
        [Memory Writing
            [{XUAT-Copilot~\cite{XUAT-Copilot}, \junweim{DroidAgent}~\cite{DroidAgent}, MAGIS~\cite{MAGIS}, Qian \et{}~\cite{Co-Learning, IterativeExperienceRefinement}, Reflexion~\cite{shinn2023reflexion}, \etc{}}]
        ]
        [Memory Reading
            [Filtering Criteria
                [Recency
                    [{ChatDev~\cite{ChatDev}, \junweim{AgileCoder}~\cite{AgileCoder}, CTC~\cite{CTC}, \etc{}}]
                ]
                [Relevance
                    [{MetaGPT~\cite{MetaGPT}, \junweim{AgileCoder}~\cite{AgileCoder}, CodePlan~\cite{bairi2023codeplan}}]
                ]
                [Similarity
                    [{Qian \et{}~\cite{Co-Learning, IterativeExperienceRefinement}, \junweim{DroidAgent}~\cite{DroidAgent}}]
                ]                
            ]
            [Reading Manners
                [Reflection
                    [{\junweim{FlowGen}~\cite{lcg}, AutoAgents~\cite{AutoAgents}}]
                ]
                [Retrieval
                    [{\junweim{DroidAgent}~\cite{DroidAgent}, Qian \et{}~\cite{Co-Learning, IterativeExperienceRefinement}}]
                ]
                [Subscription
                    [{MetaGPT~\cite{MetaGPT}, \junweim{AgileCoder}~\cite{AgileCoder}}]
                ]
            ]
        ]
    ]        
  ]
\end{forest}
\end{adjustbox}
  \caption{Taxonomy of Memory Design in LLM-based Agents for SE} 
  \label{fig:memory-tree}
\end{figure*}

The memory component is a pivotal mechanism responsible for storing the trajectories of historical thoughts, actions, and environmental observations, enabling agents to sustain coherent reasoning and address intricate tasks. In SE, complex development and maintenance tasks generally necessitate agents conducting iterative revisions, wherein historical intermediate information, \eg{} generated code and testing reports, significantly impact integrity and continuity. We then detail the implementation of memory mechanisms in SE from four perspectives: memory duration, ownership, format, and operation.
Figure~\ref{fig:memory-tree} presents the taxonomy of the memory components in existing LLM-based agents for SE.

\headt{Memory Duration.}  
Inspired by human memory systems, agent memory can be classified into short-term memory and long-term memory based on the memory duration. 

\highlight{Short-term memory}, also known as working memory~\cite{DBLP:conf/emnlp/GevaSBL21}, is integrated into agents to enhance their ability to sustain trajectories of the current ongoing task and is frequently used when multi-turn interactions are involved. 
\junwei{Specifically, it enhances the planning module by offering a reference to the historical explorations for the given task, thereby aiding the agent in tracking task progress and preventing the recurrence of the same mistakes through trial-and-error experiences.}
In SE, short-term memory primarily stores three types of information. The first type is \textit{dialog records}, which is generally used to memorize the pure dialog history among agents and is typically in the form of history summary~\cite{XUAT-Copilot, AutoAgents} and multi-turn instruction-response pairs~\cite{ChatDev, AgileCoder, CTC, MACNET}. It is straightforward to implement and can offer a thorough and detailed historical record of the task-solving process. However, the weakness is that the dialog history can be lengthy and contain irrelevant and redundant information.
The second type is \textit{Action-Observation-Critique records}. While dialog history concentrates on the thoughts and responses among agents, some works highlight the interaction between agents and the environment by memorizing the action-observation sequences. The critique information is also retained in case certain reflection mechanisms are introduced~\cite{DroidAgent}. This pattern has been adopted in SE tasks that necessitate iterative feedback from the environment, \eg{} mobile app testing~\cite{DroidAgent, GPTDroid, XUAT-Copilot, VisionDroid}, wherein operations on widgets in each turn should be memorized to facilitate the next-turn decision-making, and iterative code generation~\cite{AutoAgents, bairi2023codeplan, MetaGPT, shinn2023reflexion}, wherein the previous editing, execution, or debugging history serves as important information for code revision.
The third type is \textit{intermediate outputs}. Some agents merely store the outputs of previous turns in short-term memory to avoid overrunning the limited space, as well as being overly influenced by irrelevant or inaccurate chat history. For example, in E\&V~\cite{EV}, only intermediate analysis results are summarized to avoid inconsistency with the previously generated outputs. In SoA~\cite{SoA}, to implement a self-organized framework, each agent is equipped with memory that stores the self-generated code and unit tests. These intermediate results allow delayed test execution and code modification for agents in different layers, facilitating hierarchical collaborative code generation.

\highlight{Long-term memory}, on the other hand, is used to memorize valuable experiences of historical tasks.
\junwei{Similar to how humans draw on experience from previously completed tasks, long-term memory contributes to the planning module by providing records of historically relevant or similar tasks, which can serve as a reference for reasoning and solving the current unseen task.}
However, the entire task execution trajectory may involve extensive context, and given the limited memory space, it can be challenging to store it all completely. As a result, distilling techniques have been proposed, \eg{} trajectory summarization~\cite{DroidAgent, MetaGPT} and shortcut extraction~\cite{Co-Learning, IterativeExperienceRefinement}. These distilled records retain the task execution process in a more concise manner, thereby alleviating the burden on limited memory and prompt windows. Another approach to save long-term memory space is to selectively store vital data of each task, \eg{} the final results~\cite{AutoAgents, ChatDev, AgileCoder, CTC, MACNET},  reflections~\cite{shinn2023reflexion, AutoAgents}, and action-observations~\cite{DroidAgent, LATs, RCAgent}. Compared to complete historical trajectories, these data highlight the pivotal trace information, which can still retain the effects and feedback of previous tasks.

\headt{Memory Ownership.}
In agent systems, the memory module can be designed to serve specific agents or to serve all agents. Based on its ownership, we categorize memory into specific memory and shared memory.

\highlight{Specific memory} is an agent mechanism designed specifically for a limited group of agents. This type of memory has strict pre-defined usage regulations, only storing and serving specific agents in the workflow~\cite{XUAT-Copilot, DroidAgent, GPTDroid, AutoAgents, bairi2023codeplan, MetaGPT, ChatDev, MAGIS, shinn2023reflexion, AgileCoder, CTC, EV, SoA, Co-Learning, LATs, IterativeExperienceRefinement, MapCoder, RCAgent, MACNET, VisionDroid}. For example, in SoA~\cite{SoA}, each agent is equipped with individual memory for storing its own generated code fragments and unit tests, which will be used to evaluate the correctness of the final code and provide feedback to the agent for modification.

\highlight{Shared memory}, on the other hand, serves all agents by maintaining the record of their outputs and offering essential historical data.
In most cases, shared memory serves as a dynamic information exchange hub in the intricate SE environment, which is akin to the traditional blackboard system~\cite{craig1988blackboard}.
Generally, information stored in the shared memory is the intermediate results of previous phases, hence the agents from subsequent phases can obtain necessary information in a more convenient manner~\cite{AgentFL, MetaGPT, self-collaboration, MARE, AgileCoder}. Representative work like MetaGPT~\cite{MetaGPT} introduces a shared message pool, which saves artifacts from different agent roles, \eg{} the product requirement documents from the product manager. Another typical application of shared memory is to store comments in a decentralized debate scenario. Specifically, FlowGen\textsubscript{scrum} ~\cite{lcg} simulates the Sprint Meeting by providing a shared buffer, storing the problem and the discussion comment of all participating agents from which the product manager could extract a list of user stories.

\headt{Memory Format.}
In this section, we elaborate on the format of data stored in the memory. In SE tasks, the most commonly used storage formats include natural languages, program languages, structured messages, key-value pairs, embeddings, and trees. 
\begin{itemize}[leftmargin=*,label=-]
    \item \textit{Natural Languages.} LLM-based agents solve tasks specified in natural language, which is thus the most fundamental and prevalent data format in memory~\cite{shinn2023reflexion, XUAT-Copilot, GPTDroid, AutoAgents, MAGIS, lcg, self-collaboration, ChatDev, CTC, AgileCoder, MACNET, VisionDroid}. The advantage of raw natural language is that it allows for a more flexible storage of trajectories, thereby enhancing the universality. Moreover, raw natural language can better preserve the integrity of the original dialogue, which minimizes the loss and distortion of essential information.
    \item \textit{Programming Languages.} Some agents directly store the generated code for subsequent utilization~\cite{bairi2023codeplan, SoA}. For example, SoA~\cite{SoA} is a hierarchical code generation framework with each agent focusing on a single function implementation. It stores the generated function code and unit tests in memory for testing, modification, and aggregation.
    \item \textit{Structured Messages.} In this format, memory is organized as a list of messages with multiple attributes. The strength of this format is that it allows data to be stored in a structured manner, making it more convenient for indexing and processing. Moreover, it can store vital metadata of the message, \eg{} task type, message source and destination, so it is easier for agents to trace and subscribe to required messages, making it commonly used in \textit{shared memory}. Representative works like MetaGPT~\cite{MetaGPT} and MARE~\cite{MARE}, both wrap the artifacts of each agent as informative messages, involving the original content, instruction, task name, sender, receiver, \etc{} In E\&V~\cite{EV}, intermediate results of previous turns will be summarized and stored in JSON format.
    \item \textit{Key-value Pairs.} In this format, information received from the agents is stored in an external memory, with a key extracted for the agents to query required history memories~\cite{Co-Learning, IterativeExperienceRefinement, RCAgent}.  More specifically, in Co-Learning~\cite{Co-Learning}, shortcuts are extracted from the trajectories to construct two key-value databases: the solution-to-instruction database for the instructor, and the instruction-to-solution database for the assistant. RCAgent~\cite{RCAgent} stores the whole observation body in a key-value store, retaining a snapshot key for agents for query details.
    \item \textit{Embeddings.} In this format, the memory is embedded into a vector, which can help retrieve the most relevant task experiences. Representative work like \junweim{DroidAgent}~\cite{DroidAgent} embeds the textual history into vectors and stores them in an external embedding database. Compared to text similarity retrieval, it can further provide semantic similarity retrieval.
    \item \textit{Trees.} Some approaches construct a tree or graph for memorizing, especially in scenarios requiring flexible extension or path tracing. For example, in  LATS~\cite{LATs}, the task-solving process is modeled into a tree with each node representing a state with the instruction, the action, and the observation, and then an extended Monte Carlo Tree Search algorithm can be integrated. Similarly, Self-Repair~\cite{olausson2024selfrepair} proposes a repair tree that stores multiple generation-feedback-repair paths.
    \item \textit{Images.} Leveraging the capability of multimodal LLM, VisionDroid~\cite{VisionDroid} memorizes the testing history, including both textual descriptions and screenshots.
\end{itemize}

\headt{Memory Operations.} 
We categorize operations on memory into two main sections: memory writing and memory reading. 

\highlight{Memory Writing.}
The purpose of memory writing is to store essential information in the memory.
Information stored in memory is usually the raw task execution trajectories~\cite{ChatDev, AgileCoder,CTC, MACNET, VisionDroid}.
However, considering that the raw task trajectories might be lengthy, distilling approaches have been proposed to retain a more informative summary in the memory~\cite{XUAT-Copilot, DroidAgent, MetaGPT, MAGIS, shinn2023reflexion, Co-Learning, IterativeExperienceRefinement, EV}. For instance, in XUAT-Copilot~\cite{XUAT-Copilot}, dialog and action history are stored in working memory as summarized texts.
Moreover, Co-Learning~\cite{Co-Learning} proposes a novel distilling approach by first constructing a task execution graph and then extracting shortcuts linking non-adjacent solution nodes, which can serve as solution refinement paths for future tasks.
\junwei{These distilling strategies can generally be handled by adding an additional dialogue turn per round~\cite{DroidAgent, shinn2023reflexion, EV}, which does not introduce significant overhead compared to other multi-turn dialogue processes. On the contrary, previous research suggests that by leveraging the more concise information stored in memory, the agent can retrieve key trajectories using a smaller context window, ultimately helping to reduce token consumption~\cite{Co-Learning, MAGIS}.} 
On the other hand, the limited memory storage and prompt window size result in finite memory records. When overflow occurs, some records must be forgotten.
For example, in Reflexion~\cite{shinn2023reflexion}, the past experiences are stored in a sliding window with a maximum number of 3 to avoid exceeding the prompt window. Additionally, low-quality and rarely-used data also consume memory storage space. Previous research~\cite{Co-Learning} sets a threshold to filter out experiences with limited information. Further, an elimination mechanism based on the usage frequency is introduced to exclude rarely-used experiences\cite{IterativeExperienceRefinement}. 
\junwei{These memory elimination strategies are all implemented through simple logical scripts rather than agents, which do not introduce significant overhead.}

\highlight{Memory Reading.}
Memory reading aims at obtaining the required task history and experiences from the memory module. 
Besides directly providing the raw memory to the agent~\cite{GPTDroid, MetaGPT, MAGIS, ChatDev, AgileCoder, AgentFL, self-collaboration, CTC, MACNET, VisionDroid}, researchers prefer using three methods for obtaining relevant memory: reflection, retrieval, and subscription.
\textit{Reflection} refers to extracting pivotal experiences from the extensive trajectory memory~\cite{lcg, AutoAgents}. For example, in AutoAgents~\cite{AutoAgents}, a dynamic memory mechanism is designed to instruct an agent to extract insights from long-term memory that will serve the current action. In FlowGen~\textsubscript{scrum}~\cite{lcg}, the product manager summarizes the comments collected from all agents and extracts a list of user stories to implement.
In the \textit{retrieval} manner, the memory is retrieved based on its text or semantic similarity with the current tasks~\cite{DroidAgent, Co-Learning, IterativeExperienceRefinement}. For example, in Co-Learning~\cite{Co-Learning}, the reasoning module uses the prompt as a query to retrieve similar shortcuts from the constructed experience pool, which will serve as examples to facilitate future reasoning. In \junweim{DroidAgent}~\cite{DroidAgent}, the past tasks and widgets with similar GUI state embeddings are retrieved by comparing the cosine similarity.
Finally, the \textit{subscription} mechanism is chiefly used in shared memory. It permits agents to directly obtain required information according to their roles, without additional interaction costs with other agents, thus improving efficiency. Representative works include MetaGPT~\cite{MetaGPT} and \junweim{AgileCoder}~\cite{AgileCoder}, both adopting this kind of publish-subscribe mechanism.

Moreover, to determine whether a historical record should be integrated into the current task, some common filtering criteria have been adopted, including \textit{recency}~\cite{GPTDroid, bairi2023codeplan, MAGIS, DroidAgent, XUAT-Copilot, MetaGPT, ChatDev, AgileCoder,CTC},  \textit{relevance}~\cite{MetaGPT, AgileCoder,bairi2023codeplan}, and \textit{similarity}~\cite{Co-Learning, DroidAgent, IterativeExperienceRefinement}.
Agents integrate the most relevant and similar task experiences to provide the best reference for the current task. Additionally, the preference and weight of these factors vary across different works.
In \junweim{DroidAgent}~\cite{DroidAgent}, the planner agent considers the 20 most \textit{recent} task summaries and the 5 most \textit{similar} task knowledge items. In MAGIS~\cite{MAGIS}, the agent uses the \textit{most recent} summary of a code file to identify differences. In MetaGPT~\cite{MetaGPT} and \junweim{AgileCoder}~\cite{AgileCoder}, agents retrieve only \textit{relevant} messages from shared memory based on their roles.

\junwei{\headt{Challenges in Memory.}
The memory mechanism helps alleviate the issue of the limited context window in LLMs and is crucial for building a shared knowledge base as well as ensuring consistency in the decision-making process of different agents. However, the practical application of the memory mechanism presents several challenges: 
\begin{itemize}[leftmargin=*, label=-]
    \item \textit{Abstraction level of information.} It is non-trivial to determine the abstraction level of the information stored in memory. Storing full trajectories risks excessive context, while overly summarized data may lose crucial details.
    Moreover, different types of information may require different levels of abstraction. For example, global decisions can be more abstract, whereas implementation requires a more specific code context.
    \item \textit{Context matching.} This challenge lies in determining the required information in memory and the timing to conduct information retrieval. Excessive inclusion of irrelevant context or omission of necessary context can both significantly impact decision accuracy.
    \item \textit{Lack of fine-grained evaluation}. Similar to the planning module, the evaluation of the memory mechanism remains inadequate.
    Except for a few works that verify the effectiveness of its memory mechanisms via ablation experiments~\cite{AutoAgents, Co-Learning, IterativeExperienceRefinement}, most works merely evaluate the final task results. Inspired by the evaluation of general-purpose agents~\cite{memory_survey}, future work could focus on evaluating memory as an independent module with some numerical metrics, such as the accuracy in answering historical questions and memory-related costs.
\end{itemize}} 
\subsubsection{Perception}
\label{sec:agent:perception}
Existing LLM-based agents for SE primarily adopt two perception paradigms: textual input perception and visual input perception. 

\headt{Textual Input.}
Text can flexibly express the intent, information, and knowledge.
In SE, the majority of historical data, \eg{} documentation, code, and issues, is stored in the textual form. This alignment with the strengths of LLMs in processing natural language makes textual input the predominant form of perception for agents for SE.
Textual input in existing agents can be further categorized into natural language input (\ie{} domain-specific instructions and auxiliary information collected from the environment) and programming language input (\ie{} the code context).
For example, in NL2Code tasks~\cite{ChatDev,MetaGPT}, user requirements and function descriptions are provided as the instructions to agents.
But in some code-related tasks, \eg{} software testing~\cite{CoverUp, MuTAP} and debugging~\cite{FixAgent,RepairAgent,AgentFL,ACFIX}, the target code can also be provided for analysis. 
Specifically, in repository-level tasks such as issue-resolution~\cite{yang2024swe, AutoCodeRover, CodeR}, repository-level fault localization~\cite{AgentFL}, and code edits~\cite{bairi2023codeplan}, only a portion of code snippets are provided due to context length limitations, with further inspections achieved through navigation in the code repository.

\headt{Visual Input.}
Images represent a two-dimensional medium for storing information.
In traditional SE scenarios, there is also a portion of data presented in image form, \eg{} UML diagrams~\cite{kocc2021uml} and UI pages.  
In current SE agents, visual input is widely used in GUI testing tasks. Traditional GUI testing methods typically rely on text input, \ie{} view hierarchy files to extract widgets~\cite{su2017guided, InputBlaster, li2019humanoid}. However, this method is insufficient, as it may lose the structural semantics of the GUI when converting screenshots to text. In addition, the raw view hierarchy often contains an excessive number of tokens, and its redundancy can interfere with the agent's decision-making and reduce accuracy~\cite{VisionDroid}. On the other hand, using visual input (\ie{} screenshots) helps agents locate accessible widgets more precisely. 
Just as humans use their eyes to capture visual information, these agents integrate visual models to process and understand image data. For example, XUAT-Copilot~\cite{XUAT-Copilot} uses the SegLink++ model~\cite{SegLink++} for detecting bounding boxes and a ConvNeXts model~\cite{ConvNeXts} for text recognition. 
AXNav~\cite{AXNav} utilizes the Screen Recognition model~\cite{zhang2021screen} to analyze screenshot pixels from iOS devices and predict bounding boxes, labels, text content, and the clickability of UI elements.
In addition, VisionDroid~\cite{VisionDroid} employs multi-modal LLMs to process both the text and image information.
\junwei{By leveraging visual recognition information, these agents can gain a more intuitive understanding of the current GUI state, assess whether the goal has been achieved, and accordingly predict its next action. However, relying solely on screenshots is also limited due to the overlapping elements and the concise nature of GUI text and icons~\cite{VisionDroid}, which makes it difficult for agents to accurately and comprehensively extract information. Moreover, since the output of LLMs remains text-based, subsequent processes still rely on text. As a result, existing works all incorporate both vision input(\ie{} the screenshots) and text input (including view hierarchy files, detected elements, instructions, \etc{}) to achieve better effectiveness~\cite{XUAT-Copilot, AXNav, VisionDroid}.
}
\subsubsection{Action}~\label{sec:agent:action}
\begin{figure*}[htbp]
  \centering
  \begin{adjustbox}{width=1.0\textwidth}
    \begin{forest}
for tree={
    rounded corners,
    child anchor=west,
    parent anchor=east,
    grow'=east,
    text width=4cm,%
    draw=darkblue,
    anchor=west,
    node options={align=center},
    edge path={
      \noexpand\path[\forestoption{edge}]
        (.child anchor) -| +(-5pt,0) -- +(-5pt,0) |-
        (!u.parent anchor)\forestoption{edge label};
    },
    where n children=0{text width=6cm}{}
  }
[Action, fill=white, text width=3cm
    [Searching Tools
        [Web Searching
            [{MetaGPT~\cite{MetaGPT}, CodePori~\cite{CodePori}, ICAA~\cite{ICAA}, AgentVerse~\cite{AgentVerse}, \etc{}}]
        ]
        [Knowledge Base Searching
            [{CodeAgent~\cite{zhang2024codeagent}, ToolCoder~\cite{zhang2023toolcoder}, Co-Learning~\cite{Co-Learning}, E\&V~\cite{EV}, \etc{}}]
        ]        
    ]
    [File Operation
        [{SWE-agent~\cite{yang2024swe}, RepairAgent~\cite{RepairAgent}, MASAI~\cite{MASAI}, \etc{}}]
    ]
    [GUI Operation
        [{GPTDroid~\cite{GPTDroid}, XUAT-Copilot~\cite{XUAT-Copilot}, AXNav~\cite{AXNav}, AdbGPT~\cite{AdbGPT}}]
    ]
    [Static Program Analysis
        [Static Information Collection
            [Abstract Syntax Tree
                [{AutoCodeRover~\cite{AutoCodeRover}, \junweim{ToolGen}~\cite{TOOLGEN}, AdbGPT~\cite{AdbGPT}, TELPA~\cite{TELPA}, \etc{}}]
            ]
            [Control Flow Graph
                [{LDB~\cite{LDB}, LLM4CBI~\cite{LLM4CBI}}]
            ]
            [Call Graph
                [{TELPA~\cite{TELPA}, AutoSpec~\cite{AutoSpec}}]
            ]
            [Data Flow Graph
                [{IRIS~\cite{IRIS}, LLM4CBI~\cite{LLM4CBI}, LLM4DFA~\cite{LLM4DFA}}]
            ]
            [Code Dependency Graph
                [{AgileCoder~\cite{AgileCoder}, CodePlan~\cite{bairi2023codeplan}}]
            ]
            [Code Completion Tokens
                [{\junweim{ToolGen}~\cite{TOOLGEN}, CodePlan~\cite{bairi2023codeplan}, RRR~\cite{deshpande2024classlevelcodegenerationnatural}}]
            ]
        ]
        [Code Quality Checking
            [{RRR~\cite{deshpande2024classlevelcodegenerationnatural}, CTC~\cite{CTC}, CodeCoT~\cite{huang2024codecot}, \junweim{ACFix}~\cite{ACFIX}, \etc{}}]
        ]        
    ]
    [Dynamic Analysis
        [Method Call Trace
            [{AgentFL~\cite{AgentFL}}]
        ]
        [Runtime Values
            [{\junweim{AutoSD}~\cite{AUTOSD}, LDB~\cite{LDB}}]
        ]
        [Coverage
            [{CoverUp~\cite{CoverUp}, TELPA~\cite{TELPA}, LLM4CBI~\cite{LLM4CBI}}]
        ]
    ]
    [Testing Tools
        [Test Validation
            [{\junweim{AutoSD}~\cite{AUTOSD}, \junweim{FlowGen}~\cite{lcg}, AISD~\cite{aisd}, ClarifyGPT~\cite{ClarifyGPT}, \etc{}}]
        ]
        [Test Generation
            [{TELPA~\cite{TELPA}}]
        ]
        [Mutation Testing
            [MuTAP~\cite{MuTAP}]
        ]
    ]
    [Fault Localization Tools
        [{RepairAgent~\cite{RepairAgent}, AutoCodeRover~\cite{AutoCodeRover}}]
    ]   
    [Version Control Tools
        [{AutoDev~\cite{AutoDev}}]
    ]
  ]
\end{forest}
\end{adjustbox}
  \caption{Taxonomy of Action Components in LLM-based Agents for SE}
  \label{fig:action-tree}
\end{figure*}
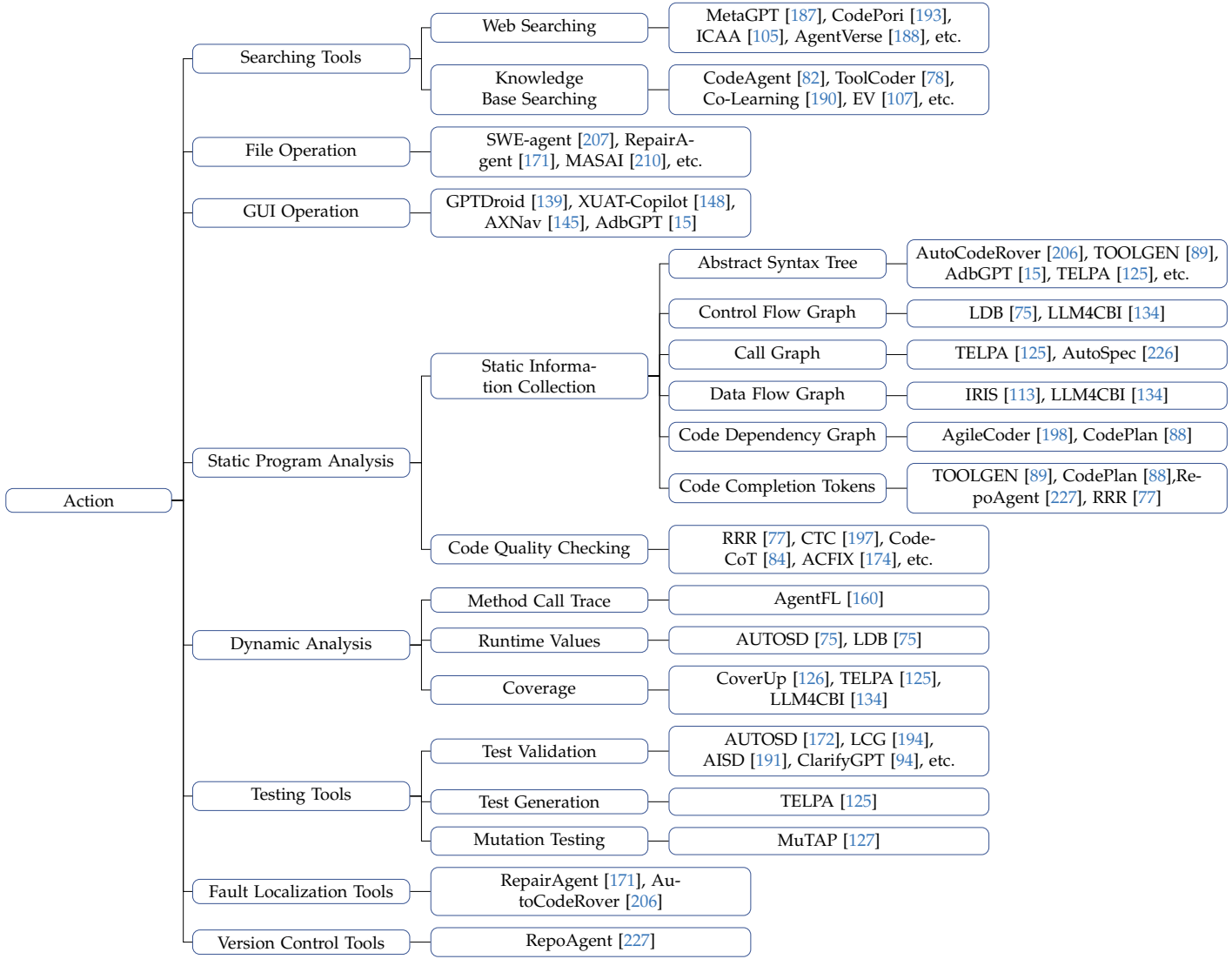

The action component of existing LLM-based agents for SE primarily involves using external tools to extend their capabilities beyond the interactive dialogue typical of standalone LLMs. Figure~\ref{fig:action-tree} summarizes the tools used in these agent systems.

\headt{Searching Tools.} 
\junwei{In SE, agents frequently use some lightweight RAG techniques to retrieve relevant information (\eg{} documentation or code snippets) that can aid in task completion. The retrieved information is either used to enrich the current context (\eg{} supplementing API documentation or background knowledge) or to provide relevant examples for in-context learning. Currently, SE agents primarily acquire information from two sources: the Web and local knowledge bases.
}

\highlight{Web Searching.}  
Online search engine tools use community and tutorial websites to offer programmers accurate and practical suggestions based on shared experiences and Q\&A. When faced with gaps in specific domain knowledge, programmers distill their needs into a query and use existing search engines (\eg{} Google, Bing, WikiSearch) to find the necessary information. Inspired by these practical experiences, some SE agents also retrieve ancillary information from the Web~\cite{MetaGPT, zhang2023toolcoder, fang2024llmagentsautonomouslyexploit, ICAA, xu2023gentopia, he2024cocost, AutoGen, CodeAct, RAT, zhang2024codeagent, ART}. For example, some agents~\cite{zhang2023toolcoder, zhang2024codeagent} use DuckDuckGo~\cite{DuckDuckGo} to search the relevant content, such as APIs. Paranjape \et~\cite{ART} employ SerpAPI~\cite{SerpApi} and extract answer box snippets when they are available or combine the Top-2 search result snippets. He \et~\cite{he2024cocost} query Google and then extract pertinent information to construct prompts for LLMs. 
Information retrieved via web search is generally used to supplement the context rather than serve as in-context learning examples.

\highlight{Local Knowledge Base Searching.}   
\junwei{
Besides using a web searching tool to externally collect the information from the Web, it is also common for existing agents to retrieve relevant knowledge from the self-established knowledge base, which includes documents~\cite{AutoGen, ICAA, LLM4Vuln, RCAgent, FromMisusetoMastery, PropertyGPT}, codebases~\cite{AutoGen, ICAA, RCAgent, EV, RepairAgent, RAT, bairi2023codeplan}, or historical experiences~\cite{DroidAgent, LLM4Vuln, ART}. 
There are various retrieval methods, including similarity-based retrieval, keyword-matching search, and model generation.  
\textit{Similarity-based retrieval} is the most traditional RAG approach, primarily consisting of \textit{sparse word-bag} and \textit{dense text embedding} methods. Both approaches vectorize code or documents and calculate the similarity between the query and segments in the knowledge base to retrieve relevant information. The sparse word-bag approach (\eg{} BM25~\cite{zhang2024codeagent, zhang2023toolcoder}) vectorizes text while partially preserving its semantics. Dense text embedding models, such as dual-encoder models, encode text into embedding vectors and compute their cosine similarity~\cite{LLM4Vuln, Co-Learning, DroidAgent, ICAA, RCAgent, deshpande2024classlevelcodegenerationnatural, AutoGen, RAT}.  On the other hand, \textit{keyword-matching retrieval} searches relevant context by matching key terms. For example, it may use keywords as keys to query a key-value database~\cite{DroidAgent, RCAgent}, directly search for elements in a code repository using keywords~\cite{RepairAgent}, or leverage source code analysis tools to match keywords~\cite{EV}.  
The final approach is to retrieve knowledge through \textit{model generation}~\cite{jiang2023selfevolve, MapCoder}. For example, MapCoder~\cite{MapCoder} treats the model itself as the data source. It instructs the retrieval agent to generate similar \textit{(problem, plan, code)} examples as in-context learning demonstrations for the planning agent.  
In summary, different works equip agents with various sources and types of knowledge bases and employ different RAG methods to retrieve relevant context. By leveraging the agent’s in-context learning capabilities, these approaches mitigate hallucinations and improve the accuracy of model-generated outputs.}

\headt{File Operation.}
As SE activities frequently access massive files, especially for the code repository and documentation, it is common for agent systems~\cite{yang2024swe, RepairAgent, AUTOSD, fang2024llmagentsautonomouslyexploit, IRIS, MASAI, AutoDev, SpecRover} to use file operations including shell commands (\eg{} Linux shell) or the code utilities (\eg{} Python \textit{os} package) for file browsing, file adding, file deleting, and file editing. For example, for file browsing, agents open files based on their paths, scroll through the contents, and jump to specific lines.

\headt{GUI Operation.} 
For SE activities related to software with a GUI, it is necessary to enable various GUI interaction operations for agent systems~\cite{GPTDroid, XUAT-Copilot, AXNav, AdbGPT, VisionDroid, InputBlaster}, including clicking, text input, scrolling, swiping, returning, and termination. 
In particular, for UI element identification, they use visual and text recognition models (\eg{} SegLink++~\cite{SegLink++}, Screen Recognition~\cite{zhang2021screen}, and ConvNeXts~\cite{ConvNeXts}), dump (\eg{} Android UIAutomator~\cite{uiautomator2}), or parse the UI view hierarchy~\cite{AdbGPT, GPTDroid}; then they simulate the testing environment using virtual Android devices (\eg{} Genymotion~\cite{Genymotion}, VirtualBox~\cite{virtualbox}, and pyvbox~\cite{pyvbox}) and autonomously execute or reply actions through tools such as Android Debug Bridge~\cite{adb}  to mimic user interactions. These actions enable agent systems to test in various GUI environments.

\headt{Static Program Analysis.}  
Static program analysis tools are widely used in agent systems for SE tasks, as they can provide more rigorous code features (\eg{} data-flow and control-flow) for LLMs, which can help agents better understand the program and tackle the relevant tasks. Existing agents primarily collect the following static information.

\begin{itemize}[leftmargin=*, label=-]
\item \textit{Abstract Syntax Tree (AST).} AST is a common representation to describe the syntactic structure of the source code and is widely used by agents. In particular, the collected ASTs help agents extract syntactic elements (\eg{} class names, method names, and variable names)~\cite{AutoCodeRover, TOOLGEN, EV, bairi2023codeplan, zhang2024codeagent, AdbGPT, deshpande2024classlevelcodegenerationnatural, AutoSpec, TELPA, MASAI, CoverUp, AgileCoder, bairi2023codeplan} and identify dependency among these code elements~\cite{RepoAgent, AutoSpec, TELPA, AgileCoder, bairi2023codeplan}.  Tree-sitter~\cite{tree-sitter} and ANTLR~\cite{ANTLR} are AST parsing tools that are widely used in existing agent systems~\cite{bairi2023codeplan, AgentFL, zhang2024codeagent, MASAI, AgileCoder,ACFIX, RepairAgent}.

\item \textit{Control Flow Graph (CFG).} LDB~\cite{LDB} uses CFG to divide a program into multiple blocks, making it easier to track intermediate variables with the help of the debugger. LLM4CBI~\cite{LLM4CBI} calculates the cyclomatic complexity based on a CFG that represents failed tests and accurately identifies high-complexity blocks of the code. These complicated code blocks would be regarded as the targets for program mutation.

\item \textit{Call Graph (CG).} TELPA~\cite{TELPA} constructs a method CG, extracting all call sequences that reach uncovered target methods. Based on these sequences, new test cases are generated to ensure comprehensive coverage of previously untested methods. AutoSpec~\cite{AutoSpec} treats loops as nodes as well, constructing an extended call graph, and it then traverses the CG from the bottom up to generate specifications.

\item \textit{Data Flow Graph (DFG).} IRIS~\cite{IRIS} constructs a data flow graph to assist taint analysis, which helps detect security vulnerabilities. LLM4CBI~\cite{LLM4CBI} performs data flow analysis to output a list of the most complex variables defined and used in the failed test, which guides test generation. LLM4DFA~\cite{LLM4DFA} leverages data flow analysis for dataflow-related bug detection.

\item \textit{Code Dependency Graph (CDG).} Agents such as AgileCoder~\cite{AgileCoder} and CodePlan~\cite{bairi2023codeplan} build a Code Dependency Graph for the entire codebase. The graph represents complex relationships between code blocks (\eg{} call relationships, inheritance, and import dependencies) and enables the accurate extraction of task-relevant context information (e.g., error traceback path for repair) within constraints of a limited prompt length. In addition, by dynamically maintaining the CDG, agents can perform incremental analysis in a more efficient way. 

\item \textit{Code Completion Tokens.} In code generation tasks, it is common for agents~\cite{TOOLGEN, bairi2023codeplan, RepoAgent, deshpande2024classlevelcodegenerationnatural,stall} to use language servers (\eg{} Jedi~\cite{Jedi} and EclipseJDTLS~\cite{EclipseJDTLS}) to collect candidate tokens at the certain position. In particular, candidate tokens returned by language servers often pass the syntactic violation (\eg{} only defined variable names are returned), which can effectively alleviate the hallucinations of standalone LLMs.

\end{itemize}

Besides, static analysis tools are also widely used by agent systems to check code quality, \eg{} syntactic correctness checking, code format checking, code complexity checking, vulnerability detection, and specifications checking. The checked results can then provide feedback or additional hints for agents to further improve code quality. In particular, existing agents~\cite{deshpande2024classlevelcodegenerationnatural, CTC, huang2024codecot, AutoDev, 3DGen} use compilers or interpreters (\eg{} GCC or Python) for syntactic correctness checking; existing agents~\cite{zhang2024codeagent, LLM4PLC} use Black~\cite{Black} and nuXmv~\cite{nuXmv} for code format checking; the agent in~\cite{LLM4CBI} uses OClint~\cite{OCLint} and srcSlice~\cite{srcSlice} for code complexity checking; existing agents~\cite{LLM4CBI, ACFIX} use static analysis tools such as Frama-C~\cite{Frama-C} and Slither~\cite{Slither} to detect vulnerabilities; the agent in~\cite{AutoSpec} uses static tools (\eg{} Frama-C) to verify the satisfiability and sufficiency of generated specifications.

\headt{Dynamic Analysis.}
In addition to static analysis, existing agents also use dynamic analysis tools to monitor program execution and collect runtime behaviors for agents. For example, AgentFL~\cite{AgentFL} uses the \textit{java.lang.instrument} package~\cite{javalanginstrument} to record all method call traces during the execution of failed tests, which can facilitate more accurate fault localization. Some agents~\cite{AUTOSD, LDB}  mimic manual debugging to set breakpoints, so as to capture runtime values for variables. The runtime values can be integrated into the prompt along with requirements to aid in defect localization. Besides, coverage also serves as important feedback for whether each code element is executed by tests or not. For example, prior agents~\cite{CoverUp, TELPA, LLM4CBI} leverage tools such as SlipCover~\cite{SlipCover}, Pynguin~\cite{Pynguin}, and Gcov~\cite{Gcov} to collect the coverage information. 

\headt{Testing Tools.} 
Test cases validate whether the software behaviors violate the specifications, and it is common for agents in SE to invoke testing tools for software validation, test generation, and mutation testing.  

\begin{itemize} [leftmargin=*, label=-]
    \item \textit{Software Validation.} Validating the software with test execution frameworks (\eg{} PyTest, unittest, or JUnit) can reveal the runtime errors and test failures, which are widely used in existing agent systems~\cite{AUTOSD, RepairAgent, lcg, zhang2024codeagent, aisd, AgentCoder, piya2023llm4tdd, INTERVENOR, ClarifyGPT, mint, tian2024testcasedriven, AutoGen, MetaGPT, Flows, jiang2023selfevolve, yang2023intercode, olausson2024selfrepair, chen2023teachinglargelanguagemodels, liventsev2023fully, shinn2023reflexion, ART, ConversationalAPR, fang2024llmagentsautonomouslyexploit, TGen, Chattester, TestPilot, WhiteFox, he2024cocost, SoA, MapCoder, TELPA, FlakyDoctor, AutoCoder2405, MASAI, AgileCoder, AlphaCodium, AutoCodeRover, FixAgent, CodeAct, AgentVerse, huang2024codecot, xu2023lemur, AutoDev, SpecRover}. The revealed execution violations can serve as feedback for agents to improve programs; otherwise, the absence of execution violations can signal the correctness of programs (e.g., proving that a plausible patch has been found for program repair agents). 
    \item \textit{Test Generation.} Although an LLM itself has promising capabilities of directly generating test code, traditional test generation tools provide complementary benefits as they are good at generating high-coverage tests in a cost-efficient way. For example, some agents~\cite{TELPA} use automated test case generation tools (\eg{} Pynguin~\cite{Pynguin}) to generate an initial set of unit test cases. Besides, for some domain-specific languages (DSLs), using existing tools to generate tests is simpler. For example, 3DGen~\cite{3DGen} converts informal specifications into executable code (\ie{} a binary format parser written in formally verified C code) through a DSL called 3D~\cite{3D}. To test the generated 3D programs, 3DGen invokes an external tool called SMT-solver Z3~\cite{de2008z3} to produce test cases. 
    \item \textit{Mutation Testing.} Some agents~\cite{MuTAP} use mutation testing tools (\eg{} MutPy~\cite{Mutpy}) to evaluate the sufficiency of test cases, as killing mutants (\ie{} exhibiting different behaviors on the mutated program than the original program) indicates the fault detection capabilities of tests. The mutation testing results can further serve as feedback for agents to enhance the tests iteratively.  
\end{itemize}

\headt{Fault Localization Tools.}
Agent systems~\cite{RepairAgent, AutoCodeRover} can invoke traditional fault localization techniques, especially spectrum-based fault localization tools (\eg{} GZoltar~\cite{Gzoltar}) to localize suspicious code elements. For example, RepairAgent~\cite{RepairAgent} invokes GZoltar to get the suspiciousness score of each code element (\ie{} the probability of being fault).

\headt{Version Control Tools.}
Version control systems manage the changes of various files in a repository, such as changes in code, configuration files, or documentation throughout the software development process. Some agents~\cite{RepoAgent, AutoDev, liu2025evodev} that manage an entire repository often leverage version control tools.

\junwei{\headt{Overhead of Tools.} The overhead of tool invocation is closely related to the tool type and the task complexity. 
For instance, method-level code generation tasks typically take only a few seconds~\cite{zhang2023toolcoder}, and thus tend to use code execution tools with lower overhead. In contrast, system-level tasks, such as mobile app testing, fault localization, and end-to-end software development, can incur time costs ranging from minutes to hours~\cite{AUTOFL, MetaGPT, ChatDev, AgileCoder, LLM4Vuln, AdbGPT, GPTDroid}, and can incorporate more time-consuming tools such as those for static analysis or fault localization.
Besides, current studies generally report the overall time overhead of agents without breaking down the overhead of tool invocations~\cite{TOOLGEN, MetaGPT, LLM4PLC, AgileCoder, AgentFL, AUTOSD, RepairAgent, ACFIX, LLM4Vuln, Chattester}, which may be attributed to two reasons: first, tool invocations are often integrated into the reasoning process and they are generally treated as a combination; second, the time overhead of commonly used tools, such as code execution and web search tools, is relatively small and can be neglected~\cite{LDB}.
For tools that may incur significant time overhead, common practices include limiting their processing time~\cite{zhang2023toolcoder, RepoUnderstander, GPTDroid} or the number of invocations~\cite{RepoUnderstander, yang2024swe, AXNav}. For example, ToolCoder~\cite{zhang2023toolcoder} limits the search delay to 0.6 seconds; SWE-agent~\cite{yang2024swe} limits the number of search results to 50, and Lingma Agent~\cite{RepoUnderstander} restricts the number of search iterations to 600 and the maximum search time to 300 seconds in the MCTS-Enhanced Repository Understanding stage.}

\junweim{
\subsubsection{Foundation LLMs.}~\label{sec:agent:basellm}
In this section, we discuss the relationship between the LLM-based agents and their foundation LLMs.
}

\junweim{
A basic observation is that due to the rapid evolution of foundation LLMs, current agent systems are not tailored to any specific LLMs. Instead, they are built upon a set of abstract capabilities that an LLM is expected to possess, thereby improving generality and adaptability. In practice, LLM-based agent systems often require the following capabilities from their foundation LLMs:}

\junweim{
\textbf{Basic Instruction-following, Planning, and Multi-turn Dialogue Capabilities.}
For LLM-based agents that accomplish tasks purely through conversational collaboration~\cite{Elicitron, MARE, ChatDev, chatunitester}, the foundation LLMs require to possess fundamental instruction-following, planning, and multi-turn dialogue (\ie{} memory) capabilities. 
Since these capabilities are core features of most LLMs, these LLM-based agents can generally operate with a variety of foundation LLMs. However, GPT-series models remain the predominant choice due to their broad accessibility and consistently strong performance.
It is also worth noting that planning can be enhanced through the CoT strategy, which, to some extent, alleviates the demand for the inherent planning ability of LLMs. In addition, some agents explicitly enhance CoT and heavily depend on CoT to boost their performance~\cite{huang2024codecot, RAT}.
}

\junweim{
\textbf{Tool-use Capability.}
In scenarios requiring autonomous tool invocation, the foundation LLM must understand tool specifications, suggest appropriate calls, and incorporate the resulting outputs into its reasoning process~\cite{AutoGen, yang2024swe, zhang2024codeagent, MARE}. For instance, CodeAgent~\cite{zhang2024codeagent} introduces five programming tools and relies on the foundation LLM to invoke them effectively for interacting with software artifacts.
}

\junweim{
\textbf{Open-source Accessibility.}
Some LLM-based agents require fine-tuning the foundation LLMs to enhance performance, adapt to custom tasks, or intervene in their decoding process~\cite{xu2023lemur, AutoCoder2405, zhang2023toolcoder, TOOLGEN, self-edit}, which restricts implementation to open-source LLMs. For instance, ToolCoder~\cite{zhang2023toolcoder} trains the foundation LLM to generate special tokens for API retrieval, while ToolGen~\cite{TOOLGEN} masks non-existent API tokens during model decoding. Such requirements inherently limit these systems to open-source LLMs.}

\junweim{
\textbf{Long Context Window.}
LLM-based agents solve problems through multi-turn interactions with the environment, which often accumulates substantial contextual information. Although memory management techniques can alleviate the dependence on extremely long context windows~\cite{shinn2023reflexion, RCAgent, Co-Learning}, the foundation LLMs' context window size and ability to comprehend complex contextual information remain critical for solving complex tasks. For example, in CodeS~\cite{CodeS}, even the largest available context window (\ie{} 200K tokens) might be insufficient for repository-level tasks due to the extensive size of code bases, documentation, and discussions. Other works on LLM-based agents have also reported errors caused by insufficient context length of base LLMs~\cite{yang2023intercode, AutoGen,AgileCoder}.
}

\junweim{
\textbf{Stronger Multi-step Reasoning Capability.}
For tasks that involve deep exploration over multiple turns (\eg{} end-to-end software development and maintenance), the foundation LLM must excel at maintaining coherent long-term context and planning subsequent steps based on historical trajectories. A typical example is that all end-to-end software maintenance agents are built on state-of-the-art closed-source models~\cite{yang2024swe, SpecRover, Agentless}, such as GPT-4, GPT-4o, and Claude-3.5-Sonnet, highlighting the strong dependence of such tasks on high-performance foundation LLMs.
}

\junweim{
In principle, the foundation LLMs can be replaced as long as they meet the agent’s fundamental requirements aforementioned. 
However, although LLM-based agents enhance the capabilities of foundation LLMs, the increasing complexity and variability of the tasks handled by such agents also place new demands and challenges on the foundation LLMs. 
In some cases, performance gaps between foundation LLMs cannot be fully bridged through the architecture and design of LLM-based agents~\cite{self-refine, he2024cocost, fang2024llmagentsautonomouslyexploit}. For example, Self-Refine~\cite{self-refine} performs well with GPT-series models but struggles with Vicuna-13B, which often fails to produce feedback in the required format, repeats previous outputs, or generates hallucinated conversations. Therefore, in practice, it is recommended to employ foundation LLMs with capabilities comparable to those employed in the original experiments.
}

\begin{figure*}[htbp]
  \centering
  \begin{adjustbox}{width=0.9\textwidth}
    \begin{forest}
for tree={
    rounded corners,
    child anchor=west,
    parent anchor=east,
    grow'=east,
    text width=4cm,%
    draw=darkblue,
    anchor=west,
    node options={align=center},
    edge path={
      \noexpand\path[\forestoption{edge}]
        (.child anchor) -| +(-5pt,0) -- +(-5pt,0) |-
        (!u.parent anchor)\forestoption{edge label};
    },
    where n children=0{text width=6cm}{}
  },
[Agent Roles, fill=white, text width=3cm
    [Manager Roles
        [Task Decomposition
            [{Manager Bot~\cite{CodePori}, Planner~\cite{AXNav, DroidAgent, 3DGen}, Manager~\cite{MAGIS, CodeR}, Scrum Master~\cite{AgileCoder, lcg}, Controller~\cite{RCAgent}, Planning Agent~\cite{MapCoder, rasheed}, Instructive Agent~\cite{IterativeExperienceRefinement}, Product Manager~\cite{MetaGPT}, Planning LLM~\cite{low-code-llm}, Summary Agent~\cite{RepoUnderstander}}]
        ]
        [Decision Making
            [{CEO \& CTO~\cite{tang2024codeagent}, Instructor~\cite{Co-Learning}, AI User~\cite{li2023camel}}]
        ]
        [Team Organization
            [{Mother Agent~\cite{SoA}, Planner \& Agent Observer~\cite{AutoAgents}}]
        ]
    ]
    [Requirement Analyzing Roles
        [{Product Manager~\cite{multi-agent-collaboration, MetaGPT, AgileCoder, aisd}, Analyst~\cite{self-collaboration}, Task Interpretation and Planing LLM~\cite{ICAA}, Requirements Engineer~\cite{lcg, sami, rasheed}, Task Specifier Agent~\cite{li2023camel}, User~\cite{Elicitron}, Stakeholder \& Collector \& Modeler \& Checker \& Documenter~\cite{MARE}}]
    ] 
    [Designer Roles
        [Software Architecture Designer
            [{Architect~\cite{lcg, MetaGPT, sami, aisd, multi-agent-collaboration}, Software Designer~\cite{rasheed}, CEO \& CTO~\cite{ChatDev}}]
        ]
        [UI/UX Designer
            [{User Experience Designer~\cite{multi-agent-collaboration}, User Interface Designer~\cite{multi-agent-collaboration}, UI/UX Designer~\cite{AgentVerse}}]
        ]
    ] 
    [Developer Roles
        [{Dev-bot~\cite{CodePori}, Writer~\cite{AutoGen}, Engineer~\cite{MetaGPT}, Programmer~\cite{ChatDev, AgentVerse, AgentCoder, AutoCoder2405, aisd, yang2024swe}, SketchFiller~\cite{CodeS},  Child Agent~\cite{SoA}, Coding Agent~\cite{MapCoder}, AI assistant~\cite{li2023camel}, Developer~\cite{lcg, MAGIS, AgileCoder,3DGen, rasheed, multi-agent-collaboration, CoverUp}, Code Learner~\cite{INTERVENOR}, Code Model~\cite{olausson2024selfrepair}, Coder~\cite{TGen, self-collaboration}, Code Generator~\cite{self-edit, sami}, Programming Assistant~\cite{LDB, MuTAP}}]
    ] 
    [Software Quality Assurance Roles
        [Code Reviewer
            [{Reviewer~\cite{tang2024codeagent, ChatDev, CTC}, QA Engineer~\cite{MAGIS}, Verification Bot~\cite{CodePori}, Senior Developer~\cite{AgileCoder}, Critic Agent~\cite{li2023camel}, Code Review Agent \& Bug Report Agent \& Code Smell Agent \& Code Optimization Agent~\cite{AIPoweredCodeReview}, Test Code Reviewer \& Source Code Reviewer~\cite{AgentFL}, Safeguard~\cite{AutoGen}, Auditor \& Critic~\cite{GPTLENS}, Consistency Checking Agent~\cite{ICAA}, Validator~\cite{ACFIX}, Detector \& Reasoner~\cite{iAudit}}]
        ]
        [Tester
            [{Crafter~\cite{FixAgent}, Tester~\cite{lcg, AgileCoder,sami, multi-agent-collaboration, rasheed, Chattester,FlakyDoctor}, Test Designer~\cite{AgentCoder}, QA Engineer~\cite{MetaGPT}, Generation LLM~\cite{WhiteFox}, Action Agent \& Evaluation Agent~\cite{AXNav}, Actor \& Observer~\cite{DroidAgent}, Operation Agent~\cite{XUAT-Copilot}, Domain Expert~\cite{3DGen}, Function-aware Explorer~\cite{VisionDroid}}]
        ]
        [Debugging Roles
            [Test Result Analysis
                [{Remediation Agent~\cite{TGen}, Feedback Model~\cite{olausson2024selfrepair}, Questioner~\cite{AutoCoder2405}, Reflector~\cite{DroidAgent}, Revisitor~\cite{FixAgent}, Verifier~\cite{CodeR}, Logic-aware Bug Detector~\cite{VisionDroid}}]
            ]
            [Test Failure Reproduction
                [{Test Template Generator \& Issue Reproducer~\cite{MASAI}, Reproducer~\cite{CodeR, SpecRover}}]
            ]
            [Repair
                [{Repairer~\cite{FixAgent}, Generator~\cite{ACFIX}, Fixer~\cite{MASAI}, Debugging Agent~\cite{MapCoder, AUTOFL}, Repair Suggestion Model \& Patch Generation Model~\cite{SRepair}, Proposer~\cite{CORE}, Localizer~\cite{FixAgent, CodeR}, Editor~\cite{CodeR, self-edit}, Patching Agent~\cite{SpecRover}}]
            ]
        ]
    ] 
    [Deployment Roles
    [Deployment Plan Agent~\cite{rasheed}]
    ]
    [Assistant Roles
        [{Assistant~\cite{CodeAct, EV, ChatRepair},Repository Custodian~\cite{MAGIS}, RepoSketcher~\cite{CodeS}, Edit Localizer~\cite{MASAI}, Retrieval Agent~\cite{MapCoder}, Report Agent~\cite{ICAA}, Reward Agent~\cite{RepoUnderstander}, Ranker~\cite{MASAI, SpecRover, CORE, iAudit}, Code Review Committee~\cite{DEIBASE}, JML Specification Generator~\cite{ma2024specgen}}]
    ] 
  ]
\end{forest}
\end{adjustbox}
  \caption{Taxonomy of Agent Roles in LLM-based Multi-agent Systems for SE}
  \label{fig:role-tree}
\end{figure*}
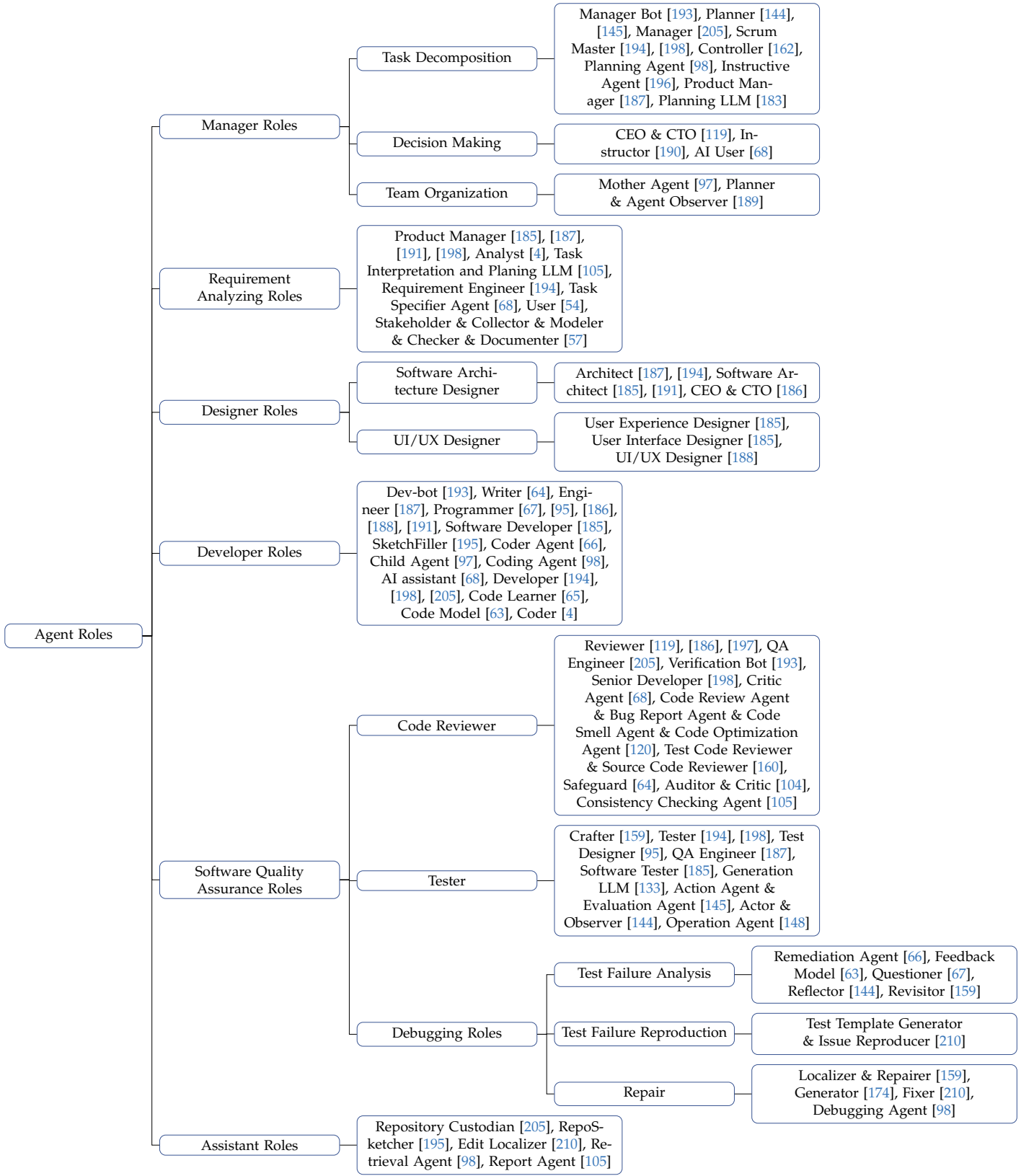

\subsection{Multi-agent System}\label{sec:agent:multiagent}
Based on our statistics, 59.7\% of existing agents for SE are multi-agent systems. 
These systems benefit from the division of specialized roles and coordination among agents, which effectively addresses the complexity of SE tasks, particularly for end-to-end activities spanning multiple phases. This section provides an overview of existing multi-agent systems for SE, with a focus on their agent roles and coordination mechanisms.

\subsubsection{Agent Roles}~\label{sec:agent:role}
In agent systems, role-playing strategy is commonly used to embed expert personas into the prompts of agents and elicit the relevant professional knowledge. 
Specifically, role assignment mainly delineates duties, available actions,  attributes, and constraints of roles. It enables agents to specialize in their corresponding tasks.
\junwei{
Theoretically, both single-agent and multi-agent systems can employ role-playing strategies. 
However, according to statistics, among the total of 50 single-agent works we have collected, only 12 works have utilized a role-playing prompt~\cite{CodeAct, EV, ChatRepair, LDB, AUTOFL, MuTAP, CoverUp, FlakyDoctor, IRIS, yang2024swe, ma2024specgen, Chattester}, which are expressed in a simple task-centered manner (\ie{} “You are a/an [TASK] expert/assistant/professional”). 
Besides, the roles used by these 12 single-agent works are covered by the roles in multi-agent works. Therefore, the roles discussed in this section can encompass all role types of current LLM-based agents in the SE field, whether single-agent or multi-agent.
} 
Figure~\ref{fig:role-tree} summarizes common agent roles in existing LLM-based agent systems for SE.

\headt{Manager Roles.}
Manager roles (such as CEO, commander, and controller), serve as the leaders of multi-agent teams. 
These roles are responsible for making decisions, planning, task decomposition and assignment, and overseeing team coordination.

\highlight{Task Decomposition.} 
To enhance the overall system performance, managers 
break down a project into manageable sub-tasks and draw up a guiding plan for developers, testers, or repairers to execute~\cite{CodePori, AXNav, MAGIS, AgileCoder, RCAgent, MapCoder, DroidAgent, IterativeExperienceRefinement, CodeR, rasheed}. 
They analyze problem statements, facilitate discussions on issues among various agent roles~\cite{lcg}, review design documents submitted by designers~\cite{MetaGPT}, summarize the global repository information~\cite{RepoUnderstander}, and incorporate related information gathered by assistants. Subsequently, they produce a specific task list or implementation blueprint, which may be presented in either natural language or as a structured workflow~\cite{low-code-llm}.

\highlight{Decision Making.}
Another task of managers is to orchestrate team collaboration and provide further guidance for task execution.
For example, the CEO and CTO in CodeAgent~\cite{tang2024codeagent} communicate with staff and make high-level decisions. Similarly, the instructor in Co-Learning~\cite{Co-Learning} and the AI user within the CAMEL system~\cite{li2023camel} provide instructions to working agents. 

\highlight{Team Organization.}
Finally, in scenarios involving dynamically derived agents, managers play a role in assembling the team.
This role is primarily designed for flexibly deciding the constitution of the agent team, \ie{} what roles are included in the team. The main benefits of including such roles are to flexibly optimize costs and better meet project demands. 
For instance, SoA~\cite{SoA} sets the mother agent, which generates new mother or child agents and designates concrete tasks (\eg{} unimplemented functions) to them. AutoAgents~\cite{AutoAgents} includes a planner agent and an observer agent, which collaborate to assemble a team for particular tasks. The planner agent is responsible for assigning existing LLM agent roles or generating new ones, while the observer agent assesses and reviews the relevant roles. These roles are represented in a structured JSON format, encapsulating details such as name, description, available tools, suggestions, and prompts to guide agent behaviors.

\headt{Requirement Analyzing Roles.}
These roles are primarily responsible for analyzing software requirements, such as translating vague and preliminary user concepts into a coherent and structured format. 
Existing agents~\cite{multi-agent-collaboration, self-collaboration, ICAA, lcg, MetaGPT, AgileCoder, rasheed} include such roles (\eg{} product manager~\cite{aisd} or task specifier~\cite{li2023camel}) to identify key requirement elements and intended objectives for a precise and organized requirement document, which may range from an elaborated task or function description~\cite{li2023camel} to a formal software requirement specification~\cite{MetaGPT}.

In addition, some agents further design more fine-grained roles for requirements analysis. For example, Elicitron~\cite{Elicitron} incorporates a set of \textit{User} agents to identify diverse user requirements by mimicking user perspectives and conducting interviews for exploring potential user needs; MARE~\cite{MARE} uses a requirements engineering team (\ie{} stakeholder, collector, modeler, checker, and documenter) to produce requirements specifications. The requirements engineering process is segmented into four sub-tasks corresponding to specific roles, seamlessly transitioning rough user ideas to precise requirement specifications. Sami \et{}~\cite{sami} employ a two-stage requirements generation approach by first instructing an agent to generate user requirements and then using another agent to prioritize user stories.

\headt{Designer Roles.}
Designer roles take input information on requirements (such as detailed task descriptions and use cases) and shape the software architecture and system integration. 
The most typical design roles are the \highlight{system architects}, who are responsible for conceptualizing and defining the high-level structure of software, \eg{} the \textit{software architect} role in agents~\cite{lcg, aisd, multi-agent-collaboration, ChatDev, MetaGPT, sami, rasheed}. They create a design document that serves as a blueprint for the subsequent stages of development, and the design document can be presented in various forms, including natural language descriptions, structured formats (\eg{} JSON for listing project architecture files), and graphical representations (\eg{} class diagrams and sequence flowcharts~\cite{MetaGPT}).
Moreover, some approaches will assign UI/UX designers to craft the visual and interactive interface~\cite{multi-agent-collaboration, AgentVerse}.

\headt{Developer Roles.}
Developers take a vital role in software development and maintenance activities, which is one of the most common roles (\eg{} also called programmer or coder) in existing agents~\cite{CodePori, AutoGen, MetaGPT, ChatDev, multi-agent-collaboration, CodeS, TGen, SoA, MapCoder, li2023camel, AgentVerse, sami} for tasks involving code generation. In accordance with software design schemes, task plans provided by other agents, or user requirements, the developer roles generate or finalize code at various levels (\ie{} from function to file and even project levels). In addition, the developer roles also engage in the code refinement process, which refines their previously generated code~\cite{lcg, AgentCoder, INTERVENOR, olausson2024selfrepair,self-collaboration, MAGIS, AutoCoder2405}. Furthermore, the developer roles can be instructed to meet more customized standards, such as elucidating their work through supplementary docstrings or adhering to particular coding criteria~\cite{aisd, AgileCoder}.

\headt{Software Quality Assurance Roles.}
Agent systems include roles dedicated to software quality assurance, similar to real-world QA teams. These roles typically encompass code reviewers, testers, and debuggers, each focused on checking and improving software quality.

\textit{Code reviewers} are responsible for identifying potential software quality issues by statically inspecting the software without execution. For example, some agents~\cite{tang2024codeagent, ChatDev, CTC, MAGIS, CodePori, AgentVerse, ACFIX} include such roles to review generated code or patches; AgileCoder~\cite{AgileCoder} and CAMEL~\cite{li2023camel} include the roles such as senior developer or critic agent to offer suggestions for enhancement; the agent in~\cite{AIPoweredCodeReview} sets up code review agent, bug report agent, code smell agent, and code optimization agent to access code quality from different aspects; AGENTFL~\cite{AgentFL} sets test code reviewer and source code reviewer to summarize code behaviour to help fault location; in addition, some agents~\cite{AutoGen,GPTLENS,ICAA} include such roles (\eg{} the auditor agent and the critic agent in GPTLens~\cite{GPTLENS} and the consistency checking agent in ~\cite{ICAA}) to detect the vulnerability or implementation issues. 

\textit{Software testers} are widely incorporated in multi-agent systems to write corresponding test cases or scripts for software verification~\cite{FixAgent, lcg, AgentCoder, MetaGPT, multi-agent-collaboration, WhiteFox, AgileCoder, AXNav, DroidAgent, XUAT-Copilot, SpecRover, sami}. For example, the tester agent in multi-agent systems~\cite{FixAgent, AgentCoder, lcg} generates test cases based on relevant code skeletons or patches, requirement documents, existing tests, or rationale for the test.
There are two special testing scenarios: one is \textit{GUI testing}, in which the tester agent generates operational action sequences~\cite{DroidAgent, XUAT-Copilot, VisionDroid}, and the other is \textit{reproduction tests} in end-to-end maintenance tasks, where issue reproduction serves as a crucial phase for localizing the issue and checking the correctness of the generated patch~\cite{CodeR, MASAI, SpecRover}.

\textit{Debuggers} help diagnose test failures or unexpected software behaviors. A typical application scenario of debugging roles is to analyze test reports and determine the correctness of the program.~\cite{TGen, olausson2024selfrepair, AutoCoder2405, DroidAgent, VisionDroid, FixAgent, INTERVENOR, SpecRover, CodeR}. 
For example, the remediation agent in TGen~\cite{TGen} and the feedback module in Self-Repair~\cite{olausson2024selfrepair} are similarly designed to analyze the test failure reports and relevant faulty code to provide explanations and suggestions. 
Furthermore, debugging roles can directly generate patches based on the detected bugs in some approaches~\cite{ACFIX, MASAI, MapCoder, SRepair, SpecRover, CORE}.

\headt{Deployment Roles.}
Deployment roles are responsible for creating deployment plans to release the software to the production environment or end-users. Rasheed~\et{}~\cite{rasheed} assign a Deployment Plan Agent to conduct this task.

\headt{Assistant Roles.}
Assistant roles primarily provide assistance for other agents. For example, the repository custodian in MAGIS~\cite{MAGIS}, the RepoSketcher in CodeS~\cite{CodeS}, the edit localizer in MASAI~\cite{MASAI}, and the Context Retrieval Agent in SpecRover~\cite{SpecRover} is designed to enhance the comprehension of the target repository architecture for the team; in addition, MapCoder~\cite{MapCoder} uses the retrieval agent to facilitate memory recall; ICAA~\cite{ICAA} introduces the report agent to convert natural language responses into formatted bug reports. The Reward agent~\cite{RepoUnderstander} measures the possibility that particular code segments contribute to a given issue.
Ranker is also a common role category, which is designed to choose the optimal results~\cite{MASAI, SpecRover, CORE}.

\subsubsection{Collaboration Mechanism}\label{sec:agent:cooperate}
\begin{figure}[htb]
    \centering
    \includegraphics[width=1.0\columnwidth]{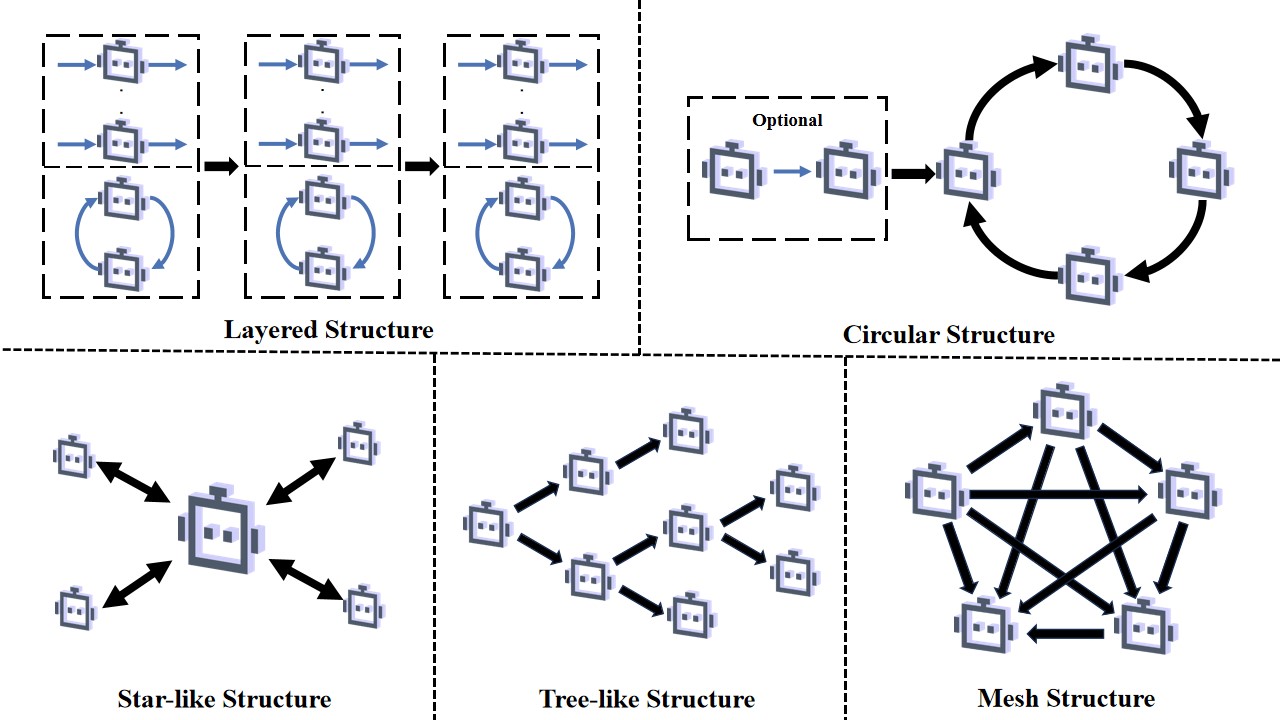}
    \caption{Multi-agent System Collaboration Mechanisms} 
    \label{fig:multi-agent-collaboration}
\end{figure}

The collaboration mechanism is essential for multi-agent systems, which can significantly impact the effectiveness and costs of the entire system. In particular, the collaborative mechanisms of existing multi-agent systems for SE tasks can be categorized into four types: layered structure, circular structure, star-like structure, tree-like structure, and mesh structure. Figure~\ref{fig:multi-agent-collaboration} illustrates each structure.

\headt{Layered Structure.} 
It is a hierarchical structure, where tasks are decomposed into several sub-stages and each is assigned to a specific agent or a group of agents selected from the agent pool. Agents between different stages collaborate sequentially, \ie{} they receive intermediate results from agents in the previous stage as input and produce their processed data to agents in the next stage. For example, the workflow within many agents~\cite{AutoCodeRover, AgentFL, ICAA, low-code-llm, WhiteFox, aisd, self-edit, LLM4DFA, SRepair, SpecRover, DEIBASE, sami, rasheed, VisionDroid} is a simple chain, where each agent focuses on its own sub-task and only interacts with adjacent agents. In addition, agents can also refer to the message produced by the previous non-adjacent agents~\cite{MetaGPT, CodeS, MapCoder}. In the sequential workflow, each sub-task can also be handled by a group of agents~\cite{Flows, MARE}. In~\cite{tang2024codeagent, ChatDev, CTC, MACNET}, each sub-task is solved by the conversation between two agent roles. FlowGen~\cite{lcg} and AgileCoder~\cite{AgileCoder} incorporate even more agents in a single stage. In addition to interactive collaboration, another scenario involves agents within the same layer working in parallel to offer their solutions. These solutions are then combined and passed down to the next layer. For example, GPTLens~\cite{GPTLENS} employs several auditors to present possible vulnerable functions individually in the generation stage. AgentForest~\cite{li2024agents} incorporates the majority voting mechanism. DyLAN~\cite{liu2023dynamic} formulates the LLM-agent collaboration structure into a multi-layered feed-forward network.

\headt{Circular Structure.}
This structure typically manifests as multi-turn dialogues between two roles or integrates the feedback mechanism within the overall collaborative processes among multiple agents. The \textit{dual-role} setup is typically implemented as a generation-validation style loop between two agents~\cite{ACFIX, AgentCoder, TGen, AutoCoder2405, olausson2024selfrepair, INTERVENOR, Multi-Role, Co-Learning, li2023camel, iAudit}, in which one agent is tasked with the primary function, such as generating code snippets or patches, while the other agent provides validation feedback, including static analysis results, test outcomes, and improvement suggestions. For example, in the INTERVENOR framework~\cite{INTERVENOR}, the code learner initiates the process by generating the initial code and subsequently engages in iterative repairs guided by the suggestions from the code teacher. 
In the \textit{multi-role} setup, the pipeline is usually a collaborative loop with iterative feedback and refinement~\cite{CodePori, shinn2023reflexion, AXNav, DroidAgent}. For example, DroidAgent~\cite{DroidAgent} designs a GUI testing loop including a Planner for task decomposition, an action module with the Actor and the Observer for execution, and a Reflector for providing task reflection and summarization to the Planner.

\headt{Star-like Structure.}
This structure is a centralized structure, where a central agent or system serves as the pivot to interact with other agents. For example, the controller agent in the RCAgent~\cite{RCAgent} framework can invoke other expert agents as a kind of tool when necessary. The commander in AutoGen~\cite{AutoGen} coordinates with the writer and the safeguard separately, to craft code and ensure safety. XUAT-Copilot~\cite{XUAT-Copilot} adopts the operation agent as the core, to receive the judgment from the inspection agent and invoke the parameter selection agent to help the action planning.
AutoDev~\cite{AutoDev} designs a scheduler agent that uses several different kinds of scheduling algorithms (\ie{} Round Robin, Token-Based, or Priority-Based) to determine the collaboration order and manner of other agents.
\junweim{MacNet}~\cite{MACNET} also tests the star topological structures.

\headt{Tree-like Structure.} 
In this structure, agents are derived gradually with the breakdown of tasks, with each agent focusing on a single task at a specific level.
For example, in SoA~\cite{SoA}, the mother agent can dynamically spawn new mother or child agents for code generation, thereby forming a tree-like collaboration structure. 
In \junweim{MacNet}~\cite{MACNET}, tree structure is one of the tested topological structures to organize the collaboration of multiple agents.

\headt{Mesh Structure.}
The mesh structure allows communication channels between agents to form a complex network, enabling them to send messages to the target agent along the network and achieve more flexible communication. For example, 3DGen~\cite{3DGen} designs an agent system based on the group chat mechanism from AutoGen~\cite{AutoGen}, which allows multiple agents (\ie{} the planner agent, 3D developer agent, and domain expert agent) to control the procedure through inter-agent conversation, \eg{} choose which agent to send messages to.
In \junweim{MacNet}~\cite{MACNET}, different agents can collaborate in a mesh structure, which means they can communicate with any other agents seamlessly.

\junwei{\headt{Performance Issues and Solutions.} The collaboration structures mentioned above may all encounter performance bottlenecks as the scale increases. Currently, most works address this issue by fixing the number of agents to match the task scale. However, for works that can dynamically derive agents~\cite{SoA, AutoAgents, AgentVerse, MAGIS}, increasing the number of agents can lead to significant performance issues. One way to address this challenge is to limit the number of derived agents. For example, AutoAgents~\cite{AutoAgents} and AgentVerse~\cite{AgentVerse}, which adopt a star-like structure, can theoretically derive an unlimited number of agents but are limited to deriving at most four agents in actual code-related tasks. In SoA~\cite{SoA}, which uses a tree structure, the depth of the tree is limited to 2.
Another solution is to limit the task scale. For example, SoA is applied to method-level code generation tasks; MAGIS derives developer agents based on the number of suspicious code files located in SWE-bench, where the average number of files to be modified is 1.7~\cite{yang2024swe}.
Moreover, a prior study~\cite{MACNET} has shown that the comprehensive performance on popular natural language understanding and code generation task benchmarks (\eg{} MMLU, HumanEval) will reach the performance saturation regardless of collaboration structure. Therefore, balancing the number of agents and the task scale is crucial for designing multi-agent systems.}

\junwei{
\subsubsection{Information Flow.}\label{sec:agent:information}
In this section, we discuss the information flow in multi-agent systems. Building on the aforementioned collaboration structures of multiple agents, we primarily focus on the most fundamental collaboration unit, which is the information exchange between two agents. Generally, there are two patterns for agents to transfer information, which are unidirectional transfer and bidirectional chat. Figure~\ref{fig:information_flow} illustrates these two communication patterns.
\begin{itemize} [leftmargin=*, label=-]
    \item \textit{Unidirectional Transfer.} This is a typical data-driven communication pattern, \ie{} the next agent will take the data produced by the preceding agent, even without being aware of the preceding agent~\cite{parsel, liu2023dynamic, INTERVENOR, AgentCoder, TGen, SoA, MapCoder, low-code-llm, multi-agent-collaboration, CORE, AutoAgents, aisd, CodePori, CodeS, AgileCoder, AutoCodeRover, RepoUnderstander, CodeR, MASAI, AUTOSD, ACFIX, ICAA, fuzz4all, AXNav, WhiteFox, XUAT-Copilot, FixAgent, LLM4DFA, SRepair, SpecRover, sami, iAudit, VisionDroid, li2024agents, GPTLENS, DEIBASE, Elicitron}. This is the most widely used pattern since it allows each agent to maintain individual context, facilitating decoupling and combination. For example, the planner or manager can focus on task decomposition, and the generated plan steps will be iteratively handed over to roles such as Developer and QA Agent, serving as their individual contexts in each round to form the top-down development pipeline~\cite{aisd, AgileCoder}.
    Depending on the implementation, the data produced by the preceding agent can be directly fed into the next agent or stored in a shared space (such as long-term memory~\cite{ChatDev, MACNET, RCAgent} or shared memory~\cite{self-collaboration, MetaGPT, MARE}) for subsequent agents to retrieve.
    \item \textit{Bidirectional Chat}. In this pattern, agents transfer information through chatting~\cite{li2023camel, olausson2024selfrepair, AutoGen, Flows, AgentVerse, Co-Learning, IterativeExperienceRefinement, MAGIS, Multi-Role, 3DGen}. With short-term memory serving as the carrier of information, these agents can share a common history of dialogues, which is particularly helpful for solving tasks through collaborative discussion.
\end{itemize}}
\junwei{
It is also worth noting that some works combine the two patterns~\cite{tang2024codeagent, ChatDev, CTC, MACNET, lcg}. For example, ChatDev~\cite{ChatDev} divides the end-to-end development tasks into different stages. Agents in different stages transfer information through the intermediate outputs, while agents in the same stage collaborate through chatting.}

\textbf{\begin{figure}[h]
    \centering
    \includegraphics[width=1.0\columnwidth]{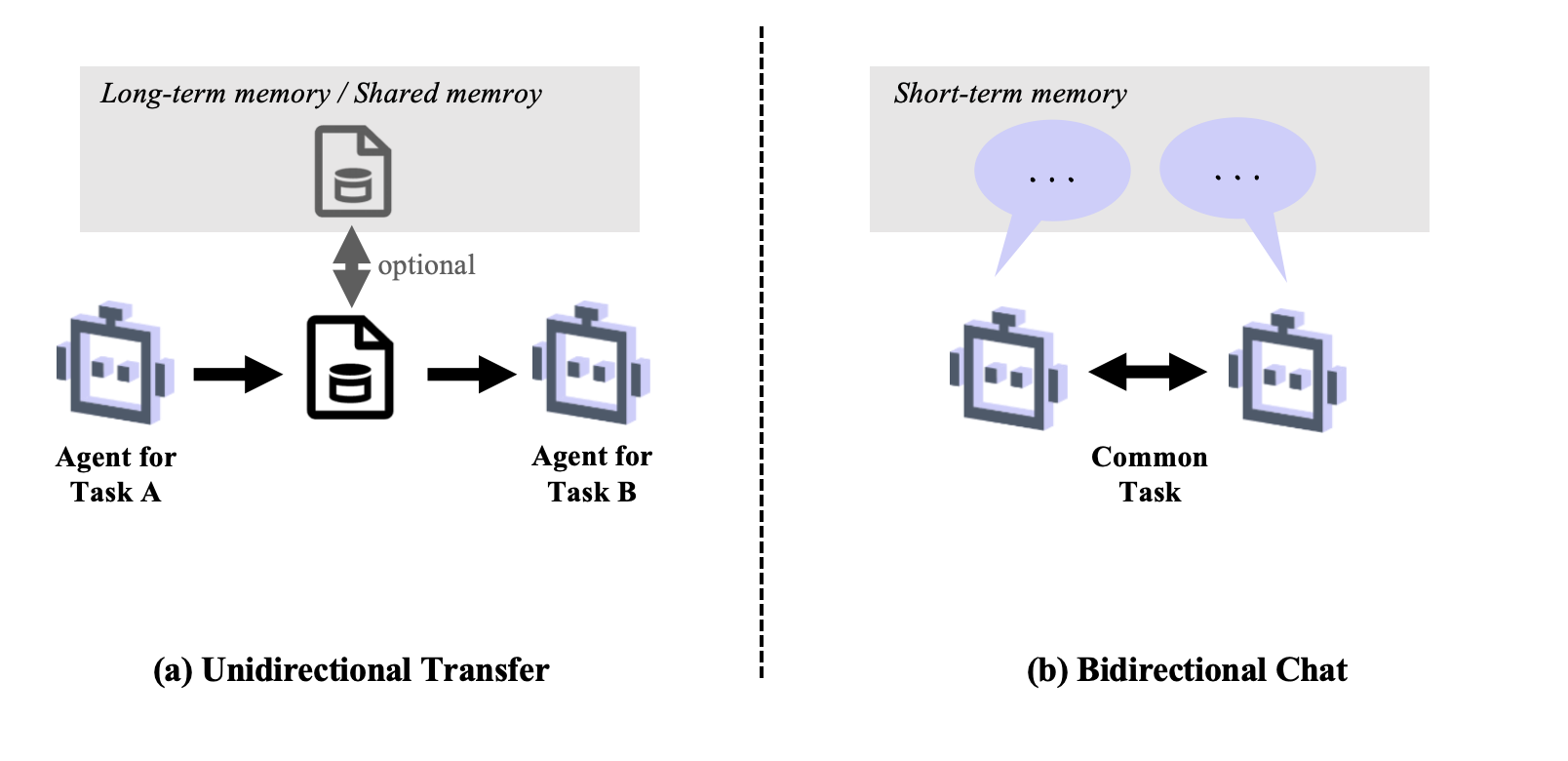}
    \caption{Two Information Flows of LLM-based Agents} 
    \label{fig:information_flow}
\end{figure}
}

\subsubsection{Real-world Applications}\label{sec:agent:application}
\junwei{In software engineering practice, there are several real-world multi-agent applications available, which are summarized in Table~\ref{tab:multiagent_apps}.} 

\begin{table}[ht]
\caption{Real-world Multi-agent Applications}~\label{tab:multiagent_apps}
\renewcommand{\arraystretch}{1.1} 
\begin{adjustbox}{width=1.0\columnwidth}
\begin{tabular}{ccccc}
\hhline
\textbf{Agent }                  & \textbf{Organization} & \textbf{Structure} & \textbf{Task}             & \makecell[c]{\textbf{Open}\\\textbf{Source}}    \\ \hhline
MultiDevin~\cite{MultiDevin}  & Cognition    & Star-like & SE tasks & $\times$        \\
Amazon Bedrock~\cite{AmazonBedrock}  & Amazon       & Star-like & Universal        & $\times$        \\
CrewAI~\cite{crewAI}          & CrewAI       & Layered   & Universal        & \checkmark      \\
Swarm~\cite{swarm}            & OpenAI       & Layered   & Universal        & \checkmark      \\ 
AgentScope~\cite{agentscope}  & Alibaba      & Layered   & Universal        & \checkmark      \\ \hhline
\end{tabular}
\end{adjustbox}
\end{table}

\junwei{Most of these products or frameworks primarily function as general-purpose multi-agent systems~\cite{AmazonBedrock, crewAI, swarm, agentscope}, with software engineering tasks as typical application scenarios. In particular, as an exclusive feature of Devin for the enterprise, MultiDevin~\cite{MultiDevin} is fundamentally engineered for software development. It enables parallel task execution using a ``manager'' Devin and up to ten ``worker'' Devins. The manager assigns tasks, integrates successful results, and merges them into a single pull request. It is ideal for independent and incremental tasks with measurable success criteria. Besides, Amazon Bedrock~\cite{AmazonBedrock} is a fully managed generative AI platform for building universal multi-agent systems. It leverages a supervisor agent and specialized sub-agents that collaborate through two core mechanisms: complex task decomposition and intelligent routing based on content-awareness.}

\junwei{Apart from these commercial products, there are several noteworthy open-source multi-agent systems. CrewAI~\cite{crewAI} allows developers to assign clear roles, goals, and backstories to individual agents using YAML files and supports both sequential and parallel task execution. Swarm~\cite{swarm} is designed with a lightweight and modular approach, focusing on exploring the basic mechanisms of multi-agent orchestration, such as task handoffs and sequential execution. AgentScope~\cite{agentscope} utilizes a message-based communication mechanism and an actor-based distributed architecture, enabling seamless migration from local to distributed environments.}
\subsection{Human-Agent Collaboration}\label{sec:agent:human}
\begin{figure}[htb]
    \centering
    \includegraphics[width=1.0\columnwidth]{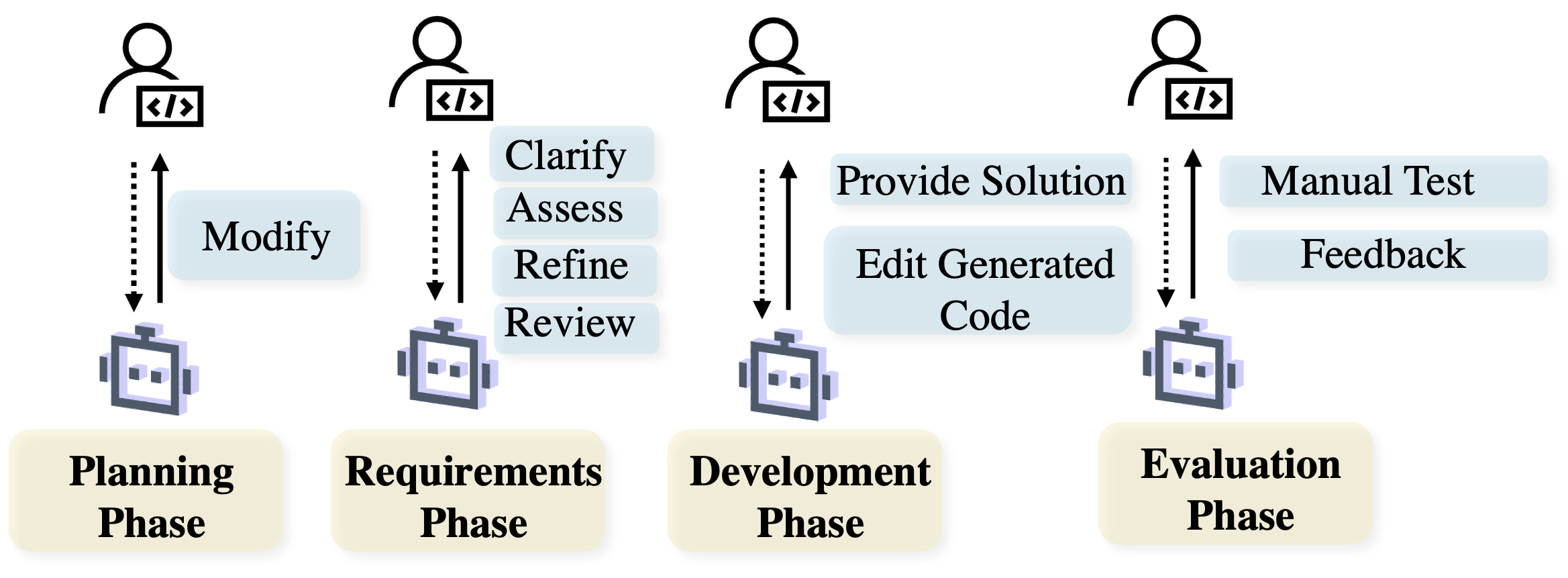}
    \caption{Human-Agent Collaboration in SE} 
    \label{fig:humanagent}
\end{figure}

While most agents aim to achieve maximum automation, where users only need to propose a request and wait for the agents to complete the task, previous studies~\cite{low-code-llm, aisd} show that LLM-based agents often encounter bottlenecks during the software development process. Therefore, some agents incorporate the human-agent cooperation paradigm to further align and enhance the agent's performance with human preferences and expertise. As summarized in Figure~\ref{fig:humanagent}, existing agents primarily include human participation in four phases: planning, requirements, development, and evaluation. 

\headt{Planning Phase.}
Some agents include human intervention in the planning stage of the agent workflow~\cite{low-code-llm, aisd, LLM4PLC, Flows}. For example, the low-code LLM platform~\cite{low-code-llm} offers a selection of predefined actions to modify auto-generated workflows, which allows the users to check and revise the workflow before execution. 
Similarly, Flows~\cite{Flows} explores the impact of having humans provide a brief natural language oracle plan during the planning phase. Experimental results show that this human-AI collaboration outperforms other AI feedback-based approaches.
However, revising the system design requires a certain level of expertise, so it is optional in some agents, such as AISD~\cite{aisd} and LLM4PLC~\cite{LLM4PLC}.

\headt{Requirements Phase.}
The initial requirements (\ie{} the task description) provided by users can be ambiguous, which can lead to a gap between the final outputs of the agent system and the user intent. Therefore, it is common for agent systems to further include manual refinement for the requirements. 
For example, ClarifyGPT~\cite{ClarifyGPT} detects ambiguous requirements through a code consistency check, generates targeted clarifying questions via an LLM, and refines the requirement based on human responses to produce the final code solution.
Similarly, AISD~\cite{aisd} enables users to assess and refine the generated use cases.

\headt{Development Phase.}
Human involvement can be included in the software development phase to correct errors introduced by the agent and guide the next steps of action.
In CodeS Extended~\cite{CodeS}, the entire code repository is generated through a three-layer sketch, allowing users to edit the generated content at each layer to ensure human involvement in refinement and optimization.
Similarly, in ART~\cite{ART}, users can enhance agent performance by offering feedback through correcting sub-step outputs(code).

\headt{Evaluation Phase.}
Human participation also serves as a post-evaluation mechanism for the outcomes produced by the agent system, which can further ensure the outputs are aligned with user intent.
For example, AISD~\cite{aisd} and Prompt Sapper~\cite{sapper} both include human intervention at the acceptance testing phase. Users can conduct manual testing of the final system, and the test reports help with the necessary refinements. 

\junwei{
\textbf{Discussion.} Overall, the exploration of human-agent collaboration in LLM-based agents within the SE domain remains limited. Existing approaches predominantly adopt an agent-centric paradigm, where human involvement is predefined as a fixed stage within the pipeline and is triggered at specific points in each iteration. For instance, in the AISD framework~\cite{aisd}, users can revise use cases, adjust plans, and validate the generated software within each development cycle, ensuring that the agent remains aligned with the original intent. Research on human-driven workflows remains an open challenge. Furthermore, current research has not explored human-agent collaboration frameworks designed for multi-person scenarios, which are common in real-world software development teams.
}

\section{Research Opportunities}~\label{sec:oppo}
This section discusses promising research directions and open problems in LLM-based agents for SE.

\textbf{Evaluation of Agents for SE.} 
Given the emergence of LLM-based agents for SE, it is crucial to develop comprehensive and rigorous evaluation frameworks, including (i) designing more diverse metrics and (ii) constructing higher-quality, more realistic benchmarks.

\textbf{\textit{Metrics.}} Current evaluations of SE agents primarily focus on their ability to solve specific tasks, such as measuring the success rate of agents on benchmarks such as SWE-bench, without delving into the intermediate states during the agent's workflow. This lack of fine-grained metrics makes it difficult to assess why agents fail in certain tasks or to what extent they fall short. Given the complexity of SE tasks, failures are common, and without deeper analysis, improving agent performance becomes challenging. 

Therefore, the design of fine-grained metrics is necessary, allowing researchers to move beyond ``black-box'' evaluations and gain insights into the agent's decision-making process and failure points.
\junwei{These fine-grained metrics have been adopted in many general-purpose agents~\cite{li2023camel, liu2024agentbench, chang2024agentboard} and embodied agents~\cite{shridhar2020alfred, shridhar2021alfworld}. For example, researchers design metrics to investigate erroneous states in agents and emphasize assessing progress rates across intermediate sub-stages rather than relying solely on final success rates~\cite{chang2024agentboard, shridhar2020alfred, shridhar2021alfworld}. 
Inspired by these works, more fine-grained metrics can also be introduced in the evaluation of SE agents, which could include error-related metrics (\eg{} the ratios of erroneous actions, the average debugging iterations, and the backtracking rate) and progress-related metrics (\eg{} the task completion rate of each phase and the overall progress rate).}
\junwei{In addition, as we discussed in Section ~\ref{sec:agent:planning} and Section~\ref{sec:agent:memory}, it is also vital to incorporate the fine-grained evaluation of individual agent modules, such as the planning module and the memory module.
Various methods and metrics have already been employed to evaluate the effectiveness of different modules in general agents~\cite{memory_survey, huang2024understanding}, \eg{} evaluating the effectiveness of the memory mechanism based on the accuracy of answering historical questions, which can serve as references for designing metrics for SE agents.}

Additionally, existing metrics heavily emphasize effectiveness, leaving trustworthy requirements such as robustness, security, and fairness underexplored. Given the flexibility and autonomy of LLM-based agents, they may exhibit unstable behavior, which can limit their practical application in real-world SE environments. Evaluating these attributes is essential for building trust in these systems.

Another critical consideration is the cost associated with these agents, particularly as they often involve lengthy workflows, frequent LLM invocations, and the management of large datasets. According to our analysis, only 46.7\% of the papers we surveyed have explicitly considered the efficiency of agents in SE tasks, incorporating quantitative analyses of time, token consumption, monetary cost, and feedback loops (\eg{} tool invocation frequency or inter-agent discussion frequency). These efficiency and computational costs are particularly important when applying agents to large-scale code repositories, complex documentation, or intricate workflows.

\textbf{\textit{Benchmarks.}} LLM-based agents significantly extend the capabilities of standalone LLMs, showing great promise in tackling more complex, end-to-end SE tasks. However, existing benchmarks used for evaluating these agents often suffer from quality issues. For instance, prior research~\cite{Agentless, swe-bench-verified} has identified that the SWE-bench benchmark includes tasks with vague or incomplete issue descriptions, reducing their relevance and applicability.

Moreover, current benchmarks are far less complex than real-world SE challenges. As outlined in Table~\ref{tab:endtoenddata}, the software generated by LLM-based agents for end-to-end development is typically small in scale (\eg{} a single function or a few files), which fails to capture the complexity of real-world projects. Furthermore, prior work \cite{swe-bench-verified} reports that most tasks in the SWE-bench benchmark (77.8\%) can be completed by an experienced engineer within an hour, further underscoring the discrepancy between benchmark tasks and real-world SE demands~\cite{rahardja2025can}.

To address these shortcomings, future research can focus on creating more realistic, high-quality benchmarks that better reflect the complexity and demands of real-world SE. These improved benchmarks will enable more accurate and meaningful evaluations of LLM-based agents' capabilities and potential.

\textbf{Human-Agent Collaboration.} 
Software development is inherently a creative process, transforming human requirements into executable software. As such, aligning agent systems with human preferences and intentions is a critical goal. While some existing agents incorporate human participation at various stages of the workflow (as discussed in Section~\ref{sec:agent:human}), there has been limited exploration of how to more thoroughly integrate human involvement throughout the entire software development life cycle. Additionally, the interaction mechanisms between agents and humans remain underexplored.

Currently, agents mainly involve humans in tasks such as requirements clarification, planning adjustments, coding assistance, or evaluation. However, extending human participation to other phases, such as architecture design, test generation, code review, and the end-to-end software maintenance process, remains largely unexplored. A deeper integration of human input across these phases could significantly enhance both the quality and adaptability of the agent’s output.

Moreover, designing effective interaction mechanisms is essential for human-agent collaboration. This includes creating user-friendly interfaces for (i) displaying relevant information, such as intermediate outputs from agents, and (ii) collecting user feedback in a streamlined way. Given the complexity of information produced during the agent's workflow, designing such interfaces presents challenges. For instance, when agents are tasked with generating or maintaining a software repository, simply presenting all code files in a flat format would be resource-intensive and inefficient. Therefore, more sophisticated methods of organizing and representing complex data are required to facilitate effective human-agent interaction.

\textbf{Perception Modality.} Most agents applied to SE tasks primarily rely on textual or visual perception. 
\junwei{However, the research on multi-modal approaches is currently insufficient. Text remains the predominant input modality, and the exploration of image inputs is still in its nascent stage, with only one work having utilized a multi-modal LLM (\ie{} VisionDroid~\cite{VisionDroid}).
This is largely because software development and maintenance activities are heavily associated with processing large volumes of textual data, such as code and documentation. }
However, there is still significant potential to explore and incorporate more diverse perception modalities into these agents.

For example, in the context of programming assistance, most LLM-powered coding agents predominantly use textual input, such as chat interfaces or Integrated Development Environment (IDE) code contexts. Alternative input formats, such as voice commands or user gestures, remain underutilized. Expanding the range of perception modalities could significantly enhance the flexibility and accessibility of coding assistants, allowing users to interact with agents in ways that better suit their individual workflows and preferences.

Furthermore, exploring diverse perception modalities may shape the future of software development and maintenance, offering new opportunities to streamline interactions and improve the efficiency of agent-driven processes.

\textbf{Applying Agents for More SE Tasks.} While existing agents have been deployed across various SE tasks, several critical phases remain underexplored. As highlighted by our analysis in Section~\ref{sec:se}, there is a lack of LLM-based agents specifically designed for tasks such as design, verification, and feature maintenance during software development and maintenance.

Developing agent systems tailored to these phases presents unique challenges. Tasks like design and verification require advanced reasoning and comprehension capabilities from the LLM-based agents, extending beyond basic code generation. These tasks demand a deeper understanding of architecture, system logic, and the ability to make informed decisions—skills that traditional LLM-controlled agents may not yet fully possess.

\textbf{Training Software-oriented LLMs for SE Agents.} LLMs are the central component controlling the ``brain'' of agent systems. Most existing agents for SE rely on LLMs trained on general-purpose data (\eg{} ChatGPT \cite{chatgpt}) or code-specific data (\eg{} Deepseek-Coder \cite{deepseekcoder} and StarCoder~\cite{lozhkov2024starcoder}). While massive code from GitHub has been leveraged to train LLMs for code, addressing complex SE tasks requires more specialized data. The reason is that software is not just about code. For example, valuable data from the whole software development life cycle, such as design, architecture, developer discussions/communications, historical code changes, and even dynamic runtime information, remain largely untapped. Incorporating such data into training could lead to the development of more powerful LLMs for software (not just for code), better suited for the unique demands of SE. These enhanced models could form the foundation for more advanced and capable agent systems designed to tackle a wider range of SE tasks.

\textbf{SE Expertise in Building Agents.} 
Incorporating well-established SE expertise into the design of agent systems is crucial. For instance, widely adopted SE techniques can be integrated as tools or sub-components of agent systems. As discussed in Section~\ref{sec:agent:action}, some existing agents already leverage SE toolkits and techniques, but many other SE tools and techniques—such as advanced debugging and testing methods—remain underutilized. Further efforts are needed to comprehensively integrate these tools and techniques into agent systems to enhance their functionality.

In addition, SE domain knowledge can guide the workflow of agent systems. As noted in Section~\ref{sec:se:development}, some agents for end-to-end software development follow traditional software process models, such as the waterfall or agile models. However, many other software process models remain unexplored. Rather than granting agents full autonomy, existing software development and maintenance methodologies can be used to partially control their workflows. For example, as revealed by the recent Agentless study \cite{Agentless} and also further confirmed by OpenAI~\cite{swe-bench-verified}, LLMs using a simplistic workflow based on traditional fault localization and program repair pipelines can even outperform other, more complex, fully autonomous agents. This suggests that leveraging domain expertise from SE can potentially help improve the effectiveness, robustness, efficiency, interpretability, and replicability of agentic solutions.

\junwei{
Another crucial direction is to leverage the quality assurance techniques in SE to build trustworthy LLM-based agents. Recently, concerns about the trustworthiness of AI have been raised~\cite{liu2023trustworthy}, including but not limited to issues of privacy~\cite{cvprGaoG000021}, security~\cite{10.1145/3735554}, fairness~\cite{ChenSHLML19}, and robustness~\cite{icse0013CLZ025}. 
Quality assurance techniques in SE (\eg{} testing and debugging) have been effectively used to evaluate and assist in addressing trustworthiness issues~\cite{tosemChenZHHS24, ahuja2020opening,ChenLZSXLLL25}, holding significant potential for building trustworthy LLM-based agents.
However, LLMs offer diverse functionalities across various domains, which necessitate task-specific designs for trustworthiness assurance. Moreover, unlike standalone LLMs, trustworthiness issues in LLM-based agents can stem from the backbone LLMs, attached modules (\eg{} memory, action), and the underlying software systems or frameworks.
These characteristics pose unique challenges for designing trustworthiness assurance techniques for LLM-based agents, requiring consideration of both individual modules and the overall system across diverse tasks.}

\junwei{\textbf{Priorities of different Research Directions.} We then discuss the priorities of the aforementioned research directions. Overall, we believe that establishing standardized and high-quality benchmarks and metrics for specific tasks should be given higher priority. On the one hand, the absence of standardized benchmarks in current works limits meaningful comparisons (especially for the end-to-end software development task); on the other hand, benchmarks play a crucial role in driving advancements in related fields~\cite{sim2003using}, as evidenced by SWE-bench, which has facilitated the development of end-to-end maintenance agents. In addition, considering the technical development in general-purpose agents~\cite{agents_fudannlp, agent_hongkongchinese, agents_Renmin}, it is feasible to explore richer human-agent collaboration patterns, more diverse perception modalities, a broader range of applicable tasks, and the integration of SE expertise into agent design. In contrast, training software-oriented LLMs for SE agents is likely a long-term endeavor, as it requires extensive data accumulation across the software engineering life cycle, which requires significant human effort and shifts in development paradigms (\eg{} documenting in more diverse formats). Additionally, it necessitates well-designed data formats and effective training methodologies. However, despite these challenges, this direction holds significant promise for realizing a general-purpose SE agent.
}

\junweim{
\section{Discussion}
\subsection{Disparity of LLM-based Agents and Standalone LLMs in SE Tasks}
LLMs serve as the central reasoning engine in LLM-based agents, enabling core tasks such as inference, analysis, and planning. Prior work has demonstrated that standalone LLMs can achieve promising results when directly applied to various SE tasks~\cite{LLM4SE1, LLM4SE2}. However, compared to LLM-based agents, standalone LLMs lack the ability to perceive environmental changes and to dynamically adjust their plans and actions, thus suffering from higher susceptibility to hallucination and non-deterministic outputs. These limitations undermine their performance in SE tasks, particularly in two dimensions: effectiveness and practicality.
}

\junweim{
From the \textbf{effectiveness} perspective, although standalone LLMs achieve satisfactory results on many single-phase SE tasks (\eg{} code generation, software testing, and program repair), their limited capabilities in handling generation hallucination and stochasticity often lead to inferior performance than LLM-based agents. For example, studies~\cite{livecodebench, zhang2024codeagent} have shown that LLM-based agents consistently achieve higher pass@1 than their foundation LLMs across code generation benchmarks of various complexity, including method-level (\eg{} HumanEval~\cite{Humaneval}), competition-level (\eg{} LiveCodeBench~\cite{livecodebench}), and repository-level (\eg{} CodeAgentBench~\cite{zhang2024codeagent}) tasks. Similar findings have also been observed in other SE tasks, such as software testing~\cite{Chattester,MuTAP}, debugging~\cite{AgentFL, AUTOFL, AUTOSD}, and static code checking~\cite{tang2024codeagent, GPTLENS}.
}

\junweim{
In terms of \textbf{practicality}, the plug-and-play nature of standalone LLMs makes them easier to adopt, but this simplicity comes at the cost of task coverage. Without the ability to interact with dynamic environments, standalone LLMs fall short in handling complex, environment-dependent tasks~\cite{LLM4SE1, LLM4SE2}, such as IT operations~\cite{mABC, D-Bot}, end-to-end software development~\cite{MetaGPT, CodeS}, and end-to-end software maintenance~\cite{yang2024swe,Agentless}. In contrast, LLM-based agents have filled this gap and been successfully applied to broader SE scenarios.
}

\junweim{
In summary, although standalone LLMs are effective for many SE tasks, their limitations in adaptability, controllability, and iterative reasoning often lead to suboptimal outcomes and narrower applications. LLM-based agents, by integrating planning, feedback, and environmental interaction, offer stronger performance and wider applicability in real-world software engineering.
}

\subsection{Threats to Validity}
\junweim{
One potential threat to the validity of our survey arises from the manual paper inspection process. Despite having two authors independently review each paper and involving a third author to resolve disagreements, the subjective judgment inherent in manual screening may still lead to relevant papers being inadvertently excluded. Such omissions could affect the comprehensiveness of our survey and potentially bias the final findings.
}

Another threats to the validity of our conclusions stem from the publication status of the collected papers.
In particular, some strategies or agent configurations are supported primarily by preprints or unpublished manuscripts, which have not undergone peer review, as follows:

\junweim{
\textbf{Multi-Agent Strategy in Requirements Engineering Lacks Sufficient Validation.}
As we mentioned in Section~\ref{sec:se:requirement}, multi-agent collaboration has emerged as a commonly adopted strategy in the requirements engineering domain. Among the four works we surveyed in this section, three employed multi-agent architectures to handle complex interactions and task decomposition ~\cite{arora2024advancing, Elicitron, MARE}. However, two works have not been published in peer-reviewed venues yet. Therefore, further empirical and peer-reviewed studies are needed to substantiate the effectiveness of multi-agent strategies in requirements engineering.
}

\junweim{
\textbf{Uncertain Efficacy of Knowledge-Enhanced Bug Detection Methods.}
Similarly, for the bug detection task described in Section~\ref{sec:se:static:bug_detection}, the use of additional knowledge from tool execution is an intriguing strategy, but it is supported by one published work~\cite{PropertyGPT} out of four~\cite{ART, LLM4Vuln, PropertyGPT, ICAA}. The approach might still be in an exploratory phase, and more future effort should be dedicated to assessing its robustness and generalizability.
}

\junweim{
\textbf{Weak Evidence for Iterative Coverage Improvements in Unit Testing.}
In the unit testing task, iterative refinement to fix compilation or execution errors and enhance fault detection is not only more widely adopted but also more frequently published, suggesting stronger community endorsement. However, iterative refinement to increase coverage is supported by fewer studies and has lower publication rates (one~\cite{CoverUp} published out of three~\cite{TELPA, AutoDev, CoverUp}). It indicates that this strategy, while promising, still requires more empirical evidence to support its effectiveness.
}

\junweim{
\textbf{Limited Validation of Visual Input in LLM-based Agents for SE.}
As discussed in Section~\ref{sec:agent:perception}, the integration of visual input into LLM-based agents for software engineering tasks has shown potential for enhancing contextual understanding and multi-modal reasoning. However, among the three works surveyed~\cite{AXNav, VisionDroid, XUAT-Copilot}, one~\cite{AXNav} has been published. The benefits and applicability of visual input in this domain remain insufficiently validated and require further systematic investigation.
}

\junweim{
\textbf{Insufficient Evidence on the Effectiveness of Specific Memory Formats.}
As outlined in Section~\ref{sec:agent:memory}, memory format plays a critical role in the coordination and long-term reasoning abilities of LLM-based agents, but the effectiveness of some memory formats has not been sufficiently validated. For example, structured messages have been adopted in MetaGPT~\cite{MetaGPT}, E\&V~\cite{EV}, and MARE~\cite{MARE}, but only MetaGPT~\cite{MetaGPT} has undergone peer-reviewed publication. Similarly, storing images in the memory has only been explored in VisionDroid~\cite{VisionDroid}, which has not been published. These gaps highlight the need for more empirical and peer-reviewed studies to substantiate the effectiveness of different memory formats in LLM-based SE agents.}

\junweim{
Overall, the uneven publication status across different tasks and strategies introduces uncertainty into our assessment of the effectiveness and generality of various LLM-agent techniques. Future research with more rigorous evaluation and wider peer-reviewed dissemination will be essential to strengthen the reliability of conclusions in this emerging field.}

\section{Conclusion}
In this paper, we have presented a comprehensive and systematic survey of \papern{} papers on LLM-based agents for SE. We analyzed the current research from both the SE and agent perspectives. From the SE perspective, we analyzed how LLM-based agents are applied across different software development and maintenance activities. From the agent perspective, we focus on the design of components in LLM-based agents for SE.  In addition, we discussed open challenges and future directions in this critical domain. 

\section*{Acknowledgement}
After drafting the initial version, we contacted the authors of the collected papers to verify the accuracy and comprehensiveness of the survey. We extend our sincere gratitude to those authors who generously provided valuable comments and feedback on the earlier draft of this paper.

\balance

\bibliographystyle{unsrt}
\bibliography{ref}

\vfill

\end{document}